\documentclass[aps,pra,10pt,twocolumn,longbibliography,nofootinbib]{revtex4-1}
\usepackage[utf8]{inputenc}
\usepackage{amsmath}
\newcommand{\eqnref}[1]{Eq.~(\ref{#1})}
\newcommand{\SO}{\mathrm{SO}}
\newcommand{\ISO}{\mathrm{ISO}}
\newcommand{\OO}{\mathrm{O}}
\newcommand{\UU}{\mathrm{U}}
\newcommand{\bZ}{\mathbb{Z}}
\newcommand{\bR}{\mathbb{R}}

\newcommand{\cH}{\mathcal{H}}
\newcommand{\cC}{\mathcal{C}}
\newcommand{\cU}{\mathcal{U}}
\newcommand{\cZ}{\mathcal{Z}}
\usepackage{amsthm}
\usepackage{amssymb}
\usepackage{tikz-cd}
\usepackage{tikz}
\usepackage{color}
\usepackage{enumerate}
\usepackage{comment}

\newtheorem{lemma}{Lemma}
\newtheorem{thm}{Theorem}

\theoremstyle{definition}

\newcommand{\ket}[1]{|#1\rangle}
\newcommand{\bra}[1]{\langle#1|}
\newcommand{\braket}[2]{\langle#1|#2\rangle}

\newcommand{\Gint}{G_{\mathrm{int}}}
\usepackage[caption=false]{subfig}

\newcommand{\Gspace}{G_{\rm spatial}}

\usepackage[colorlinks=false, linktocpage=true, linkbordercolor=blue]{hyperref}

\begin{document}
\title{Topological theory of Lieb-Schultz-Mattis theorems in quantum spin systems}
\author{Dominic V. Else}
\affiliation{Department of Physics, Massachusetts Institute of Technology,
Cambridge, MA 02139, USA}
\affiliation{Department of Physics, University of California, Santa Barbara, CA
93106, USA}

\author{Ryan Thorngren}
\affiliation{Center for Mathematical Sciences and Applications, Harvard University, Cambridge, MA 02138}
\affiliation{Department of Condensed Matter Physics, Weizmann Institute of Science, Rehovot, Israel}
\affiliation{Department of Mathematics, University of California, Berkeley, CA, 94720, USA}

\begin{abstract}
The Lieb-Schultz-Mattis (LSM) theorem states that a spin system with translation and spin rotation symmetry and half-integer spin per unit cell does not admit a gapped symmetric ground state lacking fractionalized excitations. That is, the ground state must be gapless, spontaneously break a symmetry, or be a gapped spin liquid. Thus, such systems are natural spin-liquid candidates if no ordering is found. In this work, we give a much more general criterion that determines when an LSM-type theorem holds in a spin system. For example, we consider quantum magnets with arbitrary space group symmetry and/or spin-orbit coupling.
 Our criterion is intimately connected to recent work on the general classification of topological phases with spatial symmetries and also allows for the computation of an ``anomaly'' associated with the existence of an LSM theorem. Moreover, our framework is also general enough to encompass recent works on ``SPT-LSM'' theorems where the system admits a gapped symmetric ground state without fractionalized excitations, but such a ground state must still be non-trivial in the sense of symmetry-protected topological (SPT) phases.
\end{abstract}

\maketitle

\section{Introduction}

In quantum many-body physics, there is the question of how to determine the nature of the ground state of a quantum system, given knowledge of the microscopic degrees of freedom and their Hamiltonian. Unfortunately, in general this problem is completely intractable by either analytical or numerical methods, and one is left trying to match hypotheses about the nature of the ground state to experimental observations.

In some cases, however, there exist powerful theorems that show that certain properties of the microscopic degrees of freedom (specifically, the way in which symmetries act upon them) imply highly non-trivial constraints on the nature of the ground state. An example of such a result was proven by Lieb, Schultz and Mattis (LSM) for one-dimensional systems \cite{Lieb__61}, and later generalized to higher dimensions by Oshikawa and Hastings \cite{Oshikawa_0002,Hastings_0305}. The theorem states that a system of quantum spins with translational symmetry and $\SO(3)$ spin rotation symmetry, carrying half-integer spin per unit cell, must satisfy one of the following in the thermodynamic limit: either (a) it orders at zero temperature (i.e. one of the aforementioned symmetries is spontaneously broken); or (b) the ground state is gapless; or (c) the ground state has non-trivial degeneracy in the torus. In language usual to the study of quantum magnetism, we can say that possibility (b) corresponds to the ground state being a gapless quantum spin liquid, and (c) corresponds to the ground state being a topological quantum spin liquid (with fractional excitations such as anyons). The possibility of the system being completely trivial at zero temperature, with neither spontaneous symmetry breaking nor spin liquid character, is thus disallowed.

The manifest power and utility of this theorem raises the question of when we expect similar results to hold more generally, a question which has been examined from various points of view
\cite{Affleck__86,Affleck__88,
Oshikawa_9911,Oshikawa_0002,
Parameswaran_1212,Roy_1212,
Zaletel_1410,
Watanabe_1505,Cheng_1511,Hsieh_1604,
Sulejmanpasic_1608,
Po_1703,Cho_1705,
Jian_1705,Lu_1705_04691,Huang_1705,Komargodski_1706,
Metlitski_1707,Yang_1705,Watanabe_1802,
Sulejmanpasic_1802,
Cheng_1804,Shiozaki_1810,Song_1810,
Yao_1906}. One particularly intriguing point of view is based on a connection with the theory of topological phases with symmetries \cite{Cheng_1511,Cho_1705,Jian_1705,Metlitski_1707}: one can think of a system with half-integer spin per unit cell as corresponding to the \emph{boundary} of a symmetry-protected topological (SPT) phase in one higher dimension, protected by the symmetries under consideration, namely translation symmetry and $\SO(3)$. Another closely related point of view is based on anomalies: the half-integer spin per unit cell somehow implies that the low-energy field theory describing the system must be ``anomalous'' in a certain sense. This prevents the low-energy physics from being completely trivial but also implies stronger constraints than the original LSM-Oshikawa-Hastings formulation; even a non-trivial spin liquid ground state must have the correct anomaly as dictated by the microscopic symmetry action, which places non-trivial constraints on \emph{which} spin liquid ground states are allowed \cite{Zaletel_1410,Cheng_1511} (a question on which the original LSM-Oshikawa-Hastings result was silent).

These points of view, although highly suggestive, have not so far been fully developed, mainly because theorems of LSM type always seem to involve \emph{spatial} symmetries (for example, translation symmetry in the original LSM theorem; one can also consider other spatial symmetries such as rotations, reflections, etc.), and the theory of topological phases with spatial symmetries has not historically been well understood. Nevertheless, recently a systematic theory of topological phases with spatial symmetries has emerged \cite{Song_1604,Thorngren_1612,Huang_1705,Else_1810,Song_1810}. This raises the possibility that the ``topological'' point of view on LSM-type results can now be formalized and put on a completely general footing.

In this paper, we realize this possibility. In particular, we formulate the general criterion for a quantum spin system, with a general symmetry group and representation of that symmetry on the microscopic degrees of freedom, to have a result of LSM type. We do this by precisely characterizing the anomaly that results from the microscopic symmetry action, which then implies constraints on the allowed ground states.

As a concrete application, we specialize to symmetries of particular relevance to quantum magnetism: specifically we consider the symmetry group of a crystalline quantum magnet in two or three spatial dimensions, with or without time-reversal symmetry breaking, and with or without spin-orbit coupling. In each of these cases, we find that the criterion for when a result of LSM type holds can be reduced to a simple geometrical criterion on the arrangement of spins in the unit cell, called ``lattice homotopy'', which generalizes the criterion of the same name that was conjectured (and proven in certain cases) for systems without spin-orbit coupling in Ref.~\cite{Po_1703}.

The outline of this paper is as follows. In Section \ref{sec:symrep}, we discuss a general form for the symmetry representations we intend to consider, give some examples, and introduce the notion of ``lattice homotopy'' generalizing Ref.~\cite{Po_1703}. In Section \ref{sec:defectnetworks}, we review the ``defect networks'' of Ref.~\cite{Else_1810}, and then show in Section \ref{sec:anomalymatching} that defect networks lead to an appealing physical picture for LSM-type results. In Section \ref{sec:equivarianthomology}, we translate these considerations into a concrete and computable mathematical criterion for LSM-type results based on an object called equivariant homology. In Section \ref{sec:exhaustive}, we present our results from performing exhaustive computational searches in many cases of interest for quantum magnetism, concluding that lattice homotopy completely captures the LSM criterion in these cases. In Section \ref{s:equiv-pushforward} we present some equivariant homology computations of the LSM anomaly associated to translation and point group symmetries. In Section \ref{sec:nontraditional_lsm}, we discuss how our framework also encompasses ``SPT-LSM'' theorems in which the ground state, if gapped and symmetric, is constrained to at least be in a non-trivial symmetry-protected topological (SPT) phase. In Section \ref{sec:rigorous}, we give a rigorous proof of a special case of our LSM criterion in two dimensions. Finally, in Section \ref{sec:discussion} we discuss directions for future work.

\tableofcontents

\section{Anomalous textures}
\label{sec:symrep}
Results of LSM type arise in situations where the symmetry acts projectively on sites. For example, in the original LSM result, sites carrying half-integer spin correspond to projective representations of the spin rotation group $\SO(3)$. In this section, we describe a general form of a symmetry action, and introduce an object which we call an ``anomalous texture'' to describe the projective action on sites.
(The reader who is not so interested in general formalism can skip ahead to Section \ref{subsec:texture_examples} where we discuss concrete examples of the kind of symmetries we intend to consider).

We consider a symmetry group $G$ with an associated action on $d$-dimensional space, described by a homomorphism $G \to \mathrm{ISO}(d)$, where $\mathrm{ISO}(d)$ is the set of isometries of $d$-dimensional space $\bR^d$ (this group is generated by translations, rotations, and reflections). We assume the system is composed of a collection of spins indexed by a set $\Lambda \subset \bR^d$ that is invariant under the action of $G$ (we will refer to $\Lambda$ as ``the lattice"). Its Hilbert space can be represented as a tensor product $\cH = \bigotimes_s \cH_s$, where the product is over all spins, and $\cH_s$ is the local Hilbert space of spin $s$.

 We assume that the symmetry acts on the Hilbert space by a representation of the form
\begin{equation}
\label{the_symmetry_action}
U(g) = \left(\bigotimes_s V_s(g)\right) S(g) K^{p(g)}
\end{equation}
where $V_s(g)$ is an on-site unitary acting on the spin states $\cH_s$; $S(g)$ is a unitary that acts on the whole Hilbert space by permuting the spins in accordance with the spatial action of the symmetry; and $K$ is the anti-unitary complex conjugation operator, with $\mu(g) = 0$ or $1$ depending on whether $g$ acts unitarily or anti-unitarily. Note that unitary symmetries that are orientation-reversing in space, such as reflection, still correspond to $\mu(g) = 0$.

In order for $U(g)$ to be a linear $G$-representation, i.e. $U(g_1) U(g_2) = U(g_1 g_2)$, some consistency conditions have to be satisfied. Firstly, we must have $\mu(g_1 g_2) = \mu(g_1) + \mu(g_2)$ [$\operatorname{mod} 2$]. Secondly, we must have $S(g_1 g_2) = S(g_1) S(g_2)$. Lastly, $V_s(g)$ satisfies a condition we find by computing
\begin{widetext}
\begin{align}
U(g_1) U(g_2) &= \left( \bigotimes_s V_s(g_1) \right) S(g_1) K^{\mu(g_1)} \left( \prod_s V_s(g_2) \right) S(g_2) K^{\mu(g_2)} \\
              &= \left( \bigotimes_s V_s(g_1) \right) S(g_1) \left( \bigotimes_s V_s^{*\mu(g_1)}(g_2) \right) S(g_1)^{-1} S(g_1 g_2) K^{\mu(g_1 g_2)} \\
              &= \left( \bigotimes_s \left[ V_s(g_1) V_{g_1^{-1} s}(g_2) \right] \right) S(g_1 g_2) K^{\mu(g_1 g_2)}
\end{align}
where $V^{*\mu} = V^*$ (complex conjugate) if $\mu = 1$ and $V$ if $p=0$. On the other hand, we have
\begin{align}
U(g_1 g_2) = \left( \bigotimes_s V_s(g_1 g_2) \right) S(g_1 g_2) K^{\mu(g_1 g_2)}.
\end{align}
\end{widetext}
We therefore conclude that
\begin{align}
V_s(g_1 g_2) = \omega_s(g_1, g_2) V_s(g_1) V^{*\mu(g_1)}_{g_1^{-1}s}(g_2)
\end{align}
for some phase factor $\omega_s(g_1, g_2)$.

The phase factor $\omega_s(g_1, g_2)$ is a spatially-dependent generalization of a group 2-cocycle; the latter characterizes projective representations and appears in physics to classify (1+1)-D bosonic symmetry-protected topological (SPT) phases \cite{Chen_1008,Schuch_1010}. Let us note that by expanding $V_s(g_1 g_2 g_3)$ in two different ways, $\omega_s$ must satisfy the associativity condition
\begin{equation}
\label{an_associativity}
\omega_{g_1^{-1} s}^{\sigma(g_1)} (g_2, g_3) \omega_s(g_1, g_2 g_3) = \omega_s(g_1, g_2) \omega_s(g_1 g_2, g_3),
\end{equation}
where we introduced the notation $\sigma(g) = (-1)^{\mu(g)}$.
Moreover, we had some gauge freedom in defining the operators $V_s(g)$ in the first place. We are free to choose phase factors $\beta_s(g)$ for each $g \in G$ and spin $s$, and then redefine $V_s(g) \to \beta_s(g) V_s(g)$. This has the following effect on $\omega_s(g_1, g_2)$:
\begin{equation}
\label{an_coboundary}
\omega_s(g_1, g_2) \to \omega_s(g_1, g_2) \frac{\beta_{g_1^{-1} s}^{\sigma(g_1)} (g_2) \beta_s(g_1)}{\beta_s(g_1, g_2)}.
\end{equation}

We call a lattice of sites $\Lambda$, equipped with an equivalence class of $\omega_s$ satisfying \eqnref{an_associativity}, subject to the equivalence relation \eqnref{an_coboundary}, an \emph{anomalous texture} on $\Lambda$. Under multiplication of $\omega_s$, anomalous textures on $\Lambda$ form an abelian group which we will call $H_{-2}^G(\Lambda, \UU(1))$, or equivalently $H^2_G(\Lambda,\UU(1))$\footnote{Throughout this paper, $\UU(1)$ as a coefficient group will always come equipped with a $G$-action, with the anti-unitary elements of $G$ -- but \emph{not} spatially orientation-reversing symmetries like reflection, unless they are also explicitly anti-unitary -- acting non-trivially. In cases where different $G$-actions are used, we note them explicitly.}. The reason behind these notations will become clear later.
Physically the group structure corresponds to stacking of Hilbert spaces and tensor product of symmetry operators $U(g)$.

%We note also that we can define an integer function of three group elements by
%\begin{equation}2\pi i \eta_s(g_1,g_2,g_3) = \log \omega_{g_1^{-1} s}^{\sigma(g_1)}(g_2,g_3) + \log \omega(g_1,g_2g_3)\end{equation}\begin{equation} - \log \omega(g_1,g_2) - \log \omega(g_1 g_2,g_3).\end{equation}
%When $G$ contains continuous group factors, this quantity is often more useful than $\omega_s$. By construction it satisfies a pentagon identity involving four group elements and a gauge freedom . \de{The above notation doesn't make any sense. Were you going for something else?} \de{Maybe better to introduce the integer-valued cochain at the first place where we actually use it?}

Finally, we note a useful result about anomalous textures. For every site $s$, let $G_s$ be the subgroup of $G$ that leaves $s$ invariant. We will refer to this as the ``isotropy group'' of the site. Then the on-site operators $V_s(g)$ define a projective representation of $G_s$, which defines a class in group cohomology $\cH^2(G_s, \mathrm{U}(1))$. A representative 2-cocycle can be obtained from $\omega_s$ as defined above by restricting it to $G_s$. Moreover, suppose $s$ and $s'$ are two different sites that are symmetry-related, that is there exists $g_* \in G$ such that $g_* s = s'$. Then from the associativity condition on $\omega$ one can show that the classes in $\cH^2(G_s, \UU(1))$ and $\cH^2(G_{s'}, \UU(1))$ are related by the isomorphism $\cH^2(G_s', \UU(1)) \to \cH^2(G_s, \UU(1))$ induced by the group map $G_s \to G_{s'}, g \mapsto g_* g g_*^{-1}$ (this isomorphism in group cohomology does not depend on the choice of $g_*$). Roughly speaking, this is just saying that the on-site projective action needs to be invariant under the whole action of the spatial symmetry group. Furthermore, one can show (see Appendix \ref{appendix:anomtexts}) that $H_{-2}^G(\Lambda, \UU(1))$ is in one-to-one correspondence with the sets of allowed data $[\omega_s] \in \cH^2(G_s, \UU(1))$ for each site $s$, subject to the condition of $G$ symmetry just mentioned. That is, anomalous textures on a lattice of sites $\Lambda$ just correspond to consistent assignments of on projective representations of the isotropy group at every site. Further, one needs only keep track of the projective representation at one site in each orbit. All the other projective representations are determined by symmetry.

\subsection{Examples}
\label{subsec:texture_examples}
Let us discuss some examples that we will come back to in the course of the paper:

\begin{enumerate}
\item \textbf{Quantum paramagnet without spin-orbit coupling}. For example, the Hamiltonian could be a Heisenberg interaction
\begin{equation}
\label{quantum_magnet}
H = \sum_{a,b} J_{a,b} \mathbf{S}_a \cdot \textbf{S}_b,
\end{equation}
where $S_a^{\alpha}$, $\alpha = x,y,z$, are the spin operators at position $a$, and the couplings $J_{a,b}$ respect the spatial symmetries of the lattice. This Hamiltonian has a symmetry $G = \SO(3) \times \bZ_2^T \times \Gspace$ or $\SO(3) \times \Gspace$ (depending on whether time-reversal symmetry $\bZ_2^T$ is broken), where $\SO(3)$ is the internal spin rotation symmetry, and $\Gspace$ is the discrete spatial symmetry of the lattice. The representation $U(g)$ is generated by the action of $\SO(3)$ on spins and by the permutation of spins under $\Gspace$. The on-site symmetry representations are linear $\SO(3)$ representations for integer spin, and projective representations for half-integer spins; this corresponds to the class $[\omega_s]$ for a site $s$ being either the trivial or non-trivial element of $\cH^2(\SO(3),\UU(1)) = \bZ_2$. With time reversal the half-integer spin representations are Kramers doublets with $\mathbb{T}^2  = -1$, so we obtain the diagonal element of $\cH^2(\SO(3) \times \bZ_2^T,\UU(1)) = \bZ_2 \times \bZ_2$.

\item \textbf{Exotically ordered quantum magnet}.
We can also consider spin systems where the $\SO(3)$ spin rotation symmetry is broken (either spontaneously or explicitly) down to some subgroup. In order to have non-trivial LSM, what we will need is that the integer and half-integer representations remain distinct. So ferromagnetic or antiferromagnetic order will not be sufficient (in that case $\SO(3)$ gets broken down to $\SO(2)$, which has no non-trivial projective representations), but, for example, we could consider spin-nematic order, where $\SO(3)$ is broken down to $\OO(2)$ (an out-of-plane $\pi$-rotation is preserved) [or if time-reversal is also present, then the symmetry is $\OO(2) \times \bZ_2^T$]. The projective representations are captured in group cohomology by the symmetry-reduction maps (in this case isomorphisms)
\begin{equation}
  \cH^2(\SO(3),\UU(1)) \to \cH^2(\OO(2),\UU(1)) = \bZ_2
\end{equation}
\begin{equation}
\cH^2(\SO(3) \times \bZ_2^T,\UU(1))\end{equation}\begin{equation} \to \cH^2(\OO(2) \times \bZ_2^T, \UU(1)) = \mathbb{Z}_2 \times \mathbb{Z}_2.
\end{equation}
Note that as for $\OO(2) \times \bZ_2^T$, half integer spins only realize the diagonal element of the latter group (which is projective under both $\OO(2)$ and $\bZ_2^T$).

\item \textbf{Quantum paramagnet with spin-orbit coupling}.
The Hamiltonian \eqnref{quantum_magnet} can be obtained as the effective theory for the spin degrees of freedom of a Mott insulator with one electron on each lattice site. However, suppose that the underlying electrons have spin-orbit coupling. Then the internal spin $\SO(3)$ symmetry is broken, and in general the Hamiltonian will have less symmetry than \eqnref{quantum_magnet}.
Instead, the Hamiltonian will be of the form
\begin{equation}
\label{quantum_magnet_spin_orbit}
H = \sum_{a,b} \textbf{S}_a \cdot \mathrm{J}_{ab} \cdot \textbf{S}_b,
\end{equation}
where for each $a,b$, $\mathrm{J}_{ab}$ is a $3 \times 3$ matrix.

However, we know that the laws of physics are invariant under the Euclidean group $\ISO(3)$, provided that the spins of the electrons also transform. $\ISO(3)$ is spontaneously broken to a discrete subgroup $G_s$ in a crystalline solid, so we conclude that the couplings $J_{ab}$ must be such that $H$ is still be invariant under $G_s$. Thus, the symmetry group in this case is $G = \Gspace$ or $\Gspace \times \bZ_2^T$ (depending on whether or not time-reversal symmetry is broken). We emphasize that in this case, the spatial symmetries must be taken to have an internal action on the spin degrees of freedom rather than just permuting them. For half-integer spins this internal action can lead to non-trivial projective representations of the site symmetry groups $G_s$, with classes in $\cH^2(G_s,\UU(1))$. (However, in the case where time-reversal symmetry is broken, this only occurs at sites $s$ where the site symmetry group $G_s$ remains large enough.)

\end{enumerate}

Later (see Section \ref{sec:nontraditional_lsm}), we will also consider some more exotic symmetries where the full symmetry group is a non-trivial extension of a space group by an internal symmetry.

 \subsection{Lattice homotopy}
 \label{subsec:latticehomotopy}
We define a simple equivalence relation for anomalous textures, which we call \emph{lattice homotopy}, generalizing notions in \cite{Po_1703,Song_1604,Huang_1705,Else_1809}. Recall that an anomalous texture is defined by a collection of sites $s \in \Lambda$ indexing the phase factors $\omega_s(g_1, g_2)$, with a $G$ action on $\Lambda$. We think of $\Lambda$ as an abstract set mapped into physical space $X =\mathbb{R}^d$ by a map $f : \Lambda \to X$, which is required to be equivariant with respect to the $G$ action on $\Lambda$ and on $X$, meaning $f(g\cdot s) = g \cdot f(s)$. This map may be considered part of the data of the anomalous texture.

Two anomalous textures are \emph{lattice homotopy equivalent} if they are related by a series of the following simple equivalences:
\begin{enumerate}
\item If two sites $s,s'$ sit on top of each other, that is they map to the same point in $X$ under $f$, then we can combine them into a single point $s''$ and add their corresponding phase factor data, that is, $\omega_{s''}(g_1, g_2) = \omega_{s}(g_1, g_2) \omega_{s'}(g_1, g_2)$.
\item If a site carries trivial phase factor data, i.e.\ $\omega_s(g_1,g_2) = 1$ for all $g_1, g_2 \in G$, then $s$ can be removed from $\Lambda$.
\item If two anomalous textures are related by symmetric deformation of the locations of the sites, then they are equivalent. In other words, if $h:[0,1] \times \Lambda$ is a $G$-equivariant continuous map, ie. $h(t,g\cdot s) = g \cdot h(t,s)$, then the anomalous textures defined by $f = h(0,-)$ and $f' = h(1,-)$ are equivalent. $h$ is a homotopy between the maps $f, f'$, hence the name lattice homotopy.
\end{enumerate}

\begin{figure}
\includegraphics{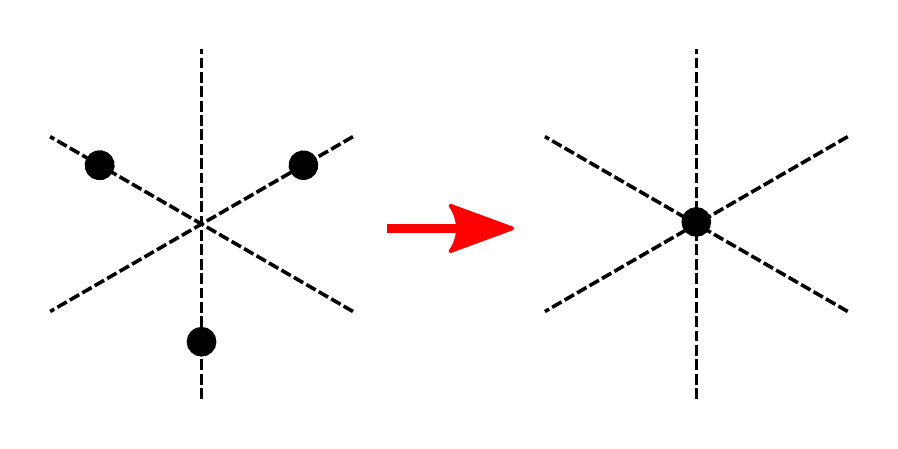}
\caption{\label{fig:fusion} An example of a fusion move. The overall symmetry group is $G = G_{\mathrm{space}} \times G_{\mathrm{int}}$, for some internal symmetry group $G_{\mathrm{int}}$, and where $G_{\mathrm{space}}$ is generated by a three-fold rotation and a reflection (i.e. $G_{\mathrm{space}} = D_3$). The site symmetry group $G_s$ for each site on the left-hand side is $\mathbb{Z}_2 \times G_{\mathrm{int}}$. Under lattice homotopy, one can fuse a $G$ orbit comprising three of these points into a single point whose site symmetry group is enlarged to $G$. We need a fusion rule giving a map $\cH^2(\mathbb{Z}_2 \times G_{\mathrm{int}}, \mathrm{U}(1)) \to \cH^2(G, \mathrm{U}(1)))$ to describe the impact of such a fusion on the anomalous texture.}
\end{figure}

The lattice homotopy equivalence relations can also be stated in a more concrete way in terms of the local data. Recall that in Section \ref{sec:symrep} we stated that the local class $[\omega_s] \in \cH^2(G_s, \mathrm{U}(1))$ for all sites $s$ is sufficient to determine the anomalous texture up to gauge freedom. Items 1 and 2 in the list above apply in the obvious way to this description of the anomalous texture. Item 3 is a bit trickier, because as we deform the locations of sites, their site symmetry groups $G_s$ can change. In particular, we need to consider \emph{fusion moves} where a collection of symmetry-related sites fuses into a single site with a larger site symmetry group \cite{Song_1604,Huang_1705,Else_1809}; an example is shown in Figure \ref{fig:fusion}. To derive the fusion rule of such a move, let $G_*$ be the enlarged site symmetry group after fusion, and let $S$ be the set of sites before fusion that are going to fuse into a single site. Then we can treat the local data on each site as defining an anomalous texture just on $S$, with symmetry $G_*$. As we know, an equivalent description of such an anomalous texture is in terms of a map $\omega : S \times G_* \times G_* \to \UU(1)$ satisfying \eqnref{an_associativity}. Then we can define an element of $\cH^2(G_*, \UU(1))$ to describe the result of the fusion, through the 2-cocycle
\begin{equation}
\omega(g_1, g_2) = \prod_s \omega_s(g_1, g_2).
\end{equation}

Now let us discuss the significance of lattice homotopy for LSM results. If we allow ourselves to add additional degrees of freedom (transforming linearly under the symmetry so that they do not affect the anomalous texture), then it is easy to see that a trivial anomalous texture, ie. one with $\omega_s = 1$ for all $s$, admits a trivial symmetric gapped ground state, in fact a product state [Such a product state ground state is not always possible if we do not add degrees of freedom; for example, consider a system built out of $S=1$ spins with $\mathrm{SO}(3)$ symmetry.] Moreover, if we again allow adding additional degrees of freedom without changing anomalous texture, then lattice homotopy equivalence can always be implemented simply by moving degrees of freedom around symmetrically. Thus, we obtain a trivial gapped symmetric ground state for any anomalous texture that is lattice homotopy equivalent to the trivial one. The case where we do not allow ourselves to add degrees of freedom is more difficult, but generally one expects that in the case of an anomalous texture that is trivial in lattice homotopy equivalence, a trivial gapped symmetric ground state can be constructed as some kind of tensor network \cite{Po_1703}.

The question now is, if an anomalous texture is \emph{not} trivial in lattice homotopy, does it necessarily lead to an LSM result? If we define an ``LSM result'' to be the statement that any symmetric gapped ground state, if it exists, is topologically ordered, we will see that the answer turns out to be yes in many cases of physical interest (see Section \ref{sec:exhaustive}), but no in general. However, see also Section \ref{sec:nontraditional_lsm} where we discuss other kinds of LSM-type results.

Finally, let us mention a simplification that occurs in many cases of physical interest. We consider a symmetry group of the form $G = G_{\mathrm{space}} \times G_{\mathrm{int}}$, where the $G_{\mathrm{int}}$ symmetry acts internally, and only the $G_{\mathrm{int}}$ acts projectively on sites. In other words, if we we decompose the classifying group for the projective representation on site $s$ as
\begin{align}
\cH^2(G_s, \UU(1))
 &= \cH^2(G_{\mathrm{space},s} \times G_{\mathrm{int}}, \UU(1))
 \\ &= \cH^2(G_{\mathrm{space,s}},\UU(1)) \\&\quad\times \cH^1(G_{\mathrm{space},s}, \cH^1(G_{\mathrm{int}}, \UU(1))) \\&\quad\times \cH^2(G_{\mathrm{int}}, \UU(1)),
\end{align}
where $G_{\mathrm{space},s}$ is the subgroup of $G_{\mathrm{space}}$ that leaves $s$ invariant, then we consider only anomalous textures resulting from the last factor. This is the relevant case for quantum magnets without spin-orbit coupling, where $G_{\mathrm{int}}$ is the spin-rotation symmetry $\SO(3)$ or time-reversal symmetry $\mathbb{Z}_2^T$.

In this case, let $P := \cH^2(G_{\mathrm{int}}, \UU(1))$ (usually $P = \mathbb{Z}_2$ for quantum magnets, which keeps track of whether the spin at site $s$ is integer or half-integer). Then an anomalous texture on a collection of sites $\Lambda$ with embedding $f : \Lambda \to X$ is just an assignment of an element $\omega \in P$ for all $s \in \Lambda$. Moreover, the lattice homotopy equivalence relations can be stated very simply:
\begin{enumerate}
\item If two sites $s,s'$ sit on top of each other, that is they map to the same point in $X$ under $f$, then we can combine them into a single point $s''$ and add their corresponding elements of $P$ using the group law for $\cH^2(G_{\rm int},\UU(1))$, corresponding to tensor product of projective representations.
\item If a site carries the trivial element of $P$, then it can be removed, since it carries a linear representation of $G_{\rm int}$.
\item If two anomalous textures are related by symmetric deformation of the locations of the sites, without changing any of the $P$-labels of the sites, then they are equivalent.
\end{enumerate}
Such an equivalence relation on anomalous textures has previously been considered in Ref.~\cite{Po_1703}. It leads to a purely geometric way to determine when there is an LSM result (in cases where this is determined by lattice homotopy equivalence), as discussed there.

\subsection{Fermionic anomalous textures}

Although we will mostly talk about bosonic systems (e.g. spins) in this paper, for completeness we also introduce anomalous textures in fermionic systems \cite{Hsieh_1604,Lu_1705_04691,Cheng_1804}. For each site $s$, we indicate by $G_{s,b} = G_{s,f}/\mathbb{Z}_2^f$ to be the bosonic isotropy group at $s$, where $G_{s,f}$ is the subgroup of the full symmetry $G^f$ (which contains fermion parity $\mathbb{Z}_2^f$) that leaves the site $s$ invariant. The anomalous texture contains the data of projective action of $G_{s,b}$ on $\cH_s$, whether any elements of $G_{s,b}$ anti-commute with fermion parity, as well as whether $s$ carries an odd number of Majorana modes. This is the same data as a 1d fermionic SPT with internal symmetry $G_s$, where $G_s$ contains fermion parity.
%As before, one can prove that the collection of this data for each orbit $[s]$ of sites uniquely determines the anomalous texture. See Appendix \ref{appendix:anomtexts}.

\section{Defect networks: review}
\label{sec:defectnetworks}

\subsection{Defect networks}
\label{subsec:defectnetworks}

In Refs.~\cite{Song_1604,Huang_1705,Else_1810,Song_1810}, a general picture of crystalline topological phases emerged based on so-called \emph{defect networks}. The starting point is a space $X$ with an action of a group $G$. For example, in the usual case of an infinite crystal, $X = \mathbb{R}^d$ and $G$ acts on $X$ by Euclidean isometries such as translation, reflection, rotation, etc. We choose a cell decomposition (for example a triangulation) of $X$, such that $G$ maps cells to cells. Furthermore, we require that for each (open) cell $\sigma$, if $g \in G$ fixes any point in $\sigma$, it fixes all of $\sigma$. The group of such elements is called the isotropy group of $\sigma$, denoted $G_\sigma$.

\begin{figure}
\includegraphics[width=7cm]{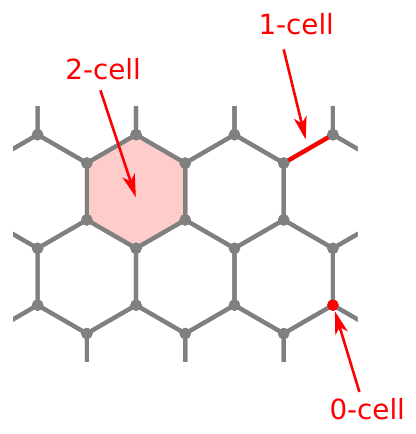}
\caption{\label{fig:cellstructure}A cell decomposition of the plane. In an invertible-defect network, 2-cells carry a 2-dimensional topological phase, 1-cells carry invertible gapped interfaces between topological phases, and 0-cells carry invertible gapped junctions between interfaces.}
\end{figure}

The idea of a defect network is that to each $d$-dimensional cell $\sigma_d$, we assign the data of a $d$-dimensional $G_{\sigma_d}$-symmetric topological phase of matter (either SPT or SET). Then, on each $d-1$-dimensional cell $\sigma_{d-1}$, we assign the data of a $G_{\sigma_{d-1}}$-symmetric interface between the $d$-dimensional phases carried on the adjoining $d$-cells. Then, on each $d-2$-dimensional cell $\sigma_{d-2}$, we assign the data of a $G_{\sigma_{d-2}}$-symmetric junction between the abutting interfaces; and so on, until we get down to 0-cells (see Figure \ref{fig:cellstructure}). There is a basic consistency condition, which states the resulting state needs to be symmetric under the whole symmetry group $G$. We already got part of the way there by requiring that the  phase/interface/junction/etc on each cell is invariant under $G_\sigma$, but since an element $g \in G$ not in $G_\sigma$ will permute the cell $\sigma$ into another cell $g\sigma$ we also require that the resulting data of a $G_{g\sigma} = g G_\sigma g^{-1}$ is related by the isomorphism $G_{\sigma} \to G_{g\sigma}$ given by $h \mapsto g h g^{-1}$ .

We can distinguish between different classes of interface defect networks. A defect network is called an \emph{invertible-defect network} if the interfaces on all the $k$-cells for $k < d$ are invertible (although the gapped phases on $d$ cells are not required to be invertible), meaning that every interface has an inverse interface such that when the interface is brought close to its inverse, they together are equivalent by a local unitary to the trivial interface. In this work, we will only ever discuss invertible-defect networks, which are already believed to be sufficient to classify ``liquid'' (eg. not fractonic) topological phases with spatial symmteries, and in fact when we say ``defect network'' without qualification, we will mean an invertible-defect network.

An invertible-defect network is further called an \emph{invertible-substrate defect network} if the top-dimensional data, i.e.\ the phase carried on $d$-cells is itself invertible (for example an SPT or a $p+ip$ superconductor, but not a phase with fractional excitations). An invertible-substrate defect network describes a crystalline topological phase which, if we forget about the symmetries, is either trivial (i.e. with symmetries it is a crystalline SPT) or an invertible topological phase. In any case it is short-range entangled (according to the definition of Kitaev \cite{KitaevIPAM}). Note that an LSM theorem is precisely the statement that no invertible topological phase symmetric ground states are allowed.

Finally, we will call a defect network \emph{$k$-skeletal} if it carries trivial data on all cells of dimension $>k$. Note that since invertible gapped $k$-dimensional interfaces between trivial $k+1$-dimensional phases are equivalent to $k$-dimensional invertible phases, for a $k$-skeletal defect network, the data on a $k$-dimensional cell $\sigma$ is always an invertible $G_\sigma$-symmetric topological phase. Moreover, since the trivial phase is certainly invertible, a $k$-skeletal defect network (for $k < d$) is always an invertible-substrate defect network.

There is an equivalence relation on defect networks described in \cite{Else_1810}. Two defect networks are equivalent if they are related by symmetric motions of the defects as well as fusion/splitting processes. Equivalently, we say they are equivalent if they are related by pumping processes applied on each cell (called ``bubble equivalences'' in Ref.~\cite{Song_1810}).

\subsection{Anomalous defect networks and anomalous textures}\label{sec:defnetworktexture}
In Refs.~\cite{Else_1810,Song_1810}, it was highlighted that a defect network can be \emph{anomalous}, meaning that it cannot be realized in a gapped system with a non-degenerate ground state, eg. as a crystalline SPT. These anomalies can be associated with a region of some dimension $r < d$. For example, a symmetry-breaking domain wall in an SPT phase typically defines an anomalous defect network with anomaly along the wall.

We define a degree-$r$ anomalous defect network as a specified configuration of interfaces on $k$-cells for all $k > r$, in which there exists an $r$-cell $\sigma$ for which no gapped symmetry-preserving junction exists between the interfaces adjoining $\sigma$. In an anomalous defect network, all the cells of dimension $\leq r$ are therefore left unspecified.

For invertible-substrate defect networks, we argued in Ref.~\onlinecite{Else_1810} that such anomalies are classified by the same data that classifies usual symmetry anomalies for the isotropy group $G_\sigma$ of the $r$-cell treated as an internal symmetry, or equivalently SPT phases in $r+1$ spatial dimensions with internal symmetry $G_\sigma$. The interpretation is that if we were to add an $r+1$-cell $\tau$ in a new direction perpendicular to $X$, with $\partial \tau = \sigma$, where $G_\sigma$ fixes $\tau$ pointwise, then we'd be able to place an $r+1$-dimensional $G_\sigma$-SPT along $\tau$ which absorbs the $G_\sigma$ anomaly induced by the abutting defects at $\sigma$.

\begin{figure}
\includegraphics[width=8cm]{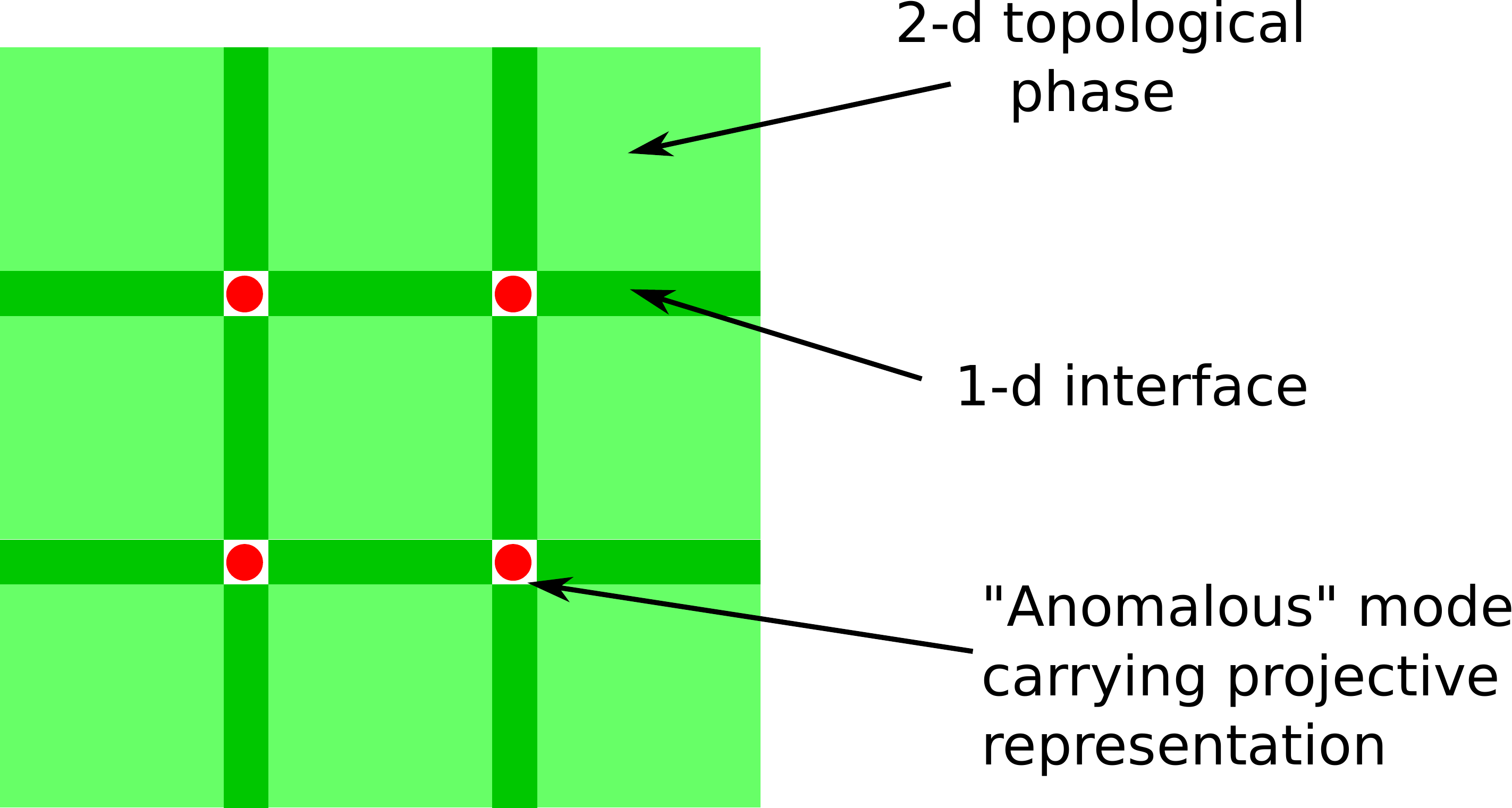}
\caption{\label{anomalous_defect_network}A degree-0 anomalous defect network in a bosonic system in two dimensions. All the 2-cells and 1-cells carry gapped invertible phases and interfaces respectively, but 0-cells (points) carry emergent degenerate modes transforming \emph{projectively} under their respective isotropy groups $G_s$.}
\end{figure}

Of particular interest to the study of LSM theorems is the case $r=0$, meaning the anomalies occur at points $s$ with isotropy group $G_s$ (see Figure \ref{anomalous_defect_network}). For bosonic systems, $G_s$ anomalies of 0d systems, or equivalently 1-dimensional $G_s$-SPTs, are classified by $\cH^2(G_s, \UU(1))$. The assignment of these anomalies is equivariant with respect to the global $G$ symmetry. Therefore, a degree-0 anomalous defect network defines an anomalous texture in the sense of Section \ref{sec:symrep}.

\begin{figure}
\includegraphics[width=5cm]{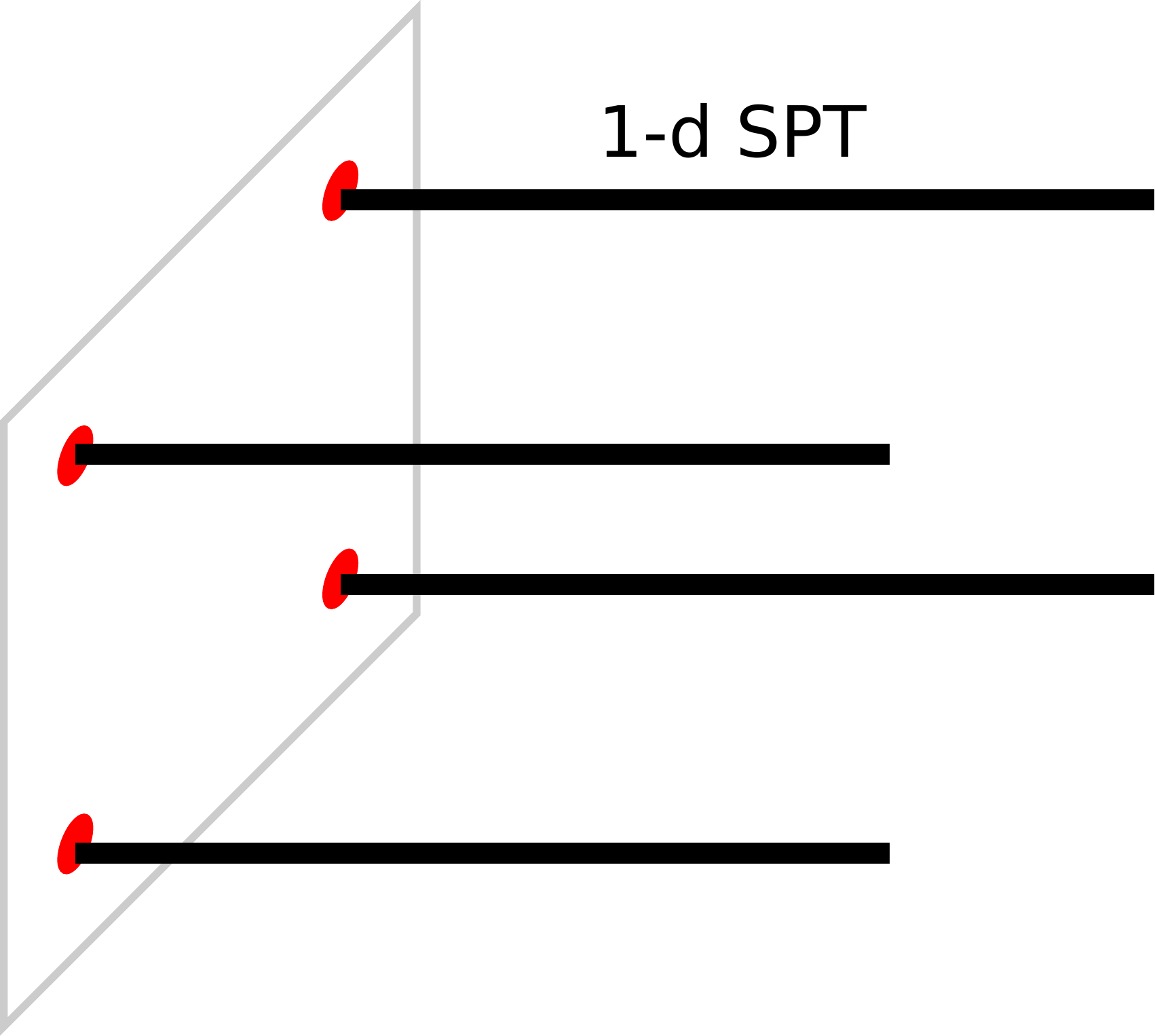}
\caption{\label{bulkboundary}An anomalous texture can appear at the \emph{boundary} of a 1-skeletal defect network in one higher dimension.}
\end{figure}

Conversely, given an anomalous texture on $\Lambda \subset X$, we can define a degree-0 anomalous (invertible-substrate) defect network on $X \times [0,\infty)$ which is non-anomalous in the bulk, but has an anomaly on the boundary characterized by the given anomalous texture. In particular, we can take the defect network to be 1-skeletal, meaning it is just some parallel arrangment of 1d SPTs (see Figure \ref{bulkboundary}). We choose these 1d SPTs so for each site $s$ on the boundary there is a 1d SPT in the bulk classified by $[\omega_s] \in \cH_s(G_s, \UU(1))$ which terminates on the boundary at the site $s$.

% The way to do this is to choose a $G$-invariant cell complex on $X$; extend it by product to $X \times [0,\infty)$ so that the $k$-cells are of the form $\sigma_k \times 0$ and $\sigma_{k-1} \times (0,\infty)$, where $\sigma_k, \sigma_{k-1}$ are cells of $X$; let $G$ act on $X \times [0,\infty)$ as a product, with the trivial action in the new coordinate; and then for each projective symmetry class $\omega_s \in \cH^2(G_s,\UU(1))$, place along the 1-cell $s \times (0,\infty)$ the corresponding 1d SPT.

More generally, one could define higher-dimensional anomalous textures so that the anomaly of an invertible-substrate defect network always corresponds to an anomalous texture of some kind. Furthermore, such anomalous textures occur at the boundaries of general crystalline SPTs. We do not consider these generalized situations in much detail, although the computational schemes we develop in Section \ref{sec:defnetclass} do apply to them.

\subsection{Bulk-boundary correspondence in terms of defect networks}\label{s:def-bulk-boundary1}

When does a spin system in $d$ dimensions with spins transforming projectively admit an invertible symmetric ground state? That is, when is there an LSM theorem?

As we have shown above, any anomalous texture sits at the boundary of a crystalline SPT in the $d+1$-dimensional space $X \times [0,\infty)$, where we have added an extra half-infinite direction.

One can show that crystalline SPTs of this special form admit invertible symmetric boundary conditions iff they are trivial, in the sense explained at the end of Section \ref{subsec:defectnetworks}. Thus, the invariants of these crystalline SPTs in $d+1$-dimensions capture LSM anomalies in $d$-dimensions.

In this paper, however, we will introduce a different (though closely related) perspective; rather than talking about crystalline SPTs in $d+1$ spatial dimensions, we will talk about anomaly cancellation in $d$ spatial dimensions, that is without adding an extra dimension to the problem. This picture will ultimately prove more powerful, and will be the subject of the next section.

\section{Anomaly matching}
\label{sec:anomalymatching}
In the previous sections, we saw that anomalous textures can arise in several different ways. In particular, they can arise both as the anomaly of a defect network, in the sense described above, and as a description of the projective representation of the microscopic degrees of freedom. This motivates the following principle, which we can think of as a form of ``UV-IR anomaly matching'':
\begin{quotation}
\item A defect network represents an allowed state for a strictly $d$-dimensional system \emph{if and only if} its anomaly cancels the anomalous texture of the microscopic degrees of freedom. Moreover, an anomalous texture will give rise to a traditional LSM theorem if and only if there are \emph{no} invertible-substrate defect networks that give rise to a matching anomaly.
\end{quotation}

%A defect network that is non-invertible in the top dimension \textbf{: define these terms} could have more general kind of anomalies, even in degree 0, not described by anomalous textures. However, as we will see, there is at least some subset of anomalies that correspond to microscopic anomalous textures.

%Finally, we note that an anomalous texture in $d$ spatial dimensions can always be thought of as the boundary of a state with spatial symmetry $G$ in $d+1$ spatial dimensions. (For example, an array of spin-1/2 particles with $SO(3)$ rotation symmetry can be thought of as the boundary of a stack of Haldane chains). If this $d+1$-dimensional state is in a non-trivial SPT bulk, it follows that it does not admit an invertible boundary; that is, if the boundary respects the symmetry, it must be in some non-invertible phase with fractional excitations. On the other hand, if the $d+1$-dimensional state is in the trivial phase in the bulk, then it must admit an invertible boundary. However, as we will see, the anomaly matching on the boundary actually contains more information, because it can place non-trivial constrains on \emph{which} invertible phases are allowed on the boundary, even when the bulk phase is trivial. For this reason, we will generally work directly in terms of the $d$-dimensional system, rather than invoking such a bulk-boundary correspondence.

Let us now discuss some concrete examples.

\subsection{Examples}

In the following, we will usually not make explicit reference to a cell decomposition of the space $X$, with the understanding that one can always be chosen in order to fit the discussion into the cellular framework introduced above.

\subsubsection{Cancellation of an anomalous texture by a 1-skeletal defect network}
\label{subsubsec:oneskeletal}
\begin{figure}
\subfloat[The microscopic anomalous texture]{\includegraphics[width=0.4\textwidth]{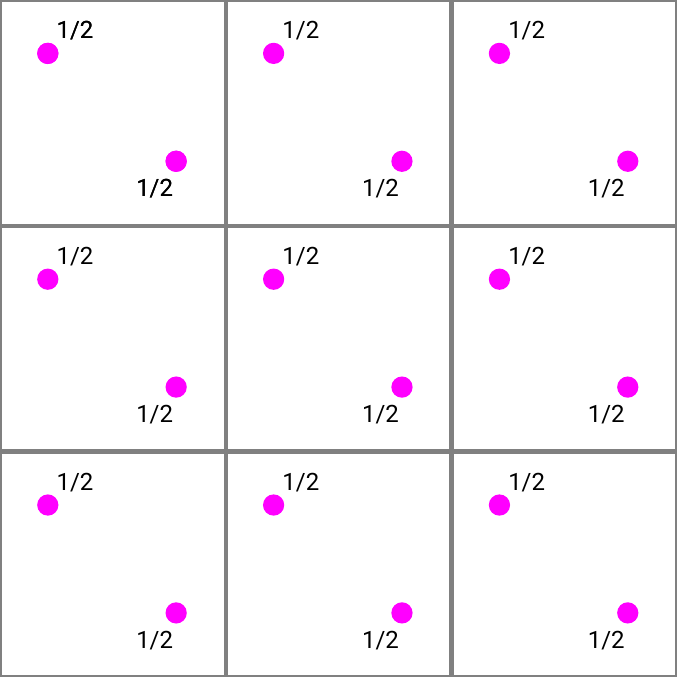}}

\subfloat[The defect network with emergent anomalous texture]
{\includegraphics[width=0.4\textwidth]{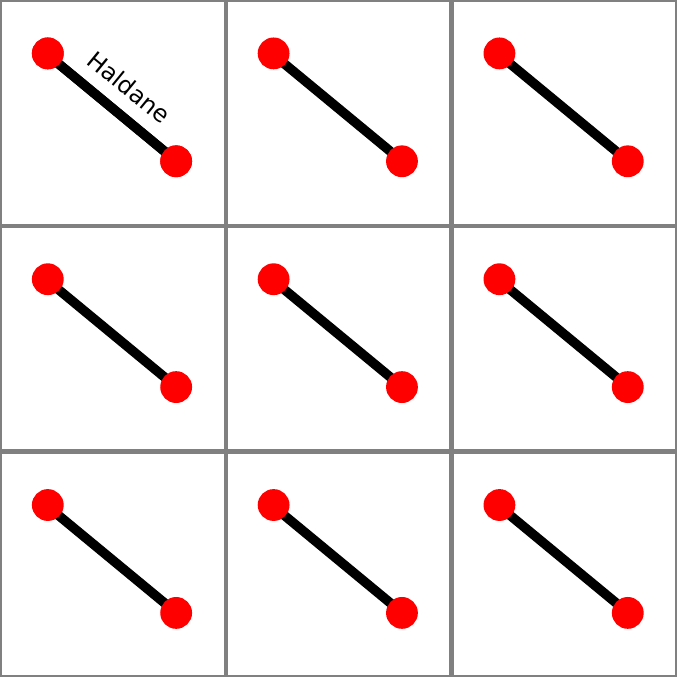}}
\caption{\label{fig:cancellation_1d}In a spin system which has $SO(3)$ spin rotation symmetry and translation symmetry, a system with two spin-half particles per unit cell gives rise to an anomalous texture which can be cancelled by a collection of Haldane chains.}
\end{figure}

The simplest example of cancellation of anomalous textures occurs in a spin system with $\SO(3)$ symmetry and translation symmetry and two spin-half particles per unit cell. According to our definitions, so long as the two spin-halfs sit at different locations in the unit cell, this system has non-trivial anomalous texture in the sense that $\omega_s$ defines a non-trivial element of $H_{-2}^G(\Lambda, \UU(1))$. Nevertheless, it is clear that we do not have an LSM theorem, since this Hilbert space admits a gapped symmetric invertible ground state, in which the spin-half particles are paired into singlets. We can rephrase this from the point of view of anomalous texture cancellation if we add extra integer-spin degrees of freedom into the unit cell (which does not affect the anomalous texture) and replace the singlets with 1d Haldane chains \cite{Haldane__83,Affleck__88,Pollmann_0909}, as shown in Figure \ref{fig:cancellation_1d}. Because the Haldane chains have boundaries, there are emergent spin-1/2 degrees of freedom associated with the endpoints, which we interpret as an emergent anomalous texture. This emergent anomalous texture cancels with the anomalous texture of the microscopic spin-1/2 degrees of freedom, allowing for a gapped nondegenerate symmetric ground state. This is an example of a microscopic anomalous texture which can be cancelled by a 1-skeletal defect network, since the Haldane chains are 1-dimensional.

Note that the above argument could easily be repeated in a fermionic system, with the spin-1/2's replaced with  microscopic Majorana zero modes, and the Haldane chains replaced with Kitaev chains \cite{Kitaev_0010}.

It might seem contrived in this example to think of the singlets as miniature Haldane chains. Moreover, this is an example of anomalous texture which is trivialized by the lattice homotopy equivalence relations described in Section \ref{subsec:latticehomotopy} (generally, any anomalous texture which is lattice homotopy equivalent to the trivial one can be trivialized by 1d wires, as we show in Appendix \ref{appendix:smoothstates}). Nevertheless, we chose to present this example in order to illustrate the analogy with the more non-trivial kinds of anomalous texture cancellations discussed below.

\subsubsection{Cancellation by an invertible-substrate defect network}
\label{subsubsec:highercancellation}
\begin{figure}
\subfloat[The microscopic anomalous texture]{\includegraphics[width=0.2\textwidth]{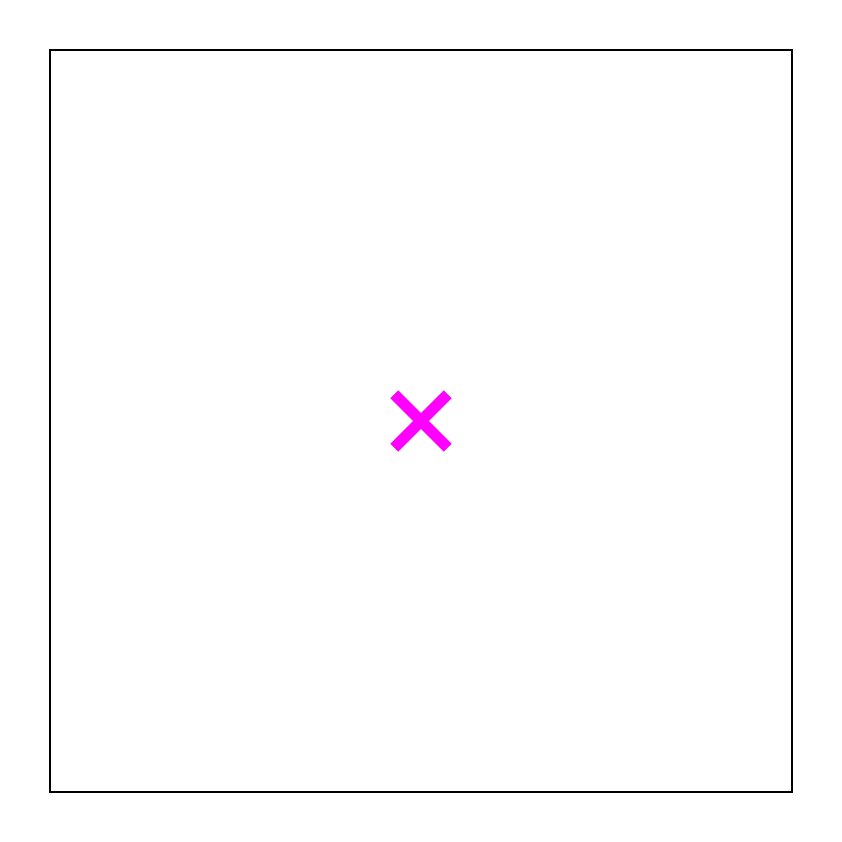}}
\quad\quad
\subfloat[The defect network with emergent anomalous texture]
{\includegraphics[width=0.2\textwidth]{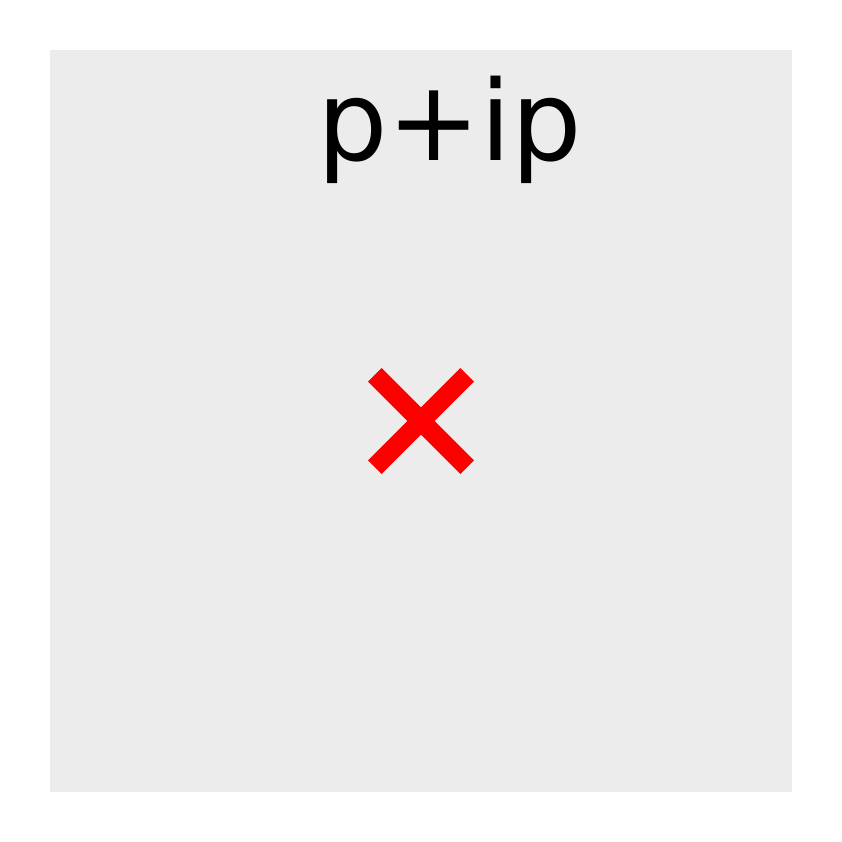}}
\caption{\label{fig:ppip_cancellation}In a fermionic system in (2+1)-D with $C_2$ spatial rotation symmetry, a microscopic MZM at the origin can be cancelled by a $(p+ip)$ superconductor.}
\end{figure}

In the previous section, we discussed an anomalous texture which can be cancelled by a 1-skeletal defect network. Next we will discuss an anomalous texture which cannot be cancelled by a 1-skeletal defect network, but can be by an invertible-substrate defect network. We discuss fermionic systems here because the simplest analogous case we know of for bosonic systems, the magnetic translations example discussed in Section \ref{sec:nontraditional_lsm}, is substantially more complicated. Our example is related by a bulk-boundary correspondence to a 3d defect network defining a trivial crystalline SPT state discussed in Refs.~\cite{Else_1810,Freed_1901}.

The state which will have the desired anomalous texture in the emergent sense is a $(p+ip)$ superconductor with $C_2$ rotation symmetry. For our purposes, it will be sufficient to consider a continuum Hamiltonian which has a $(p+ip)$-superconductor as its ground state, namely:

\begin{multline}
H = \int d^2 \textbf{r} \biggl[ \Psi^{\dagger}\left(-\frac{1}{2m}\nabla^2 - \mu\right) \Psi \\+ \Delta \Psi^{\dagger} (\partial_x + i \partial_y) \Psi^{\dagger} + h.c. \biggr],
\end{multline}
where $\Psi(\textbf{r})$ is a fermionic field, and $\Delta, m$ and $\mu$ are constants. The pairing term is not rotationally invariant, as can be seen by writing it in polar coordinates $(r,\theta)$:
\begin{equation}
\label{pplusip_pairing}
\int d^2 \textbf{r} \Psi^{\dagger}(\partial_x + i \partial_y) \Psi^{\dagger}
= \int r dr d\theta \, e^{i\theta} \Psi^{\dagger} (\partial_r + i r \partial_\theta) \Psi^{\dagger}.
\end{equation}
What, then, are we to do if we want to construct our $(p+ip)$ superconductor to be $C_2$ invariant? In fact, we can make \eqnref{pplusip_pairing} rotationally invariant if we redefine $\Psi^{\dagger} \to e^{i\theta/2} \Psi^{\dagger}$. The problem is that this is effectively introducing a $\pi$ vortex (flux of fermion parity) at the origin, and we know that this binds a Majorana zero mode (MZM) \cite{Read_9906}. Hence, we conclude that the $C_2$ invariant $(p+ip)$-superconductor has an emergent anomalous texture characterized by a single MZM at the origin. Therefore, if we construct a system which microscopically has a MZM at the origin, then the anomalous textures can cancel, and the two MZMs couple to form a non-degenerate ground state (see Figure \ref{fig:ppip_cancellation}). Observe that this anomalous texture \emph{cannot} be cancelled by a 1-skeletal defect network, because there is no way to add Kitaev chains to cancel the MZM while preserving the $C_2$ symmetry. However, the $C_2$ invariant $(p+ip)$ superconductor corresponds to an invertible-substrate defect network that cancels the anomalous texture.

Similarly, one can also show that if we restore (lattice) translational symmetry, so that the full symmetrfy group is the wallpaper group $p2$, then we can show that the $(p+ip)$ superconductor with $p2$ symmetry has an emergent anomalous texture with a MZM at each rotation center (of which there are 4 per unit cell). Therefore, to get a non-degenerate ground state we need a lattice system which microscopically has a MZM at each rotation center.

%By definition, a system with a fragile pre-anomaly exists on the boundary of a \emph{trivial} SPT state, which means there exists a symmetric local unitary that transforms the bulk into a product state. We can apply a restricted version of this local unitary to the $d+1$-dimensional system with $d$-dimensional boundary. This defines a locality-preserving unitary $\cH \to \cH' \to \cH$, where $\cH'$ is a microscopic Hilbert space with pre-anomaly, and $\cH$ is the microscopic space without pre-anomaly. With no pre-anomaly, it is easy to find a product state $\ket{\psi}$ in $\cH$ that is invariant under the symmetry. But a locality preserving unitary cannot create intrinsic topological order, so we see that the ground state of $\ket{\psi}$ must be an invertible state.

\subsubsection{Cancellation by a non-invertible state}
\begin{figure}
\subfloat[The microscopic anomalous texture]{\includegraphics[width=0.4\textwidth]{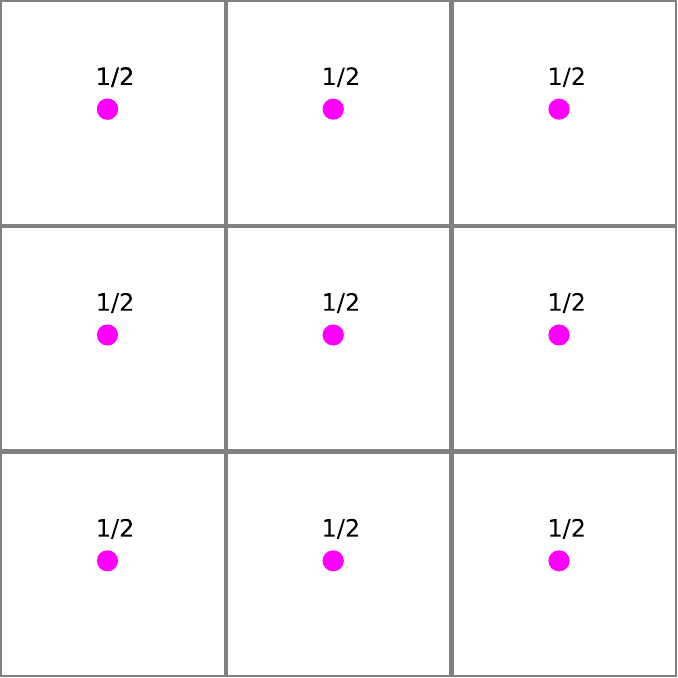}}

\subfloat[The defect network with emergent anomalous texture]
{\includegraphics[width=0.4\textwidth]{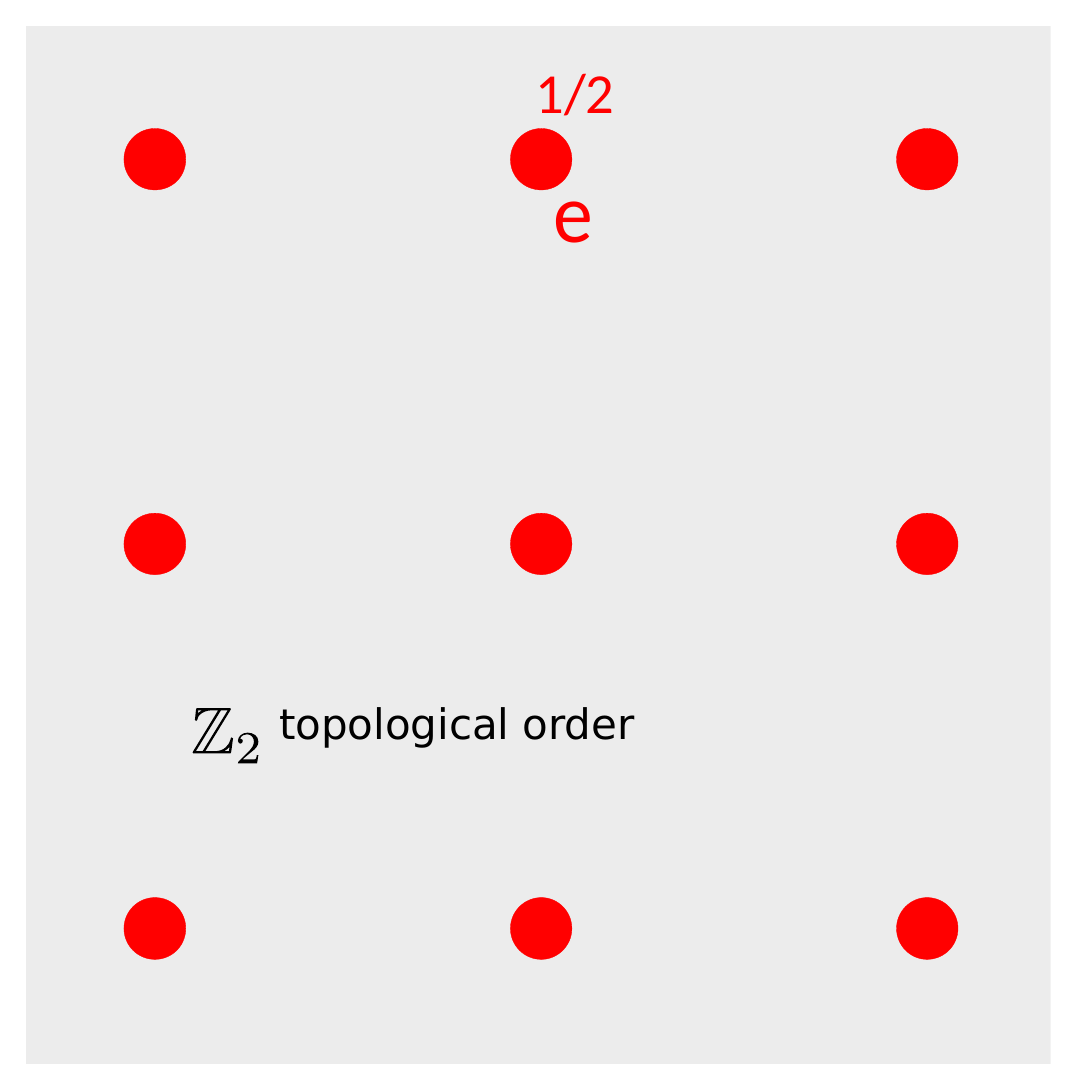}}
\caption{\label{fig:tc_cancellation}In a spin system which has $SO(3)$ spin rotation symmetry and translation symmetry, a system with one spin-half particle per unit cell gives rise to an anomalous texture which can be cancelled by a $\bZ_2$ topological order (i.e. the topological order of the toric code) with an $e$ particle carrying half-integer spin located in each unit cell.}
\end{figure}

Next we consider an example of an anomalous texture which cannot be cancelled by any invertible-substrate defect network, but can be by a general invertible-defect network, ie. one where the top-dimensional cells carry a non-invertible phase but all defects are invertible.

The anomalous texture corresponds to a spin system with discrete translation symmetry in two dimensions, with a spin-1/2 per unit cell. The statement that such a system cannot have an invertible symmetric ground state is, of course, the original LSM theorem in 2d. On the other hand, a symmetric ground state can exist if it has non-invertible topological order with fractionalized excitations. It was shown in Ref.~\cite{Zaletel_1410}, however, that there are still non-trivial constraints on which non-invertible topological orders are allowed.

We can interpret the constraints of Ref.~\cite{Zaletel_1410} as corresponding to the requirement that the anomalous texture be cancelled. Ref.~\cite{Zaletel_1410} showed that the translation symmetry and spin rotation symmetry must be fractionalized on anyons; such a fractionalization can be interpreted as saying that there is an anyon present in each unit cell in the ground state, and that anyons carry fractional spin, respectively. Furthermore, Ref.~\cite{Zaletel_1410} showed that the specific anyon which is present in each unit cell in the ground state must carry half-integer spin. We can interpret this as an anomalous texture cancellation condition: the microscopic spin-1/2 in each unit cell and the emergent half-integer spin per unit cell due to the fractionalization must combine to form an integer spin (see Figure \ref{fig:tc_cancellation}). Indeed, if this cancellation did not occur, then in the defect network picture there would be a massive ground-state degeneracy due to the half-integer spins in each unit cell. Note that in the defect network picture, due to the assumption that the unit cell size is much larger than the correlation length, these spins would not be able to couple to each other.

\subsection{Connection with lattice homotopy}
\label{sec:anomaly_cancellation_vs_lattice_homotopy}
As we have already mentioned in Section \ref{subsubsec:oneskeletal}, the lattice homotopy equivalence relation is closely connected with the idea of anomaly cancellation. Specifically, an anomalous texture is equivalent to the trivial texture in lattice homotopy if and only if the anomaly can be cancelled by a 1-skeletal defect network. In general, one can take this as the \emph{definition} of lattice homotopy, but for bosonic systems we show in Section \ref{sec:equivarianthomology} that this also agrees with the concrete formulation of lattice homotopy from Section \ref{subsec:latticehomotopy}. The fact that this is not the most general invertible-substrate defect network is precisely why an anomalous texture can be non-trivial in lattice homotopy but still not result in a traditional LSM theorem which guarantees non-invertible ground states. On the other hand, in such case we will generally have an ``SPT-LSM'' theorem which constrains what kind of invertible ground states one can have. We discuss such results in more detail in Section \ref{sec:nontraditional_lsm}.

\subsection{Connection with the bulk-boundary correspondence}
\label{subsec:cancellatoin_bulkboundary}

As we mentioned in Section \ref{s:def-bulk-boundary1}, the LSM theorem may also be formulated in terms of a bulk-boundary correspondence, where the anomalous texture sits at the boundary of a crystalline SPT in the $d+1$-dimensional space $X \times [0,\infty)$, where we have added an extra half-infinite direction. Thus, the LSM constraints come from the invariants of these crystalline SPTs.

The two approaches are equivalent. In particular, in Appendix \ref{appendix:smoothstates}, we prove that an anomalous texture in $d$ spatial dimensions can be cancelled by an invertible-substrate defect network if and only if its associated $d+1$-dimensional crystalline SPT is trivial according to the defect network equivalence relation.

\section{Calculations in bosonic systems using equivariant homology}
\label{sec:equivarianthomology}
So far, we have given a very appealing set of physical pictures, but we have not yet explained how to \emph{compute} anything in this picture. In this section, we develop a computational method which applies in most cases of interest for LSM theorems in spin systems and which easily may be computerized. Specifically, we consider bosonic systems, and we restrict to invertible-substrate defect networks. Moreover, we consider only a certain subset of invertible-substrate defect networks, which we call ``in-cohomology''. Roughly, this corresponds to requiring that the data on each $k$-cell $\Sigma$ correspond to an in-cohomology SPT phase -- that is, one constructed from a class in group cohomology $\cH^{k+1}(G_\Sigma, \mathrm{U}(1))$. This is not quite the precise statement, because in general the data on a $k$-cell in an invertible-substrate defect network is a \emph{torsor} over $G_\Sigma$ SPT phases, hence cannot be canonically identified with an element of $\cH^{k+1}(G_\Sigma, \mathrm{U}(1))$. Nevertheless, there is a well-defined notion of the data on a $k$-cell being in-cohomology, as we discuss in Appendix \ref{appendix:smoothstates}.

The phases constructed from classes in $\cH^{d+1}(G, \mathrm{U}(1))$ are known not to be the most general bosonic invertible topological phases with $G$ symmetry \cite{Vishwanath_1209,Wang_1302,Burnell_1302,Kapustin_1403}. For example, for $d=2$ they do not include the so-called Kitaev $E_8$ state \cite{Kitaev_0506,Lu_1205}, which exists even for $G=1$. Accordingly, in-cohomology defect networks are only a subset of all possible invertible-substrate defect networks. Further, working with group cohomology invariants restricts the type of deformations we can apply to our states. We discuss how this affects our LSM criterion later in Section \ref{subsubsec:evaluating}.

For in-cohomology defect networks, the pictures we have previously introduced can be expressed in a relatively concrete mathematical way, and one which allows for explicit computations, in terms of \emph{equivariant homology} (see also Refs.~\cite{Shiozaki_1810,Freed_1901}).

Equivariant homology is something that is defined in terms of a space $X$ and a group $G$ acting on it. It is a kind of generalization of cellular homology of $X$, agreeing with it for $G = 1$. It is also closely related to group cohomology (\emph{not} group homology) of $G$. Cellular homology is a standard notion from elementary algebraic topology, and group cohomology is by now familiar to physicists through the bosonic SPT classification. We will first review both of these notions, before moving on to equivariant homology.

\subsection{Cellular homology}
\label{subsec:cellular_homology}
Let $X$ be a space with a cell decomposition (specifically a regular CW complex) and $A$ be some additive abelian group. A \emph{(cellular) $k$-chain} is a formal linear combination (with coefficients in $A$) of oriented $k$-cells $\sigma$, such that if $\bar \sigma$ is $\sigma$ with the opposite orientation, then $\bar\sigma = -\sigma$. For our purposes, since we will want to consider non-compact spaces such as $X = \mathbb{R}^n$, we will allow infinite linear combinations. This gives rise to what is known as ``Borel-Moore'' homology\footnote{This differs from ordinary homology. In particular, for us $H_n(\bR^n,\bZ) = \bZ$ and $H_0(\bR^n,\bZ) = 0$.}. These generate an abelian group denoted $C_k(X,A)$ where the empty $k$-chain is the identity and the oriented $k$-cells with weight 1 generate over $A$. An orientation of $\sigma$ determines an orientation of the $(k-1)$-cells $\tau \subset \partial \sigma$, and we use this to define a linear map called the \emph{boundary operator}
 \begin{equation}\partial:C_k(X,A) \to C_{k-1}(X,A)\end{equation}
 on generators as
 \begin{equation}\partial \sigma = \sum_{\tau \subset \partial \sigma} \tau,\end{equation}
 where each $\tau$ is given the orientation induced by $\sigma$. A $k$-chain $V$ with $\partial V = 0$ is called a $k$-cycle and the group of $k$-cycles is denoted $Z_k(X,A)$. Likewise the image under $\partial$ of $C_{k+1}(X,A)$ is called the group of boundaries, and denoted $B_k(X,A)$. We have $\partial^2 = 0$ so $B_k(X,A) \leq Z_k(X,A)$. We obtain a group
 \begin{equation}H_k(X,A) = Z_k(X,A)/B_k(X,A)\end{equation}
 called the $k$th homology of $X$ with coefficients in $A$.

\subsection{Group cohomology}
\label{subsec:groupcohomology}

Now let $M$ be some additive abelian group with an action of $G$ (sometimes abbreviated as a $G$-module).

A \emph{$G$ $k$-cochain with values in $M$} is a map
\begin{equation}\alpha:G \times \stackrel{k}{\cdots} \times G \to M.\end{equation}
(In the case where $G$ and $M$ are continuous we require cochains to be measurable functions \cite{Chen_1106}; see Appendix \ref{subsec:coefficients} for more details).
These form a group $\cC^k(G,M)$ under addition of values, where the constant map to $0 \in M$ is the additive identity in $\cC^k(G,M)$. We define a linear map called the \emph{(group) coboundary operator}
\begin{equation}\delta : \cC^k(G,M) \to \cC^{k+1}(G,M)\end{equation}
according to
\begin{multline}
\label{group_coboundary_first}
\left(\delta\alpha\right)(g_1,\dots,g_{k+1})
\\ = g_1 \alpha(g_2,\dots,g_{k+1})
\\ + \sum_{i=1}^p (-1)^{i} \alpha(g_1,\dots,g_{i-1},g_i g_{i+1},g_{i+2},\dots,g_{k+1})
\\+  (-1)^{k+1} \alpha(g_1,\dots,g_k),
\end{multline}
where in the first term we use the action of $G$ on $M$. A $k$-cochain $\alpha$ with $\delta\alpha = 0$ is called a $k$-cocycle and the group of $k$-cocycles is denoted $\cZ^k(G,M)$. We have $\delta^2 = 0$. We denote the group of $k$-coboundaries (i.e. the image of $\cC_{k-1}(G,M)$ under $\delta$) by $\mathcal{B}^k(G,M)$. The \emph{$k$th group cohomology with coefficients in $M$} is defined as
\begin{equation}\cH^k(G,M) = \cZ^k(G,M)/\mathcal{B}^k(G,M).\end{equation}

In some cases, it will be more convenient to use \emph{homogeneous} cochains (whereas the cochains defined above are called \emph{inhomogeneous}). A homogeneous $G$ $k$-cochain with values in $M$ is a map
\begin{equation}
\nu : G \times \stackrel{k+1}{\cdots} \times G \to M
\end{equation}
which satisfies the homogeneity condition
\begin{equation}
g \nu(g_1, \cdots, g_{k+1}) = \nu(gg_1, \cdots, gg_{k+1}).
\end{equation}
Homogeneous cochains are in one-to-one correspondence with inhomogeneous cochains: an homogeneous cochain can be constructed from an inhomogeneous cochain according to
\begin{equation}
\nu(g_1, \cdots, g_{k+1}) = g_1 \alpha(g_1^{-1} g_2, g_2^{-1} g_3, \cdots, g_{k}^{-1} g_{k+1}),
\end{equation}
while an inhomogeneous cochain can be constructed from a homogeneous cochain according to
\begin{equation}
\alpha(g_1, \cdots, g_{k}) = \nu(1, \hat{g}_1, \cdots, \hat{g}_k),
\end{equation}
where $\hat{g}_l = g_1 g_2 \cdots g_l$.
 In terms of the homogeneous cochains, the coboundary operator \eqnref{group_coboundary_first} becomes
\begin{multline}
\label{eq:homogeneous_coboundary}
(\delta \nu)(g_1, \cdots, g_{k+2}) \\= \sum_{i=0}^{k+2}(-1)^i \nu(g_1, \cdots, g_{i-1}, g_{i+1}, \cdots, g_{k+2}),
\end{multline}
that is, for each term of the sum, each of $g_1, \cdots, g_{k+2}$ are included in the argument of $\nu$ except $g_i$. One sees that the correspondence between homogeneous and inhomogeneous $k$-cochains respects $\delta$.

\subsection{Equivariant homology}
\label{subsec:equivarianthomology}

\begin{figure}
\includegraphics[width=7cm]{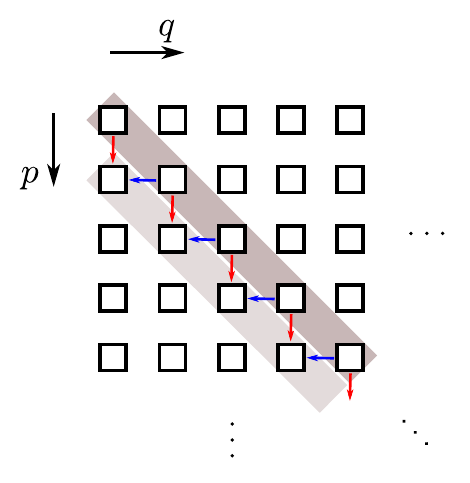}
\caption{\label{fig:doublecomplex}A graphical representation of the double complex. Each square represents an abelian group $\mathcal{Q}_q^p = \cC^p(G, \cC_q(X,A))$. The group of equivariant $r$-chains, ${\cC_r^G(X, A)}$, is the direct sum along the $r$-th diagonal (highlighted). The equivariant boundary operator $D$ goes from $\cC_r^G(X, A) \to \cC_{r+1}^G(X,A)$ and is constructed out of the horizontal maps $\partial$ (shown in blue) and vertical maps $\delta$ (shown in red) that link the $r$-th and $r+1$-th diagonals.}
\end{figure}

Now let us suppose our space $X$ with a cell decomposition admits an action of $G$ such that for each $g \in G$, each $k$-cell $\sigma$ maps bijectively to another $k$-cell $g\sigma$. Likewise if $\sigma$ has an orientation, $g\sigma$ receives an orientation as well. We extend this by $A$-linearity to a $G$-action on $C_k(X,A)$. Taking $M = C_p(X,A)$ we form the group
\begin{equation}\mathcal{Q}^p_q = \cC^{p}(G, C_q(X,A)).\end{equation}
We typically organize these groups in a grid in the plane where $q$ runs along the $x$-axis and $p$ along the $y$-axis (see Figure \ref{fig:doublecomplex}). Accordingly, the boundary operator on cellular chains defines \emph{horizontal} maps
\begin{multline}
\partial_{p,q}: \mathcal{Q}^p_q = \cC^p(G,C_q(X,A)) \\ \xrightarrow{\partial} \cC^p(G,C_{q-1}(X,A)) = \mathcal{Q}^p_{q-1},
\end{multline}
where $\partial$ acts on the values of the cochains, while the coboundary operator on group cochains gives \emph{vertical} maps
\begin{multline}\delta_{p,q}: \mathcal{Q}^p_q = \cC^p(G,C_p(X,A))\\ \xrightarrow{\delta} \cC^{p+1}(G,C_q(X,A))= \mathcal{Q}^{p+1}_q.\end{multline}
These maps satisfy $\partial_{p,q} \partial_{p,q+1} = 0$, $\delta_{p+1,q} \delta_{p,q} = 0$, and
\begin{equation}\partial_{p+1,q+1} \delta_{p,q+1} = \delta_{p,q} \partial_{p,q+1} \quad {\rm as\ maps}\quad \mathcal{Q}^p_{q+1} \to \mathcal{Q}^{p+1}_q.\end{equation}
This whole structure is known as a \emph{double complex}. We will sometimes drop the degrees on $\delta_{p,q}$ and $\partial_{p,q}$ when they can be inferred from context or when the formulas hold at arbitrary degree. For instance, we can write the above as
\begin{equation}\partial \delta = \delta \partial.\end{equation}

An \emph{equivariant $r$-chain} is defined to be an element of the direct sum
 \begin{equation}
 \mathcal{Q}_r := \bigoplus_{k=0}^{d} \mathcal{Q}^{k-r}_k = \bigoplus_{k=0}^d \cC^{k-r}(G,C_k(X,A)).
 \end{equation}
This direct sum is along the slope $-1$ diagonals of the double complex (see Figure \ref{fig:doublecomplex}). We write $C_r^G(X, A)$ for the abelian group of equivariant $r$-chains. Note that the target of $\partial_{k+1-r,k+1}$ from the $k+1$st piece, namely $\mathcal{Q}^{k+1-r}_{k}$, a summand of $\mathcal{Q}_{r-1}$, is the same as the target of $\delta_{k-r,k}$ from the $k$th piece (see Figure \ref{fig:doublecomplex}). This allows us to define the \emph{total boundary}
\begin{equation}D_r : \mathcal{Q}_r \to \mathcal{Q}_{r-1}\end{equation}
on equivariant chains according to
\begin{equation}
\label{equivariant_boundary}
(D_r \beta)_k = \partial_{k-r+1,k+1}(\beta)_{k+1} + (-1)^{r} \delta_{k-r,k} (\beta)_k,
\end{equation}
or simply
\begin{equation}D = \partial + (-1)^r \delta,\end{equation}
where $(\alpha)_j$ denotes the projection onto the $\mathcal{Q}^{j-r}_j$ summand of $\mathcal{Q}_r$. The signs are chosen so $D^2 = 0$. An equivariant chain in the kernel of $D$ is called an equivariant cycle, the group of these denoted $Z_r^G(X, A) \subset C_r^G(X,A)$; and one in the image is called an equivariant boundary, the group of these denoted $B_r^G(X,A)\subset C_r^G(X,A)$.

We define the \emph{equivariant homology}
\begin{equation}
H_r^G(X, A) := Z_r^G(X, A) / B_r^G(X,A).
\end{equation}
As an aside, we note there is another kind of equivariant homology (sometimes called ``Borel equivariant homology''), which is quite different from this one. The version we have presented is a special case of the kind of equivariant homology discussed in Ref.~\cite{Freed_1901} (despite the fact that in that reference their formulation was referred to as ``Borel equivariant homology'' in a non-standard usage). See also Ref.~\cite{Shiozaki_1810}; that work assumed a mathematical object called an ``equivariant spectrum'' that was there left unspecified. The equivariant homology of this work and Ref.~\cite{Freed_1901} can be viewed as corresponding to a \emph{particular choice} of equivariant spectrum.
% Our indexing convention is closer to the second reference.

\subsection{Anomalous textures as equivariant chains}
\label{anomalous_texture_equivariant_chains}
The data of an anomalous texture, as introduced in Section \ref{sec:symrep}, is conveniently packaged as an equivariant $-2$-cycle. In particular, if we compare the definition of equivariant homology with the definition of anomalous texture, we find that anomalous textures as defined in Section \ref{sec:symrep} are classified by the equivariant homology $H_{-2}^G(\Lambda, \mathrm{U}(1))$, where where $\Lambda$ is interpreted as a space containing only 0-cells; indeed we already introduced this notation earlier. The data $\omega_s$ discussed in Section \ref{sec:symrep}  corresponds to an element of $\mathcal{C}^2(G, C_0(\Lambda, \UU(1))) = C_{-2}^G(\Lambda,\UU(1))$; the associativity condition \eqnref{an_associativity} corresponds to restricting to equivariant cycles in $Z_{-2}(\Lambda,\UU(1))$, and the gauge freedom \eqnref{an_coboundary} corresponds to modding out by equivariant boundaries in $B_{-2}(\Lambda,\UU(1))$.

 We assume that we have chosen the cell decomposition of $X$ such that all the points in $\Lambda$ map into vertices of $X$.
Then there is a natural inclusion map $Z_{-2}^G(\Lambda,\UU(1)) \to \mathcal{Z}^2(G, C_0(X,\UU(1))) \leq Z_{-2}^G(X,\UU(1))$. We find that an anomalous textures that gives an $\omega, \omega' \in \mathcal{Z}^2(G, C_0(X,\UU(1))$ is trivial if and only if there exists $\lambda \in \cZ^2(G,C_1(X,\UU(1)))$ such that $\partial \lambda = \omega$.

We can write this in a compact way by introducing the $1$-skeleton $X_1$ of $X$, which is the union of all the 0- and 1-cells of $X$. Then we have that $\mathcal{Z}^2(G, C_1(X,\UU(1))) = \mathcal{Z}^2(G, C_1(X_1,\UU(1)) = C_{-1}^G(X_1,\UU(1)))$. So we see that an anomalous texture is trivial in lattice homotopy equivalence if and only if it gives trivial map class under the inclusion map $H_{-2}^G(\Lambda, \UU(1)) \to H_{-2}^G(X_1,\UU(1))$. As we discuss in Appendix \ref{appendix:spectralsequences}, this can also be understood in the context of a spectral sequence.

As we have already alluded to, however, an anomalous texture that is non-trivial in lattice homotopy equivalence might still be cancellable by an invertible-substrate defect network which is nontrivial on higher dimensional  cells. The above formulation already hints what we need to do to find the general criterion, namely study the image of the anomalous texture in $H_{-2}^G(X,\UU(1))$ rather than $H_{-2}(X_1,\UU(1))$. In the next subsection, we will make this precise.

% \de{Isn't this paragraph just the same as Theorem 2? Maybe we don't need it?}
% The strongest equivalence relation we consider for anomalous textures is given by tensoring with an anomalous defect network. We argue (: where?) that this equivalence relation is $\omega \sim \omega + D\lambda$ where $\lambda = (\lambda_0, \cdots, \lambda_d) \in C_{-1}^G(X,\UU(1))$ is an arbitrary equivariant $-1$-chain. The equivalence class of $\omega$ under this is its image in equivariant homology $[\omega] \in H_{-2}^G(X,\UU(1))$ and captures the in-cohomology LSM anomaly (cf. ).
%
\subsection{General classification of defect networks and anomalous textures}\label{sec:defnetclass}

Equivariant homology allows us to compactly express the equivalence classes of defect networks described in Section \ref{subsec:defectnetworks}, where all defects involved are described by group cohomology. We refer to this as the in-cohomology equivalence relation.
\begin{thm}
\label{thm:firsthomologythm}
The in-cohomology equivalence classes of $G$-symmetric non-anomalous invertible-substrate defect networks on a space $X$ are in one-to-one correspondence with classes in equivariant homology $H_{-1}^G( X, \UU(1))$.
\end{thm}
In Section \ref{subsec:relationship} we relate this to the ``Crystalline Equivalence Principle" of Ref.~\cite{Thorngren_1612}, which offers an isomorphic classification.

A closely related result is
\begin{thm}
\label{thm:secondhomologythm}
Let $[\omega] \in H_{-2}^G(\Lambda, \UU(1))$ be an anomalous texture. Then the anomalous texture can be cancelled by an invertible-substrate in-cohomology defect network on $X$ if and only if the image of $[\omega]$ is trivial under the map $H_{-2}^G(\Lambda, \UU(1)) \to H_{-2}^G(X, \UU(1))$ induced by the inclusion $\Lambda \to X$.
\end{thm}

The relationship between Theorems \ref{thm:firsthomologythm} and \ref{thm:secondhomologythm} is an aspect of the bulk-boundary correspondence for crystalline SPTs. Indeed, in Section \ref{sec:defnetworktexture} we described how to associate to an anomalous texture on $X$ an anomalous defect network on $X \times \bR_{\ge 0}$ which realizes the anomalous texture on its boundary. This is equivalently captured by an anomaly-free defect network on $X \times \bR$. We show in Appendix \ref{appendix:smoothstates} that this construction defines an isomorphism
\begin{equation}H_{-2}^G(X,\UU(1)) \simeq H_{-1}^G(X \times \bR,\UU(1))\end{equation}
such that the invariant of the anomalous texture in Theorem \ref{thm:secondhomologythm} is mapped to the invariant of the defect network in Theorem \ref{thm:firsthomologythm}. In other words, an anomalous texture can be cancelled by an invertible state if and only if it defines a trivial SPT in one higher dimension, as expected.

Finally, let us observe that by replacing $X \to X_1$ (the 1-skeleton of $X$), we find that an anomalous texture can be cancelled by a 1-skeletal defect network if and only if it is trivial in lattice homotopy, as expected.

There are two ways one can think about deriving Theorems \ref{thm:firsthomologythm} and \ref{thm:secondhomologythm}. First of all, we show in Appendix \ref{appendix:lattice_wfns} that every element of $Z_{-1}^G(X,\mathrm{U}(1))$ can be used to construct a concrete gapped symmetric lattice wavefunction, and that wavefunctions constructed from two equivariant chains that correspond to the same class in $H_{-1}^G(X, \mathrm{U}(1))$ can be related by a symmetric finite-depth quantum circuit (that is, they are in the same SPT phase). Moreover, we also show in Appendix \ref{appendix:lattice_wfns} that if an anomalous texture defines a trivial class in $H_{-2}^G(X,\mathrm{U}(1))$, which means it can be written as a boundary of a chain in $C_{-1}^G(X,\mathrm{U}(1))$, then this chain tells us how to construct a gapped symmetric lattice wavefunction in the presence of the microscopic anomalous texture. Such considerations, though suggestive, do not completely establish Theorems \ref{thm:firsthomologythm} and \ref{thm:secondhomologythm}; for example, they do not prove the ``only if'' direction in Theorem \ref{thm:secondhomologythm}. However, in Appendix \ref{appendix:smoothstates} we prove Theorems \ref{thm:firsthomologythm} and \ref{thm:secondhomologythm} in full through a more abstract argument based on the in-cohomology model of defect networks.

\subsubsection*{Evaluating the ``in-cohomology'' assumption}
\label{subsubsec:evaluating}
Theorems \ref{thm:firsthomologythm} and \ref{thm:secondhomologythm} are statements about the in-cohomology model of defect networks. So in general one might need to ask whether results derived based on Theorems \ref{thm:firsthomologythm} still apply if one talks about general invertible-substrate defect networks rather than in-cohomology defect networks. (Of course, the portions of Theorems \ref{thm:firsthomologythm} and \ref{thm:secondhomologythm} that can be established by the explicit lattice constructions of Appendix \ref{appendix:lattice_wfns} do not require any further justification.) Firstly, it is clear that in general one can consider defect networks built from beyond-cohomology components, and therefore the classification from Theorem \ref{thm:firsthomologythm} is not complete. However, one can also ask whether a non-trivial class in the in-cohomology model, ie. in $H_{-1}^G(X,\mathrm{U}(1))$, can ever become trivial as an invertible-substrate defect network (because the class of deformations one is allowed to consider in invertible-substrate defect networks is larger).

However, we do not expect this to happen, at least in low spatial dimensions. The reason is that one can show \cite{Thorngren_1612,Else_1810} that the classification of invertible-substrate defect networks obeys the Crystalline Equivalence Principle -- that is, there is a one-to-one correspondence between the classification of invertible-substrate defect networks with spatial symmetry $G$ and the classification of invertible phases with \emph{internal} symmetry $G$. Moreover, one can similarly show that the classification of in-cohomology defect networks with spatial symmetry $G$ is in one-to-one correspondence with the classification of in-cohomology SPTs with internal symmetry $G$. But with internal symmetries it is believed \cite{Kapustin_1403} that a phase that looks non-trivial in group cohomology is always non-trivial in the true classification, at least in spatial dimension $d < 7$.

 For similar reason, we expect that if an anomalous texture cannot be cancelled by an in-cohomology defect network (the condition for which is given by Theorem \ref{thm:secondhomologythm}), then it cannot be cancelled by an invertible-substrate defect network either. The reason is that, following the discussion of Appendix \ref{appendix:smoothstates}, even without making in the in-cohomology assumption we still expect there to be a map from anomalous textures in $d$ spatial dimensions into SPT phases in $d+1$ spatial dimensions, such that the anomalous texture can be cancelled by an invertible-substrate defect network in $d$ dimensions if and only if the SPT phase in $d+1$ spatial dimensions is trivial. Then the result follows from our discussion in the preceding paragraph provided that $d < 6$.

%All the results of this section have been based on the assumption that in-cohomology defect networks correctly model invertible-substrate defect networks. This seems to be true at least in one sense; every equivariant cycle does define a defect network state, and every equivariant boundary does define a deformation relating different defect network states; one can show this by explicit construction (for example by generalizing the construction of Section \ref{subsec:pumps}) \de{: or do we want to write out in full generality somehwere?} However, there is still the possibility of more general invertible-substrate defect networks and deformations that are not captured by the in-cohomology model.

\subsection{A simplification for direct product symmetry groups}
\label{subsec:eqhomkunneth}
As we mentioned in Section \ref{subsec:latticehomotopy}, there is a simplification in cases where the symmetry decomposes as $G = \Gspace \times G_{\mathrm{int}}$ for some internal symmetry $G_{\mathrm{int}}$.
Then the data associated with a site $s$ in the anomalous texture can be decomposed using the K\"unneth formula as
\begin{align}
\cH^2(G_s, \UU(1))
 &= \cH^2(G_{\mathrm{spatial},s} \times G_{\mathrm{int}}, \UU(1))
 \\ &= \cH^2(G_{\mathrm{spatial,s}},\UU(1)) \nonumber\\&\quad\times \cH^1(G_{\mathrm{spatial},s}, \cH^1(G_{\mathrm{int}}, \UU(1))) \nonumber\\&\quad\times \cH^2(G_{\mathrm{int}}, \UU(1)) \label{topsitekunneth},
\end{align}
where $G_{\mathrm{spatial},s}$ is the subgroup of $\Gspace$ that leaves $s$ fixed.

There is also a K\"unneth formula for the equivariant homology (see Appendix \ref{sec:kunneth}), which here reads:
\begin{multline}
H_{-2}^G(X, \mathrm{U}(1)) \\
 = \bigoplus_{k=0}^{d+2} H_{-2+k}^{\Gspace}(X, \cH^k(G_{\mathrm{int}}, \mathrm{U}(1))).
 \label{tophomology_kunneth}
\end{multline}
Under the map from anomalous textures into equivariant homology, the first, second and third factors in \eqnref{topsitekunneth} map into the $k=0,1,2$ factors in \eqnref{tophomology_kunneth}. Importantly, this means that we can consider the three factors in \eqnref{sitekunneth} \emph{separately}: there is a traditional LSM theorem (i.e. the anomalous texture cannot be cancelled by an invertible-substrate defect network) if and only if the image of any one of the three factors in \eqnref{tophomology_kunneth} is non-trivial in the respective equivariant homology.

\subsection{Relationship with anomalies/SPT phases with internal symmetries}
\label{subsec:relationship}
An important property of equivariant homology is that it reduces to group cohomology in the case where $X$ is $\mathbb{R}^d$. We have
\begin{equation}
H_k^G(\mathbb{R}^d, A) \cong \cH^{d-k}(G, A^{\mathrm{or}}).
\end{equation}
Here $A^{\mathrm{or}}$ is the $G$-module whose $G$-action is given by $g * a = s(g) g . a$, where $(g . a)$ is the $G$-action for $A$, and $s(g) = - 1$ if and only if $g$ has orientation-reversing action on $\mathbb{R}^d$, otherwise $s(g) = 1$. In particular, in-cohomology crystalline SPTs on $\mathbb{R}^d$ are classified by
\begin{equation}H_{-1}^G(\mathbb{R}^d, \UU(1)) \cong \cH^{d+1}(G,\UU(1)^{\mathrm{or}}),\end{equation}
which is the same as the classification of in-cohomology SPTs with \emph{internal} symmetry $G$, but where the orientation-reversing unitary elements of $G$ correspond to anti-unitary symmetries. This is an example of the ``Crystalline Equivalence Principle'' of Ref.~\cite{Thorngren_1612}.

Further, the in-cohomology LSM anomaly of an anomalous texture on $\bR^d$ is an element of
\begin{equation}H_{-2}^G(\bR^d,\UU(1)) \cong \cH^{d+1}(G,\UU(1)^{\rm or}),\end{equation}
which also classifies in-cohomology anomalies for an internal symmetry $G$.

For some purposes (e.g.\ for gapless systems as we briefly mention in Section \ref{sec:discussion}), it will be convenient to have an explicit construction of the map from anomalous textures into $\cH^{d+2}(G, \UU(1)^{\mathrm{or}})$, which we will now provide. Recall that an anomalous texture on a lattice $\Lambda$ can be described by an element of $H_{-2}^G(\Lambda, \UU(1))$.

For greater generality, let us construct a map from $H_{-m}^G(\Lambda, A)$ into $\cH^{d+m}(G, A^{\mathrm{or}})$. We call this map the equivariant pushforward. Let
\[\Gamma_{p,q} = \{\omega \in \cC^p(G,C_q(\bR^d,A))\ |\ \delta \omega = \partial \omega = 0\}.\]
Recall that an element of $H_{-m}^G(\Lambda, A)$ gives rise to an element of $\cZ^m(G, C_0(X,A)) \leq \Gamma_{m,0}$. Then if $q < d$, then for any $\omega \in \Gamma_{p,q}$ we can write $\omega = \partial \alpha$ for some $\alpha \in \cC^p(G, C_{q+1}(X,A))$ (here we used the fact\footnote{We are using non-compactly supported (i.e.\ Borel-Moore) homology as discussed in Section \ref{subsec:cellular_homology}, so the only non-trivial homology group of $\mathbb{R}^d$ is $H_d(\mathbb{R}^d, A) = A$.} that $H_q(\mathbb{R}^d,A) = 0$ for $q < d$). Then we see that $\partial \delta \alpha = \delta \partial \alpha = 0$ and $\delta \delta \alpha = 0$. So $\delta \alpha \in \Gamma_{p+1,q+1}$. So if we start from $\Gamma_{m,0}$, we just need to apply this procedure iteratively to obtain an element $\omega \in \Gamma_{d+m,d}$. Now we use the fact that $H_d(\mathbb{R}^d,A) = A$. Concretely, this corresponds to the fact that a closed $d$-chain on $\mathbb{R}^d$ is just a superposition of all the $d$-cells on $\mathbb{R}^d$ with the same coefficient. So we obtain an element of $\cC^{m+d}(G,A)$. The fact that $\delta \omega  = 0$ ensures that this gives a class in $\cH^{m+d}(G,A^{\mathrm{or}})$ [the reason why we have coefficients in $A^{\mathrm{or}}$ is that orientation-reversing elements of $G$ act non-trivially on $H_d(\mathbb{R}^d,A)$].  One can check that the element of $\cH^{m+d}(G,A^{\mathrm{or}})$ so obtained does not depend on any of the arbitrary choices [i.e. choice of representative in $\cZ^{p}(G, C_0(X,A))$ for a class in $H_{-2}(\Lambda,\UU(1))$, and, at each iteration, choice of $\alpha$ such that $\delta \alpha = \omega$] made along the way. For further detail, the relationship to spectral sequence calculations, and a generalization, see Appendix \ref{appendix:spectralsequences}.

% For the polyhedral point groups, there is not such a nice description of the cohomology, but the 2-torsion condition on $e(G)$ is a strict one. ()

%\subsection{Computing the Anomaly}

%Now we specialize to $X = \bR^d$. In this case there is an isomorphism,
%\begin{equation}\cH_{-2}^G(\bR^d,\UU(1)) = \cH^{d+2}(G,Z_d(\bR^d,\UU(1))) = \cH^{d+2}(G,\UU(1)^{tw}),\end{equation}
%a manifestation of the Crystalling Equivalence Principle of Ref.~\onlinecite{}, where $\UU(1)^{tw}$ denotes coefficients twisted by anti-unitary as well as orientation-reversing elements of $G$. Since group cohomology is a bit easier to understand than equivariant homology, we would like to compute $LSM(\alpha)$ as an element of $\cH^{d+2}(G,\UU(1)^{tw})$. This will also help to reveal the physics of the theorem.

\section{Exhaustive computations for quantum magnets}
\label{sec:exhaustive}

As mentioned, the equivariant homology formulation of the problem introduced in the previous section allows for explicit calculations on a computer. We give more details about the algorithms in Appendix \ref{appendix:computation}; the code we used to implement them can be found at \cite{thecode}. Here we will state the results that we have obtained using this technique. Specifically, we have searched for cases in which the lattice homotopy equivalence relation discussed in Section \ref{subsec:latticehomotopy} fails to give the correct criterion for a traditional LSM theorem, that is, there is an anomalous texture which is non-trivial in the lattice homotopy sense, but nevertheless admits an invertible ground state, as represented by an invertible-substrate defect network. Our key result is that there is \emph{no such anomalous texture} for any of the symmetry groups that are relevant for quantum magnets. Specifically, our exhaustive computational search has ruled out such a possibility in the following cases, where $G_{\mathrm{space}}$ is any of the 17 wallpaper groups in two spatial dimensions or any of the 230 space groups in three spatial dimensions:

\begin{enumerate}
\item \label{nosoc} $G = G_{\mathrm{space}} \times \mathrm{G}_{\mathrm{int}}$, where $G_{\mathrm{int}}$ is any group such that $\cH^2(G_{\mathrm{int}},\UU(1)) = \mathbb{Z}_2$, and the anomalous texture just corresponds to putting projective representations of $G_{\mathrm{int}}$ on sites.

\item \label{soc_tr} $G = G_{\mathrm{space}} \times \mathbb{Z}_2^T$, and the microscopic degrees of freedom giving rise to the anomalous texture are spins which transform like a spin-orbit-coupled electron spin under spatial symmetry and time-reversal.

\item \label{soc_notr} $G = G_{\mathrm{space}}$, with no internal symmetry (for example, quantum magnets with spin-orbit coupling and broken time-reversal symmetry). Here there is no restriction on the anomalous textures considered.
\end{enumerate}

Cases \ref{nosoc} (for any choice of $G_{\mathrm{int}}$) and \ref{soc_tr} in fact are covered by a single computation any given $G_{\mathrm{space}}$, by exploiting the K\"unneth decomposition for the equivariant homology, as discussed in Appendix \ref{appendix:computation}.

In particular, this covers all the possible symmetries of quantum magnets discussed in Section \ref{subsec:texture_examples}. Therefore, for quantum magnets we have the result that there is a traditional LSM theorem if and only if the anomalous texture is nontrivial in the lattice homotopy sense. Moreover, since LSM theorems that enforce non-trivial SPT phases always come from anomalous textures that are non-trivial in the lattice homotopy sense, as discussed in Sections \ref{sec:anomaly_cancellation_vs_lattice_homotopy} and \ref{sec:nontraditional_lsm}, we conclude that there are no such results for the symmetry groups considered here.

\section{Equivariant pushforward and LSM theorems for translations and point groups}\label{s:equiv-pushforward}

In this section, we present some results presenting the LSM anomaly associated to $G = G_{\rm int} \times \bZ^d$ (internal symmetry times translations) and to $G = G_{\rm int} \times G_{\rm  pt}$ (internal symmetry times point group) as elements of the corresponding group cohomology $\cH^{d+2}(G,\UU(1))$. This is equivalent to computing the equivariant pushforward of the anomalous texture as decribed in Section \ref{subsec:relationship}. The explicit calculations of the descent sequence can be found in Appendices \ref{a:classic-lsm} and \ref{a:point-group}, respectively.

\subsection{Translation symmetry and the classic LSM theorem}

In the case that all spatial symmetries are translations, $G = G_{\rm int} \times \bZ^d$, the descent sequence takes a particularly simple iterative form we describe in Appendix \ref{a:classic-lsm}. Here we just describe the result of the calculation.

Let us suppose all of $\Lambda$ is a single $\bZ^d$ orbit of a site $s$. The isotropy group of $s$ is $G_{\rm int}$ and an anomalous texture defines a class
\begin{equation}\alpha = \omega_s|_{G_{\rm int}} \in \cH^2(G_{\rm int},\UU(1)),\end{equation}
which we can also gives a class in $\cH^2(G, \UU(1))$ by the projection $G \to \Gint$.
We also define cocycles
\begin{equation}\tau_j \in \cH^1(\bZ^d,\bZ) \cong \mathbb{Z}^d, \quad j = 1, \cdots, d \end{equation}
which have $\tau_j(e_k) = \delta_{jk}$, where $ \{ e_k \} $ are the generators of $\mathbb{Z}^d$. (This is sufficient to determine $\tau_j$ uniquely, by linearity.) Again, by projection $G \to \bZ^d$, we obtain corresponding $\tau_j \in \cH^1(G,\mathbb{Z})$.
 We find the LSM anomaly is
\begin{equation}\label{e:classic-lsm}
  \tau_1 \cup \cdots \cup \tau_d \cup \alpha \in \cH^{d+2}(G,\UU(1)),
\end{equation}
where $\cup$ is the so-called ``cup product'' on cohomology \cite{Brown}. This means that our anomalous texture is equivalent in equivariant homology to
\begin{equation}\tau_1 \cup \cdots \cup \tau_d \cup \alpha \cdot [\bR^d] \in \cZ^{d+2}(\bZ^d \times G_{\rm int},\cZ_d(\bR^d,\UU(1))),\end{equation}
where $[\bR^d] \in Z_d(\bR^d,\bZ)$ is the fundamental cycle of $\bR^d$.

Another way to say this is that the K\"unneth formula shows that
\begin{equation}
\cH^{d+2}(G, \UU(1)) \cong \cH^d(\mathbb{Z}^d, \cH^2(\Gint,\UU(1))) \times \cdots,
\end{equation}
where we ignore the other factors. Moreover, we have
\begin{equation}
\cH^d(\mathbb{Z}^d, \cH^2(\Gint,\UU(1))) \cong \cH^2(\Gint, \UU(1)),
\end{equation}
which reflects the total projective class of $\Gint$ per unit cell.

\subsection{Point group symmetry}

Next we study the case where the spatial symmetry is just a point group $G_{\rm pt}$ acting on $\bR^d$ (which always leaves the origin fixed). In that case, by lattice homotopy equivalence (see Section \ref{subsec:latticehomotopy}) one can always concentrate the anomalous texture at the origin. Again the details of the descent sequence can be found in Appendix \ref{a:point-group}, here we just describe the results.

We split the symmetry group as $G = G_{\rm int} \times G_{\rm pt},$ where $G_{\rm pt}$ is the point group. The isotropy group of the origin is the entire group, but we restrict our attention to anomalous textures where only $G_{\rm int}$ acts projectively, which is captured by a class
\begin{equation}\alpha \in \cH^2(G_{\rm int},\UU(1)).\end{equation}
The action of $G_{\rm pt}$ is linear on $\bR^d$ so it defines an element called the Euler class
\begin{equation}e(G_{\rm pt}) \in \cH^d(G_{\rm pt},\bZ^{\rm or}).\end{equation}
We find the LSM anomaly is
\begin{equation}\label{e:pointgrouplsmbody}
e(G_{\rm pt}) \cup \alpha \in \cH^{d+2}(G,\UU(1)^{\rm or}).
\end{equation}
In other words, if we use the K\"unneth formula to write
\begin{equation}
\cH^{d+2}(G, \mathrm{U}(1)^{\mathrm{or}}) = \cH^d(G_{\rm pt}, \cH^2(\Gint, \UU(1))^{\mathrm{or}}) \times ...
\end{equation}
then the anomalous texture maps into the first factor, and the resulting class is induced from the Euler class by the  homomorphism on coefficients
\begin{equation}
\sigma_\alpha : \mathbb{Z}^{\mathrm{or}} \to \cH^2(\Gint, \mathrm{U}(1))^{\mathrm{or}}, \, m \mapsto m\alpha.
\end{equation}

We proceed to describe the Euler class for several point groups in $d = 1, 2, 3$, beginning with some general facts about Euler classes.

An immediate corollary of the formula \eqref{e:pointgrouplsmbody} is that if the point group $G$ preserves an axis, then the LSM anomaly is trivial. Indeed, in this case we can in lattice homotopy send the projective representation at the origin along this axis to infinity symmetrically.

More generally, we can define an Euler class $e(V) \in \cH^k(G,\bZ^{\det V})$ for any $k$-dimensional linear $G$-representation $V$, where the superscript $\det V$ denotes twisting by the determinant $\det V \in \cH^1(G,\bZ_2)$ of the representation. If the representation of $G$ on $\bR^d$ may be written as a direct sum $V_1 \oplus V_2$ of $G$-representations, then
\begin{equation}\label{e:whitneysum}
e(V_1 \oplus V_2) = e(V_1) \cup e(V_2).
\end{equation}
This helps in the computation of the LSM anomaly for simple point groups.

% Furthermore, if the $G$ action on $\bR^d$ factors through $G \xrightarrow{\rho} G_{pt}$, then
% \begin{equation}e(G) = \pi^* e(G_{pt}).\end{equation}
% It suffices therefore to compute the Euler class only for faithful group representations, which have been classified by crystallographers.

Finally, we note that in odd dimensions, all Euler classes are 2-torsion, meaning $2e(V) = 0$. This strongly constrains the behavior of point group LSM theorems in $d = 3$. This 2-torsion phenomenon of the point-group LSM anomaly can be seen in lattice homotopy, by choosing a $G_{\rm pt}$-invariant polyhedron encircling the origin, one can bring in from infinity a copy of a projective $G_{\rm int}$-representation $\alpha$ along a ray passing through the centroid of each $0,2,\ldots$-cell of the polyhedron and a $-\alpha$ along a ray passing through the centroid of each $1,3,\ldots$-cell of the polyhedron. By Euler's formula that relates the number of vertices, edges and faces of a polyhedron (or its higher-dimensional analog), in odd dimensions the result will change the projective representation at the origin by $2\alpha$. Thus in odd dimensions, the anomalous texture $2\alpha$ is anomaly-free for any $\alpha$.

Now we compute the Euler class for all point groups in $d = 1,2,3$, barring the polyhedral point groups in $d = 3$, whose group cohomology does not have a simple form. For $d = 1$, the only nontrivial point group is the reflection group $D_1 \simeq \bZ_2$. If we write the generator of $\cH^1(D_1,\bZ_2)$ as $r$, then we get a natural lift to
\begin{equation}e(D_1) = r \in \cH^1(D_1,\bZ^r),\end{equation}
by embedding $\bZ_2 \sim [0,2) \subset \bZ$, and where $\bZ^r$ indicates twisted coefficients, ie. $\bZ^r$ has differential
\begin{equation}\alpha \mapsto d\alpha - 2 r \cup \alpha.\end{equation}
The subscript of $\bZ^r$ also indicates that $\det V = r \in \cH^1(D_1,\bZ_2)$ for this representation.

In $d = 2$ there are two infinite families of point groups, cyclic $C_n$ and dihedral $D_n$. We have
\begin{equation}e(C_n) = \frac{d\alpha}{n} \in \cH^2(C_n,\bZ),\end{equation}
where $\alpha \in \cH^1(C_n,\bZ_n)$ is a generator, lifted to $\cC^1(C_n,\bZ)$ by the embedding $\bZ_n \sim [0,n) \subset \bZ$ and likewise
\begin{equation}e(D_n) = \frac{d\alpha - 2 r\cup \alpha}{n} \in \cH^2(D_n,\bZ^r),\end{equation}
where $r \in \cH^1(D_n,\bZ_2)$ is the generator corresponding to a reflection and $\alpha \in \cH^1(D_n,\bZ_n^r)$ is a generator corresponding to a rotation, both suitably lifted. Both of the Euler classes are thus a sort of Bockstein operation.

In $d = 3$ there are several infinite families of so-called axial point groups and one finite family of so-called polyhedral point groups. First we discuss the axial point groups. We indicate them by their corresponding Frieze group. All of their 3d representations split into a sum of a rank 1 bundle (along the axis) and a rank 2 bundle (perpendicular to the axis). We have:
\begin{equation}e(p1) = 0\end{equation}
\begin{equation}e(p11g) = \alpha \cup \frac{d\alpha}{2n} \in \cH^3(\bZ_{2n}^\alpha,\bZ^\alpha)\end{equation}
\begin{equation}e(p11m) = r \cup \frac{d\alpha}{n} \in \cH^3(\bZ_n^\alpha \times D_1^r,
\bZ^r)\end{equation}
\begin{equation}e(p1m1) = 0\end{equation}
\begin{equation}e(p211) = r \cup \frac{d\alpha - 2r \cup \alpha}{n} \in \cH^3(D_n^{\alpha,r},\bZ)\end{equation}
\begin{equation}e(p2mg) = \alpha \cup \frac{d\alpha - 2r \cup \alpha}{n} \in \cH^3(D_n^{r,\alpha},\bZ^{r + \alpha})\end{equation}
\begin{equation}e(p2mm) = \tilde r \cup \frac{d\alpha - 2 r \cup \alpha}{n} \in \cH^3(D_1^{\tilde r} \times D_n^{\alpha,r},\bZ^{r + \tilde r})\end{equation}
We have indicated the degree-1 generators of cohomology by the superscripts, see our discussion about $d=2$ above. The two that vanish, $p1$ and $p1m1$, fix the axis. For the others, the first term appearing in the Euler class indicates the $\bZ_2$ quotient of $G$ which reflects the axis. We note that only $p211$ is the only chiral group with a nonvanishing Euler class.

\section{SPT-LSM theorems}
\label{sec:nontraditional_lsm}
One advantage of our general framework is that it can capture not just when an LSM anomaly can be trivialized, but also how it can be trivialized. Thus we can actually state stronger constraints on the ground state than the traditional LSM theorem does. As we discussed in Section \ref{sec:anomalymatching}, the general statement is that the ground state needs to be described by a defect network that can cancel the microscopic anomalous texture. There are cases where, even though invertible ground states are possible, they always must be \emph{non-trivial} SPT phases \cite{Lu_1705_04691,Yang_1705}, ie. we cannot have a completely trivial ground state without breaking the symmetry explicitly or spontaneously. We call such a result an SPT-LSM theorem.

For bosonic systems, we can use the equivariant homology framework of Section \ref{sec:equivarianthomology} to determine the nature of the possible ground states. Recall that an anomalous texture corresponds to an equivariant $-2$-cycle $\alpha \in Z_{-2}^G(X, \UU(1))$. This anomalous texture is trivial in equivariant homology iff $\alpha = D\beta$ for some equivariant $-1$-chain $\beta$. In this case, $\beta$ tells us about the defect network that cancels the anomaly. For example, $\beta$ represents a $r$-skeletal defect network if its components $\beta_k$ are zero for $k > r$.

If the anomaly can be cancelled by an invertible-substrate defect network, but not any $k$-skeletal defect network for $k < d$, then the top-dimensional cells must carry non-trivial phases. Recall that these top-dimensional cells carry invertible phases with symmetry $G_{\mathrm{int}}$, the subgroup of internal symmetries, since if a symmetry leaves top-dimensional cells invariant it must be internal. In this case, we have a ``strong-SPT LSM theorem'': if the ground state is invertible, it must at least be a non-trivial $G_{\mathrm{int}}$-SPT. An example of such a result occurs in fermionic systems with $C_2$ rotation symmetry and a microscopic Majorana zero mode at the origin, as discussed in Section \ref{subsubsec:highercancellation}. In bosonic systems, it can occur in the presence of magnetic translations or non-trivial extensions of a point group symmetry by an internal symmetry, as discussed below. In particular, we will be able to recover the result of Ref.~\cite{Yang_1705} from our general framework.

One can also envision ``crystalline-SPT-LSM'' theorems, where the anomalous texture enforces at the minimum that the ground state be a non-trivial crystalline SPT. However, there is a subtlety in what ``non-trivial'' means in this context. Usually the trivial phase is the one that contains a product state, but in the presence of non-trivial anomalous texture, there \emph{are} no strict product state ground states. What one can instead do is try to diagnose a non-trivial crystalline SPT though the ``higher-order'', i.e. subdimensional, gapless modes that are enforced on the boundary by the symmetry \cite{Benalcazar_1611,Dubinkin_1807,You_1807,Rasmussen_1809}. Note that 0-dimensional gapless modes, which in bosonic systems are just characterized by projective representations, are not particularly well-defined in the presence of a bulk anomalous texture, since we can just push projective representations from the bulk onto the boundary. Nevertheless, we expect that higher-dimensional gapless modes remain well defined, so that there can be LSM theorems that guarantee non-trivial $k$-th order SPTs for $k < d$. This corresponds to cases where the anomalous texture can be cancelled by a $(d-k+1)$-skeletal defect network.

In the remainder of this section we will go into more detail on strong SPT-LSM theorems.

\subsection{Strong SPT-LSM with magnetic translations}

As a first concrete example, we consider a generic lattice $\Lambda \subset \bR^d$ with translation group $\bZ^d$ and internal symmetry group $G_{\rm int}$. We assume for simplicity that $\Lambda$ consists of a single $\bZ^d$-orbit. Thus, an anomalous texture on $\Lambda$ is determined by the class of projective representation of $G_{\rm int}$ at any of the sites, which we denote $\alpha \in \cH^2(G_{\rm int},\UU(1))$. Recall from Section \ref{s:equiv-pushforward} that in the case where $G = \bZ^d \times G_{\rm int}$, we can represent the corresponding LSM anomaly in group cohomology by
\begin{equation}\label{e:lsm-anom-2}
  \tau_1 \cup \cdots \cup \tau_d \cup \alpha \in \cH^{d+2}(\bZ^d \times G_{\rm int},\UU(1)),
\end{equation}
where $\tau_j \in \cH^1(\mathbb{Z}^d, \UU(1))$ gets interpreted as an element of $\cH^1(G, \UU(1))$ through the projection $G \to \mathbb{Z}^d$, and similarly for $\alpha$.

Now suppose we consider the case where $G$ fails to be a direct product. That is, we make a modification to the group multiplication law such that we have a non-trivial central extension
\begin{equation}\label{e:mag-symm-ext}
  G_{\rm int} \to G_\beta \to \bZ^d,
\end{equation}
where
\begin{equation}\beta \in \cH^2(\bZ^d,Z(G_{\rm int}))\end{equation}
classifies the extension. Here $Z(G_{\rm int}) \subset G_{\rm int}$ denotes the center of $G_{\rm int}$. $\beta$ is determined by a choice of commutation relations $\hat t_1 \hat t_2 \hat t_1^{-1} \hat t_2^{-1} \in Z(G_{\rm int})$ for each pair of translation generators. On the other hand, it determines a commutation relation by
\begin{equation}\hat t_1 \hat t_2 \hat t_1^{-1} \hat t_2^{-1} = \beta(t_1,t_2)\beta(-t_1,-t_2).\end{equation}
We thus interpret $\hat t_1, \hat t_2 \hat t_1^{-1} \hat t_2^{-1}$ as a $G_{\rm int}$-flux $\Phi_{ij} \in G_{\rm int}$ going through the plaquette spanned by $t_1$ and $t_2$. For this reason, a translations which exist in an extension as above are referred to as magnetic translations. We can write a cocycle representative for $\beta$ as
\begin{equation}\beta = \sum_{1 \le i<j \le d} \Phi_{ij} \tau_i \cup \tau_j.\end{equation}

Since the magnetic symmetry has the same isotropy group, namely $G_{\rm int}$ for each site as $G_0$ had, our original anomalous texture may be considered an anomalous texture also for $G_\beta$. In general, the descent sequence could yield an anomaly different from \eqref{e:lsm-anom-2}. However, there is a large enough set of symmetry classes to illustrate the SPT-LSM phenomenon where we can arrive at the same anomaly.

\subsubsection*{Split internal symmetry and SPT-LSM}

To this end, we momentarily restrict our attention to internal symmetries of the form
\begin{equation}G_{\rm int} = G_{\rm proj} \times G_{\rm flux}.\end{equation}
We further assume that in our anomalous texture, only $G_{\rm proj}$ acts projectively, by $\alpha \in \cH^2(G_{\rm proj},\UU(1))$, and that the magnetic symmetry has $\Phi_{ij} \in Z(G_{\rm flux})$ for all $i,j$. This guarantees that $\alpha$ has an extension to $\cH^2(G_\beta,\UU(1))$. Using this extension in the spectral sequence of Appendix \ref{a:classic-lsm}, we find the same LSM anomaly cocycle \eqref{e:lsm-anom-2}.

Let $m$ be the order of $\alpha$, that is the smallest positive integer such that $m\alpha = 0 \in \cH^2(G_{\rm proj},\UU(1))$. Some examples are:
\begin{itemize}
  \item A lattice of spin-1/2s with $G_{\rm proj} = SO(3)$ has $m = 2$.
  \item A lattice of $SU(n)$ fundamentals with $G_{\rm proj} = PSU(n)$ has $m = n$.
  \item A lattice of Kramers doublets with $G_{\rm proj} = \bZ_2^T$ has $m = 2$.
\end{itemize}
We assume for the rest of the discussion that $m$ is finite, as it is for all compact Lie groups. We can take $\alpha$ such that $m\alpha \in \bZ$.

To simplify the discussion we now focus on $d = 2$ for which
\begin{equation}\beta = \Phi \tau_1 \cup \tau_2,\end{equation}
where $\Phi \in Z(G_{\rm flux})$ represents the flux-per-plaquette. In this case, there is an SPT-LSM theorem if we can find a homomorphism
\begin{equation}\label{e:ffunction}
  f:G_{\rm flux} \to \bZ_m
\end{equation}
such that
\begin{equation}\label{e:ffunctioncond}
  f(\Phi) = 1 \mod m.
\end{equation}
Indeed, a universal property of the extension cocycle is that there is a 1-cochain $A \in \cC^1(G_\beta,Z(G_{\rm int}))$ with
\begin{equation}\label{eqnmagtransflux}
  \delta A = \beta = \Phi \tau_1 \cup \tau_2.
\end{equation}
Using $f(A)$ (with arbitrary extension of $f$ to $G_{\rm proj}$) we can write the LSM anomaly as a coboundary
\begin{equation}\label{e:lsmanomtriv}
  \tau_1 \cup \tau_2 \cup \alpha = \delta \left( f(A) \cup \alpha\right) \mod 1.
\end{equation}
Indeed, by the linearity of $f$,
\begin{equation}\delta f(A) = f(\delta A) = f(\Phi) \tau_1 \cup \tau_2 = \tau_1 \cup \tau_2 \mod m.\end{equation}
Thus $f(A) \cup \alpha$ defines an invertible defect network which cancels the LSM anomaly of the texture.

The cochain $A$ has the additional property that when restricted to $G_{\rm int}$, it is the tautological 1-cocycle which generates $\cH^1(G_{\rm int},G_{\rm int})$. Thus, when we look at our invertible defect network as a $G_{\rm int}$ defect network, it is anomaly-free, and defines the $G_{\rm int}$-SPT
\begin{equation}\label{e:symmenfSPT}
  f(A) \cup \alpha \in \cH^3(G_{\rm int},\UU(1)).
\end{equation}

There is some ambiguity in the SPT \eqref{e:symmenfSPT} given by different choices of the function $f$ in \eqref{e:ffunction}. For instance, the choice of
\begin{equation}f|_{G_{\rm proj}}:G_{\rm proj} \to \bZ_m\end{equation}
is completely arbitrary, since it does not affect the value of $f$ on $\Phi \in G_{\rm flux}$. This can shift the pure-$G_{\rm proj}$ SPT class in \eqref{e:symmenfSPT}, and by choosing $f|_{G_{\rm proj}}$ to be the zero map, we can arrange our defect network to be in the trivial pure-$G_{\rm proj}$ SPT and still cancel the LSM anomaly. On the other hand, we cannot choose $f|_{G_{\rm flux}}$ to be the zero map, because of the condition \eqref{e:ffunctioncond}. Thus, we always find ourselves in some non-trivial mixed $G_{\rm proj} \times G_{\rm flux}$ SPT.

The simplest examples are made by taking $G_{\rm flux} = \bZ_m$. Then we can choose an $f:\bZ_m \to \bZ_m$ satisfying $f(\Phi) = 1$ iff $\Phi$ is coprime to $m$. Taking $G_{\rm proj} = SO(3)$ and a lattice of spin-1/2s, with $G_{\rm flux} = \bZ_2$ and a $\pi$-flux per plaquette, the LSM anomaly is trivialized and we find the mixed $\bZ_2 \times SO(3)$ SPT with class
\begin{equation}\frac{1}{2} A \cup w_2(SO(3)) \in \cH^3(\bZ_2 \times SO(3),\UU(1)).\end{equation}
This SPT is characterized by the $\bZ_2$ $\pi$-flux carrying a spin-1/2. Since there is a $\pi$-flux per plaquette, we can pair them into singlets with the spin-1/2s on the sites in a translation-invariant manner, hence the LSM anomaly is trivialized. All the examples constructed this way have a similar intuitive picture, because of the form of \eqref{e:symmenfSPT}.

\subsubsection*{More general approach}

In this section, we consider a more general setup, where there are internal symmetries which act both projectively and are involved in the magnetic translations. In this case, it is not as simple to use the method in Appendx \ref{a:classic-lsm} to compute the group cohomology anomaly. However, one can still show that only
\begin{equation}\label{eqnclassiclsm3}
  \tau_1 \cup \cdots \cup \tau_d \cup \alpha \in \cH^{d+2}(G_\beta,\UU(1))
\end{equation}
satisfies the condition that breaking any translation symmetry trivializes the anomaly and compactifying on a torus of one unit cell yields the projective symmetry action of $G_{\rm int}$ in class $\alpha \in \cH^2(G_{\rm int},\UU(1))$.

Note that $\alpha$ does not necessarily extend to a class in $\cH^2(G_\beta,\UU(1))$, but intuitively for any lift $\hat \alpha$, $\delta \hat \alpha$ is proportional to $\beta$, and $\tau_1 \cup \cdots \cup \tau_d \cup \beta = 0$. More rigorously, one can check that because $\cH^*(\bZ^d,A)$ vanishes above degree $d$ with any coefficient group $A$, there is nowhere for the differentials in the Lyndon-Hoschild-Serre (LHS) spectral sequence emanating from \eqref{eqnclassiclsm3} to land, so \eqref{eqnclassiclsm3} forms a cocycle. However, there are differentials which can make \eqref{eqnclassiclsm3} exact, in particular, coming from \eqref{eqnmagtransflux}. In this case, we can have an SPT-LSM theorem.

An example in $d = 2$ with $G_{\rm int} = \bZ_2^C \times \bZ_2^T$, where $\bZ_2^T$ is an order-2 time reversal symmetry, and $\bZ_2^C$ is an order-2 unitary symmetry, has projective symmetry class
\begin{equation}\frac{1}{2} A \cup A \in \cH^2(\bZ_2 \times \bZ_2^T,\UU(1)),\end{equation}
where $A$ is the $\bZ_2$ cohomology generator. This projective symmetry class can be realized as
\begin{equation}T^2 = 1 \qquad C^2 = -1 \qquad CTCT = 1,\end{equation}
eg. in a 2d Hilbert space with $T$ given by complex conjugation and $C = i \sigma^y$. We consider a magnetic translation with a $\bZ_2$ $\pi$-flux per unit cell. The modification \eqref{eqnmagtransflux} from this extension is
\begin{equation}\delta A = \tau_1 \cup \tau_2.\end{equation}
Thus, the anomaly \eqref{eqnclassiclsm3} is exact, given by
\begin{equation}\delta\left( \frac{1}{2} A \cup A \cup A\right) = \frac{3}{2} A \cup A \cup \tau_1 \cup \tau_2 = \frac{1}{2} A \cup A \cup \tau_1 \cup \tau_2,\end{equation}
where we cancelled an integer piece $A^2 \tau_1 \tau_2$. This is the only SPT phase available in this symmetry class, so it is forced by anomaly cancellation. Note that the $\frac{1}{2} A^3$ SPT we found may be protected just by the unitary $\bZ_2$, but without time reversal there is no anomaly, so we could also trivialize the anomalous texture without introducing a nontrivial SPT, eg. by rephasing the $C$ operator to $\sigma^y$. We note that by a change of basis $T \mapsto CT$, we can re-express this example in terms of a split symmetry with a $CT$-Kramers doublet per site. In that sense, the approach taken here for this example was unnecessary, but we hope it illustrates the general nature of these computations.

We end this section with a proof of Theorem-II from \cite{Yang_1705}. Let $\omega \in \cZ^3(G_{\rm int},\UU(1))$. Using $A \in \cC^1(G_\beta,Z(G_{\rm int}))$, described above, we define an extension of $\omega$ to $G_\beta$ by
\begin{equation}\hat \omega(g_1,g_2,g_3) = \omega(A(g_1),A(g_2),A(g_3)).\end{equation}
Since $Z(G_{\rm int})$ is abelian, $[\omega]$ has a representative by a tri-linear cocycle. One can use this tri-linearity and the cocycle equation to derive a ``chain rule"
\begin{equation}\delta \hat \omega = \omega(\delta A,A,A) - \omega(A,\delta A,A) +
\omega(A,A,\delta A).\end{equation}
Using the expression \eqref{eqnmagtransflux} for $\delta A$, we can write
\begin{equation}(\delta \hat\omega)(g_1,g_2,g_3,g_4)\end{equation}\begin{equation} = \omega(\Phi,g_3,g_4) \tau_1(g_1) \tau_2(g_2) - \omega(g_1,\Phi,g_4) \tau_1(g_2) \tau_2(g_3)\end{equation}\begin{equation} + \omega(g_1,g_2,\Phi) \tau_1(g_3) \tau_2(g_4).\end{equation}
Observe that this is precisely what we computed above. Then, we use the fact that the cohomology $\cH^2(\bZ^2,\cH^2(G_{\rm int},\UU(1)))$ is generated by cup products, so there is a universal cohomology operation we can add to the above so that
\begin{equation}\delta\left( \hat \omega + \cdots \right) = \end{equation}\begin{equation}(\omega(\Phi,g_1,g_2) - \omega(g_1,\Phi,g_2) + \omega(g_1,g_2,\Phi)) \tau_1(g_2) \tau_2(g_3).\end{equation}
Thus, we can trivialize the LSM anomaly with the $G_{\rm int}$-SPT $\omega$ and magnetic flux $\Phi \in G_{\rm int}$ if
\begin{equation}\omega(\Phi,g_1,g_2) - \omega(g_1,\Phi,g_2) + \omega(g_1,g_2,\Phi) = -\alpha\end{equation}
in $\cH^2(G_{\rm int},\UU(1))$. The converse follows from the fact that there is only one differential in the LHS spectral sequence, from $H^3(G_{\rm int},\UU(1)) \to H^2(\bZ^2,H^2(G_{\rm int},\UU(1)))$, which can trivialize the LSM anomaly.

\subsection{Strong SPT-LSM for point groups}

Another simple case to consider is SPT-LSM theorems for an anomalous texture occupying a single point. For $G_0 = G_{\rm pt} \times G_{\rm int}$, with only $G_{\rm int}$ acting projectively, the LSM anomaly again a simple form in \eqref{e:pointgrouplsmbody}:
\begin{equation}e(G_{\rm pt}) \cup \alpha,\end{equation}
where $\alpha \in \cH^2(G_{\rm int},\UU(1))$ describes the projective representation of the internal symmetries. The strategy to find an SPT-LSM theorem is the same as with translations: we look for an extension
\begin{equation}G_{\rm int} \to G_\beta \to G_{\rm pt}\end{equation}
such that the Euler class $e(G_{\rm pt})$ becomes exact in $\cH^d(G_\beta,\bZ_m)$, where $m$ is the order of $\alpha$, as before.

The simplest cases to study are where the internal symmetry splits into
\begin{equation}G_{\rm int} = G_{\rm spin} \times G_{\rm proj},\end{equation}
where now $G_{\rm spin}$ plays the role of $G_{\rm flux}$ in the translation case, so named because it is the part of the symmetry group involved in spin-orbit coupling. That is, we take our extension cocycle $\beta$ to be valued in $Z(G_{\rm spin})$, so $G_{\rm proj}$ remains as a split factor in $G_\beta$. Meanwhile we also assume that only $G_{\rm proj}$ acts projectively in the anomalous texture, so that we may use the analysis of Appendix \ref{a:point-group} to derive the same LSM anomaly for $G_\beta$.

To construct some examples, we begin with $d = 2$. The Euler class for the cyclic point groups may be written
\begin{equation}e(C_n) = \frac{dA}{n},\end{equation}
where $A \in \cH^1(\bZ_n,\bZ_n)$ is the generator of the cohomology ring. The LSM anomaly associated to a projective $G_{\rm proj}$ representation with class $\alpha \in \cH^2(G_{\rm proj})$ siting at the rotation center can thus be expressed in group cohomology as
\begin{equation}\frac{dA}{n} \cup \alpha.\end{equation}
We're looking for extensions of $C_n$ by $G_{\rm spin}$ such that $dA/n$ is exact modulo the order $m$ of $\alpha$. As with translations, the simplest choice to take $G_{\rm spin} = \bZ_m$ and $dA/n \in \cH^2(C_n,\bZ_m)$ to classify the extension, ie. a $2\pi$ rotation amounts to a $\bZ_m$ generator. In this case, we find a mixed $G_{\rm spin} \times G_{\rm flux}$ SPT with class
\begin{equation}C \cup \alpha,\end{equation}
where $C \in \cH^1(G_{\rm spin}, G_{\rm spin})$ is the cohomology ring generator. All the examples of $\alpha$'s from the previous section may thus be ported into this setting. As before, there is an ambiguity which leads to the symmetry-enforced SPTs forming a torsor over $\cH^3(G_{\rm spin},\UU(1))$, in this case an ambiguity by the SPTs
\begin{equation}\frac{k}{m} C^3 \in \cH^3(G_{\rm spin},\UU(1))\end{equation}
for $k \in \bZ_m$.

There is also a simple family of examples in $d = 3$, where we make use of the Wu formula
\begin{equation}e(G_{\rm pt}) = Sq^1 w_2(G_{\rm pt}),\end{equation}
where $w_2(G_{\rm pt})$ is the 2nd Stiefel-Whitney class of the action of $G_{\rm pt}$ on $\bR^3$, and $Sq^1$ is the first Steenrod square, or the Bockstein. We take $G_{\rm flux} = \bZ_2$ and $\beta = w_2(G_{\rm pt})$, which means $G_\beta$ acts in a half-spin representation on the anomalous texture. As before, we obtain a cochain $A \in \cC^1(G_{\beta},\bZ_2)$ with $\delta A = w_2(G_{\rm pt})$. Then, by the linearity of $Sq^1$,
\begin{equation}\delta Sq^1 A = Sq^1 w_2(G_{\rm pt}) = e(G_{\rm pt}).\end{equation}
This leads us a to symmetry-enforced mixed SPT with class
\begin{equation}Sq^1 A \cup \alpha \in \cH^4(\bZ_2 \times G_{\rm proj},\UU(1)),\end{equation}
where $\alpha \in \cH^2(G_{\rm proj},\UU(1))$ is a projective symmetry class.

\section{Rigorous status of our results}
\label{sec:rigorous}
Throughout this paper, we have stated results assuming that the general framework of Refs.~\cite{Thorngren_1612,Else_1810} holds. However, this framework is based on various assumptions.
 Therefore, it would be nice to find proofs of our LSM criterion from first principles, independently of this framework.

 First of all, we point out that the arguments of Section I of Ref.~\cite{Else_1810} (which did not require any assumptions about the framework classifying crystalline topological phases) showed that if there is a symmetric gapped ground state whose correlations go strictly to zero at distances greater than some scale $\xi$ smaller than the size of the cells introduced in Section \ref{subsec:defectnetworks}, then it must be deformable to a defect network form. Less rigorously, one can argue that the same ought to be true for a state with finite correlation length $\xi$ that is much less than the size of the cells. By similar arguments, one finds that if such a state exists in the presence of a microscopic anomalous texture, it must be deformable to a degree-0 defect network with whose emergent anomalous texture cancels the microscopic anomalous texture. However, given that a typical ground state in a realistic system would not satisfy such a condition on the correlation length, such a conclusion may not be very satisfactory. Moreover, the statement that bosonic invertible-substrate defect networks and their anomalies are described by equivariant homology, as discussed in Section \ref{sec:equivarianthomology}, does depend on the ``in-cohomology'' model of defect networks, which is difficult to justify from first principles.

In the remainder of this section, we will give a partial proof, starting from first principles, of the LSM criterion in terms of equivariant homology from Section \ref{sec:equivarianthomology}, for spatial dimensions $d \leq 2$.

\subsection{Assumptions and statement of the theorem}
\label{sec:assumptions_and_statement}
We consider the case where the full symmetry group $G$ decomposes as the product $G = G_{\mathrm{int}} \times \Gspace$, where $G_{\mathrm{int}}$ acts internally, and $\Gspace$ acts on $X = \mathbb{R}^d$. Recall that in this case we can exploit the K\"unneth formula for equivariant homology, as discussed in Section \ref{subsec:eqhomkunneth}. Let $P := \cH^2(G_{\mathrm{int}}, \mathrm{U}(1))$. We want to show that if the anomalous texture leads to a non-trivial element in the $H_0^G(\mathbb{R}^d, P)$ factor of \eqnref{tophomology_kunneth}, then a gapped symmetric ground state cannot be an invertible state. (In particular, this will cover cases 1 and 2 in the relevant symmetries for quantum magnets discussed in Section \ref{subsec:texture_examples}).

We now formally state the following assumptions which we believe to be eminently reasonable:

\begin{description}
\item[Assumption 1] \label{assumption:first} The space of $G_{\mathrm{int}}$-symmetric gapped ground states in $d$ spatial dimensions that are in the trivial phase if the symmetry is neglected can be formalized as a topological space $\Omega_d$, such that the classification of $G_{\mathrm{int}}$-symmetric SPT phases corresponds to $\pi_0(\Omega_d)$. Moreover, there is a $\Gspace$ action on $\Omega_d$ corresponding to the action of $\Gspace$ on states.

\item[Assumption 2]
The group cohomology classification of SPT phases in (1+1)-D with $\Gint$ symmetry in one spatial dimension is correct, at least as a partial classification, in the sense that there is a homomorphism
\begin{equation}
\label{eq:Pmap}
\pi_0(\Omega_1) \to \cH^2(\Gint, \UU(1)) = P,
\end{equation}
such that the representative ground states constructed, for example, in Ref.~\cite{Chen_1106} map to the appropriate elements of $P$.

\item[Assumption 3] \label{assumption:pumps} There is a map from $\pi_{1}(\Omega_2) \to \pi_0 (\Omega_{1})$, and in particular (via Assumption 1) a map $\pi_1(\Omega_2) \to P$. This represents the (1+1)-D SPT phase pumped to the boundary when the bulk state goes through a loop \cite{Else_1602}.

%\item[Assumption 2] There is an isomorphism
%\begin{equation}
%\label{eq:Pmap}
%\pi_0(\Omega_1) \cong \cH^2(\Gint, \UU(1)) = P,
%\end{equation}
% In other words, the group cohomology classification of (1+1)-D SPTs is correct.

%\item[Assumption 3] \label{assumption:pumps} There is an isomorphism $\pi_1(\Omega_2) \cong \pi_0(\Omega_1)$. This represents the (1+1)-D SPT phase pumped to the boundary when the bulk state goes through a loop [: citations].

%\item[Assumption 3] \label{assumption:last} There is a continuous function $\otimes : \Omega_d \times \Omega_d \to \Omega_d$ (physically, it represents stacking), and there is an element $\psi_* \in \Omega_d$ (the ``trivial state'') and continuous maps $f,g : [0,1] \times \Omega_d \to \Omega_d$ such that $f(0,\omega) = g(0,\omega) = \omega$, $f(1,\omega) = \psi_* \otimes \omega$, $g(1,\omega) = \omega \otimes \psi_*$ for all $\omega \in \Omega_d$. This is just saying that tensoring with a trivial state is equivalent to doing nothing up to some deformation (where the deformation can itself be chosen continuously as a function of $\omega$).

\end{description}

Now we can state our main theorem:
\begin{thm}
\label{thm:mainthm}
Consider a system with symmetry $G = \Gspace \times G_{\mathrm{int}}$, where $G_{\mathrm{int}}$ is finite, and a representation of this symmetry with an anomalous texture in $d \leq 2$ spatial dimensions that is non-trivial only in the third factor of \eqnref{topsitekunneth}, and leads to a non-trivial element in equivariant homology $H_0^G(\mathbb{R}^d, P)$.
Then, given Assumptions 1--3, any gapped symmetric ground state $\ket{\Psi}$ cannot be in the trivial $G_{\mathrm{int}}$ SPT phase.
\end{thm}

We can immediately strengthen this result in a few ways. Firstly, the restriction of finite $G_{\mathrm{int}}$ obviously does not prevent us from considering continuous internal symmetry groups such as $\SO(3)$, so long as there is a subgroup that gives the same $\cH^2(G_{\mathrm{int}}, \mathrm{U}(1))$; for example, for $G_{\mathrm{int}} = \SO(3)$ we can just consider the $\mathbb{Z}_2 \times \mathbb{Z}_2$ subgroup generated by $\pi$ rotations about two orthogonal axes.

Secondly, we can rule out other phases than just the trivial $G_{\mathrm{int}}$ SPT phase. Suppose that that the ground state is a non-trivial $G_{\mathrm{int}}$ SPT phase (or invertible phase). Furthermore, suppose that the inverse SPT phase can also be realized in a different system with $\Gspace \times G_{\mathrm{int}}$ symmetry, with spins \emph{not} carrying any projective representations of $G_{\mathrm{int}}$ (that is, trivial anomalous texture). For group-cohomology phases, the construction of Ref.~\cite{Chen_1106} (combined with the observation one can always find a $\Gspace$-invariant triangulation and branching structure \cite{Thorngren_1612}) gives an explicit such realization; we conjecture that it is always possible for any bosonic SPT (or invertible) phase. (However, the analogous statement is not true for fermions, as the $p+ip$ superconductor example from Section \ref{subsubsec:highercancellation} demonstrates.) Then we can adjoin this inverse SPT without changing the anomalous texture, and the resulting state is now in the trivial $G_{\mathrm{int}}$ SPT phase. Therefore, to rule out any invertible ground state (satisfying the condition mentioned), it is sufficient to rule out a ground state that is in the trivial $G_{\mathrm{int}}$ SPT phase.

Secondly, we can lift the assumption that the anomalous texture is non-trivial only in the third factor of \eqnref{topsitekunneth}. To see this, suppose we have a ground state $\ket{\Psi}$ that is invariant under a representation of $\Gspace \times G_{\mathrm{int}}$ (possibly with anomalous texture). Let $U(g)$ be the representation of $G_{\mathrm{int}}$ and $S(g)$ be the representation of $\Gspace$. Then observe that the complex conjugated state $\ket{\Psi^*}$ is invariant under the complex conjugated representation, which has the opposite anomalous texture. Now consider layering two copies of $\ket{\Psi}$ and one copy of $\ket{\Psi^*}$, i.e.\ consider the state $\ket{\Psi}_{ABC} = \ket{\Psi}_A \otimes \ket{\Psi}_B \otimes \ket{\Psi^*}_C$. This state is invariant under $U_A(g), U_B(g)$ and $U_C^{*}(g)$, for $g \in G_{\mathrm{int}}$ where $U_A(g)$, etc. denotes the representation of $G_{\mathrm{int}}$ acting on the respective layer, and similarly it is invariant under $S_A(g), S_B(g), S_C^{*}(g)$ for $g \in G_{\mathrm{int}}$. In particular, $\ket{\Psi}_{ABC}$ is invariant under the representation of $\Gspace \times G_{\mathrm{int}}$ generated by $S_A(g) S_C^{*}(g)$, $g \in \Gspace$, and $U_A(g) U_B(g) U_C^{*}(g)$, $g \in G_{\mathrm{int}}$. The anomalous texture of this new representation is the same as the original one in the third factor of \eqnref{topsitekunneth}, but the other two factors have been cancelled off. Moreover, if $\ket{\Psi}$ is invertible then so is $\ket{\Psi}_{ABC}$.

The rest of Section \ref{sec:rigorous} will be devoted to a proof of Theorem \ref{thm:mainthm}.

\subsection{Defining the Hilbert space and symmetry}
\label{subsec:hilbert}

For concreteness, for a given an anomalous texture $p$ we will want to consider a ``canonical'' representation $U_p(g)$ with that anomalous texture. This canonical representation has the property that $U_p(g) \otimes U_p(g)^*$ admits a symmetric product state ground state. To prove the theorem for general anomalous textures, we will need to consider tensor products $U_p(g) \otimes U_0'(g)$, where $U_0'(g)$ is some representation of $G$ that admits a product state symmetric ground state. Let us show that it is sufficient to prove Theorem \ref{thm:mainthm} for representations of this form. Suppose that $U_p'(g)$ is some representation of $G$ with anomalous texture $p$. The tensor product representation $U_p'(g) \otimes U_p(g)^*$ defines a trivial anomalous texture, so it admits a symmetric product ground state, perhaps after enlarging the on-site Hilbert space by another representation defining a trivial anomalous texture. We denote the result $U_0'(g)$.

Then we can consider the tensor product representation $U_p''(g) = U_p(g) \otimes U_0'(g)$. We see that $U_p''(g)$ still has the anomalous texture $p$. Moreover, if $U_p'(g)$ admits a symmetric gapped ground state that is in the trivial $G_{\mathrm{int}}$ phase, then so does $U_p'(g) \otimes U_p^{*}(g) \otimes U_p(g)$ and therefore so does $U_p''(g)$. In what follows, for brevity we will ignore the possibility of a non-trivial $U_0'(g)$, but it is easy to fix the arguments to take it into account.

Next we will define our canonical representation $U_p(g)$.
We assume that we have a triangulation of the space $X$ with branching structure, ie. an orientation of all edges so that no face forms a cycle, such that $\Gspace$ acts on the triangulation in such a way that the branching structure is invariant under the $\Gspace$ action. This is a special case of a cell decomposition as introduced in Section \ref{subsec:defectnetworks}, and we require the same condition on the $G$ action that we mentioned there. We assume that the Hilbert space of the system corresponds to a spin carrying a $|\Gint|$-dimensional Hilbert space at each 0-cell in the triangulation. (We can also have some additional degrees of freedom not transforming under $\Gint$ at each site; these will not affect the argument). We assume that the $\Gspace$ action is just by permutation of the Hilbert space of the sites, and we write this action as $S(g)$, $g \in \Gspace$.

For each element of $\pi \in P = \cH^2(G_{\mathrm{int}}, \UU(1))$, we let $\omega_\pi(g_1, g_2)$ be a corresponding 2-cocycle. We can always choose $\omega_\pi$ so that it is multiplicative in $\pi$, i.e. $\omega_{\pi_1 + \pi_2} = \omega_{\pi_1} \omega_{\pi_2}$. We will also want to work with the corresponding homogeneous cocycle $\nu_\pi(g_1, g_2, g_3)$, which satisfies the homogeneous cocycle condition $\delta \nu_\pi=0$ with $\delta$ given by \ref{eq:homogeneous_coboundary}. Moreover, we define a collection of unitary operators $u_\pi(g)$, $g \in G_{\mathrm{int}}$, acting on $\mathbb{C}^{|G_{\rm int}|}$ according to
\begin{equation}
u_\pi(g) \ket{h} = \omega_\pi(g,h) \ket{gh}.
\end{equation}
One can check that $u_\pi(g)$ is a projective representation of $G_{\rm int}$. Indeed, we have [recall the definitions of $\mu(g)$, $\sigma(g)$ from Section \ref{sec:symrep}]:
\begin{align}
& u_\pi (g) K^{\mu(g)} u_\pi (h) K^{\mu(h)} \ket{k}
\\ &=  \omega_\pi^{\sigma(g)}(h,k) K^{\mu(g)} u_\pi(g) \ket{hk} \\
                        &= \omega_\pi^{\sigma(g)}(g, hk) \omega_\pi(h,k) \ket{ghk}, \\
                        &= \omega_\pi(g,h) \omega_\pi(gh, k) \ket{ghk}
\end{align}
using the cocycle condition, whereas
\begin{align}
u_\pi(gh) K^{\mu(gh)} \ket{k} &=  \omega_\pi(gh,k) \ket{ghk}.
\end{align}
 So we conclude that
\begin{equation}
u_\pi(g) K^{\mu(g)} u_\pi(h) K^{\mu(h)} = \omega_\pi(g,h) u_\pi(gh) K^{\mu(gh)},
\end{equation}
and we see that the class of the projective representation precisely corresponds to $\pi$. Then, for an anomalous texture, which is represented by a $0$-chain $p \in C_0(X,P)$, we define the Hilbert space
\begin{equation}\label{eqncanonhilb}
\bigotimes_{s \in X_0} \mathbb{C}^{|G_{\rm int}|}
\end{equation}
with $G_{\mathrm{int}}$ acting by
\begin{equation}
\cU_p(g) = \left(\bigotimes_{s \in X_0} u_{p_s}(g) \right) K^{\mu(g)},
\end{equation}
Thus the canonical action of $G = G_{\mathrm{int}} \times \Gspace$, i.e.\ $U_p(g)$, is generated by $\cU_p(g)$, $g \in G_{\mathrm{int}}$, and $S(g)$, $g \in \Gspace$.

\subsection{Chains, states, and pumps}
\label{subsec:pumps}

\begin{figure}
\input{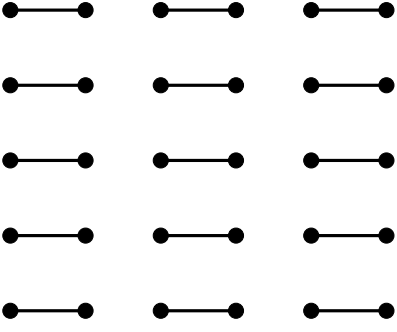}
\caption{\label{fig:chains0}
A 1-chain $\alpha \in C_1(\mathbb{R}^d,P)$ intuitively corresponds to spatial arrangements of 1-D SPTs classified by $P = \cH^2(G_{\rm int},\UU(1))$, ie. a 1-skeletal defect network. Such a configuration is represented by the wavefunction $\ket{\Psi_\alpha}$. This wavefunction is symmetric under $\Gint$ in the presence of an anomalous texture described by the boundary $\partial \alpha \in C_0(\mathbb{R}^d,P)$.
}
\end{figure}

\begin{figure}
\input{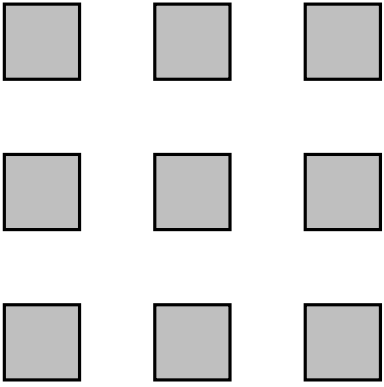}
\caption{\label{fig:chains1}
If a 1-chain $\alpha' \in C_1(\mathbb{R}^d,P)$  can be expressed as $\alpha' = \partial \beta$ for some $\beta \in C_2(\mathbb{R}^d,P)$, then it represents a configuration which can be created out of the trivial product state. Roughly, we can imagine that on each 2-cell $\sigma$, we pump the 1-D SPT described by $\beta_\sigma$ onto the boundary of the 2-cell. For example, this figure depicts a configuration of 1-D SPTs (on the boundary of the shaded squares) being created out of the vacuum by a 2-D pump on the shaded squares. More generally, two defect networks $\alpha$, $\alpha' \in C_1(\mathbb{R}^d,P)$ can be related by such a pumping process whenever $\alpha - \alpha' = \partial \beta$.
}
\end{figure}

In this section, we will prove some technical results that will be useful later on. Specifically, we give a construction showing that any $1$-chain $\alpha \in C_1(\mathbb{R}^d,P)$ gives a state $\ket{\Psi_\alpha}$ defined on the degrees of freedom introduced in the previous subsection, which is $\Gint$-symmetric with the anomalous texture $\partial \alpha \in C_0(X,P)$. Then we show that any $2$-chain $\beta \in C_1(\mathbb{R}^d,P)$ gives a continuous family of states connecting $\ket{\Psi_\alpha}$ and $\ket{\Psi_{\alpha + \partial \beta}}$.

\newcommand{\gvec}{\vec{g}}
\begin{lemma}
\label{lem:first}
\emph{\textbf{Wavefunction}} For every $1$-chain $\alpha \in C_1(\mathbb{R}^d,P)$, there is a corresponding state $\ket{\Psi_\alpha}$ in the Hilbert space \eqref{eqncanonhilb} above, such that:
\begin{enumerate}
 \item $\cU_{\partial\alpha}(g) \ket{\Psi_\alpha} = \ket{\Psi_\alpha}$ for all $g \in G_{\mathrm{int}}$, and
 \item $\ket{\Psi_{g\alpha}} = S(g) \ket{\Psi_\alpha}$ for all $g \in \Gspace$.
 \end{enumerate}
The physical interpretation is shown in Figure \ref{fig:chains0}.
\begin{proof}
In fact, this is a special case of the construction from Appendix \ref{appendix:lattice_wfns}, but for completeness we give the proof here as well.
Here and in what follows, we will use the notation $\ket{\gvec}$ to denote $\ket{g_1, \cdots, g_N}$, that is, a basis state in the Hilbert space of the whole system, where $N$ is the number of vertices.

Let $\nu_{\alpha_l}(g_1,g_2,g_3)$ be the homogeneous 2-cocycle associated to $\alpha_l \in \cZ^2(\Gint,\UU(1))$ where $l$ is an oriented edge in $X_1$. We define
\begin{equation}
\label{eq:Psialpha}
\ket{\Psi_\alpha} = \sum_{\gvec} \left( \prod_{ s_1 s_2 = l \in X_1}  \nu_{\alpha_l}(1, g_{s_1}, g_{s_2}) \right) \ket{\gvec},
\end{equation}
where the product is over all edges $l$ oriented from $s_1$ to $s_2$. Then we have
\begin{align}
& \bra{\vec{g}} U_{\partial\alpha}(g) \ket{\Psi_\alpha} \\
 &=
\left( \prod_{s \in X_0} \nu_{(\partial\alpha)_s}(1, g, g_s) \right) \left( \prod_{s_1 s_2 = l \in X_1} \nu_{\alpha_l}(g,  g_{s_1}, g_{s_2}) \right)  \\
&= \prod_{s_1 s_2 = l \in X_1} \nu_{\alpha_l}(1, g, g_{s_1}) \nu_{\alpha_l}^{-1}(1, g, g_{s_2}) \nu_{\alpha_l}(g, g_{s_1}, g_{s_2}) \\
&= \prod_{s_1 s_2 = l \in X_1} \nu_{\alpha_l}(1, g_{s_1}, g_{s_2}) \\
&= \braket{\gvec}{\Psi_\alpha}.
\end{align}
The statement for the $\Gspace$ permutation action is clear.
\end{proof}
\end{lemma}

\begin{lemma} \emph{\textbf{Grouping}}
\label{lem:grouping}
Let $\omega \in C_0(\mathbb{R}^d,P)$ be invariant under the action of $\Gspace$ on $\mathbb{R}^d$, and suppose that $\Gspace$ includes a $\mathbb{Z} \times \mathbb{Z}$ translation subgroup. Then there exists an $\alpha \in C_1(\mathbb{R}^d,P)$ with $\partial\alpha = \omega$ and a grouping of sites into finite sets such that $\ket{\Psi_{\alpha}}$ is a product state over the grouped sites.
\begin{proof}
First we can try just grouping vertices of the triangulation according to which translation unit cell they fall in. Let $\mathcal{S}$ be the set of vertices within a single unit cell, and define $\omega_{\mathrm{tot}} = \sum_{s \in \mathcal{S}} \omega_s$. Now recall that we assumed $\Gint$ is finite, from which it follows that $P$ is finite. Therefore, there must be some integer $k$ such that $k\omega_{\mathrm{tot}} = 0$. Hence, we simply group $k$ unit cells together and  now we have that $\omega_{\mathrm{tot}} = \sum_{s \in \mathcal{S}} \omega_s = 0$ for each new group $\mathcal{S}$. Our choice of grouping then ensures that we can choose an $\alpha \in C_1(\mathbb{R}^d,P)$ such that $\alpha_l = 0$ for any link $l$ hat connects two sites within different groups. Examining the form of the wavefunction \eqnref{eq:Psialpha} then reveals that it is then a product state between the groups.
\end{proof}

\end{lemma}

%\begin{lemma}
%Suppose we have $\alpha, \alpha' \in C_1(\mathbb{R}^1,P)$. with $\partial (\alpha - \alpha') = 0$. Then we must have $(\alpha - \alpha')_l = p$ for all 1-simplices $l$, for some fixed $p \in P$. if $p \neq 0$, then $\ket{\Psi_\alpha}$ and $\ket{\Psi_\alpha'}$ are in two different SPT phases with respect to $\Gint$.
%\begin{proof}
%For example, one can group sites in order to make the action on each site non-projective, and then check the projective action on the low-lying states at a boundary.
%\end{proof}
%\end{lemma}

\begin{lemma}\emph{\textbf{Wavefunction Deformation}}
\label{lem:second}
For every 2-chain $\beta \in C_2(\mathbb{R}^d,P)$ and 1-chain $\alpha \in C_1(\mathbb{R}^d,P)$ there is a corresponding family of states $\ket{\Phi_\beta(t)}$, $t \in [0,1]$ satisfying the following properties:
\begin{enumerate}
\item \label{beta_startingpt} $\ket{\Phi_\beta^\alpha(0)} = \ket{\Psi_\alpha}$.
\item \label{beta_endingpt} $\ket{\Phi_\beta^\alpha(1)} = \ket{\Psi_{\alpha + d\beta}}$.
\item \label{beta_internal} $\cU_{d\alpha}(g) \ket{\Phi_\beta^\alpha(t)} = \ket{\Phi_\beta^\alpha(t)}$ for any $g \in \Gint$.
\item \label{beta_spatial} $S(g) \ket{\Phi_\beta^\alpha(t)} = \ket{\Phi_{g\beta}^{g\alpha}(t)}$ for any $g \in \Gspace$.
\end{enumerate}
The physical interpretation is shown in Figure \ref{fig:chains1}.

\begin{proof}
First of all, observe that
\begin{align}
& \frac{\braket{\gvec}{\Psi_{\alpha + d\beta}}}{\braket{\gvec}{\Psi_{\alpha}}}\\
&= \prod_{s_1 s_2 = l \in X_1} \nu_{(d\beta)_l}(1,g_{s_1}, g_{s_2}) \label{eq:firstlemma}\\
 &= \prod_{s_1 s_2 s_3 = \sigma \in X_2} \nu_{\beta_\sigma}(1,g_{s_1}, g_{s_2}) \nu_{\beta_\sigma}(1, g_{s_2}, g_{s_3}) \nu_{-\beta_\sigma}(1, g_{s_1}, g_{s_3}) \\
 &= \prod_{s_1 s_2 s_3 = \sigma \in X_2} \nu_{\beta_\sigma}(g_{s_1}, g_{s_2}, g_{s_3}). \label{eq:lastlemma}
 \end{align}

 Now we can define a state $\ket{\Phi_\beta^\alpha(t)}$ according to
\begin{equation}
\label{eq:somewavefunction}
\frac{\braket{\gvec}{\Phi_\beta^\alpha(t)}}{\braket{\gvec}{\Psi_\alpha}}
= \prod_{\{ s_1, s_2, s_3\} \in \sigma } e^{t \log \nu_{\beta_\sigma}(g_{s_1}, g_{s_2}, g_{s_3})}
\end{equation}
where we choose some fixed branch of the complex logarithm.
\end{proof}
\end{lemma}

\begin{lemma}\emph{\textbf{Pumping}}
\label{lem:composition}
Let $\beta_1, \beta_2, \ldots, \beta_n \in C_2(\mathbb{R}^2,P)$, and $\alpha \in C_1(\mathbb{R}^2,P)$, with $\sum_{i=1}^n \partial \beta_k = 0$. Then $\beta_{\mathrm{tot}} := \sum_{k=1}^n \beta_k$ must have $\beta_\sigma = p$ for all $\sigma$, for some fixed $p \in P$. Now consider the loop obtained by composing the paths
\begin{equation}
\label{eq:somepaths}
\ket{\Phi^\alpha_{\beta_1}}, \ket{\Phi^{\alpha + \partial \beta_1}_{\beta_2}}, \cdots, \ket{\Phi^{\alpha + \partial \beta_1 + \cdots + \partial \beta_{k-1}}_{\beta_k}}
\end{equation}
Then this loop, which determines an element of $\pi_1(\Omega_2)$, maps into $p$ via the map $\pi_1(\Omega_2) \to P$ of Assumption \ref{assumption:pumps}. More concretely, in the presence of a boundary this loop pumps the 1-D SPT phase classified by $p$ onto the boundary. This is in agreement with the intuitive picture of Figure \ref{fig:chains1}.

\begin{proof}
 Now consider some region $Y$ in $\mathbb{R}^d$ comprising some subset of the vertices, links and triangles in the triangulation. We can define the truncated state $\ket{\Psi_\alpha^{(Y)}}$ according to \eqnref{eq:Psialpha} (but restricted to $Y$), and similarly we truncate the states \eqnref{eq:somewavefunction} by restricting to $Y$. Then we compose the paths \eqnref{eq:somepaths} as before, and we find that the endpoint state is given by
\begin{align}
\frac{\braket{\gvec}{\Psi_\alpha^{\prime(Y)}}}{\braket{\gvec}{\Psi_\alpha^{(Y)}}}
&= \prod_{s_1, s_2, s_3 = \sigma \in Y_2} \nu_p(g_{s_1}, g_{s_2}, g_{s_3}). \\
&= \prod_{s_1 s_2 = l \in (\delta Y)_1} \nu_{p}(1, g_{s_1}, g_{s_2}), \label{eq:endptwfn}
\end{align}
where we get to the second line by following a calculation similar to Eqs.~(\ref{eq:firstlemma}) to (\ref{eq:lastlemma}) in reverse. We recognize \eqnref{eq:endptwfn} as the wavefunction of a 1-D SPT classified by $p$ \cite{Chen_1106}. In other words, if we bring in additional degrees of freedom on the boundary, transforming nonprojectively under $\Gint$, and initially in a product state, then $\ket{\Psi_\alpha^{\prime}(Y)}$ is equivalent by a $\Gint$-symmetric finite-depth quantum circuit on the boundary to $\ket{\Psi_\alpha(Y)} \otimes \ket{\psi_{p}(\partial Y)}$, where $\ket{\psi_{p}(\partial Y)}$ is the ground state of a 1-D SPT classified by $p$ on the boundary.
\end{proof}
\end{lemma}

\subsection{Homotopy theoretic obstructions to a $\Gspace$-invariant ground state}
\label{subsec:obstructions}
Consider some arbitrary $\psi \in \Omega_d$ (not necessarily invariant under $\Gspace$, though it is invariant under $\Gint$ by the definition of $\Omega_d$). We will define a series of ``obstructions'' that, as we will see, prevent $\psi$ from being deformed, in the presence of the $\Gint$ symmetry, to a state that is invariant under $\Gspace$.

Let us first establish some notation. A \emph{pointed space} is a pair $(A,a)$, where $A$ is a space and $a \in A$ is some choice of ``basepoint''. A \emph{based map} between based spaces $(A,a)$ and $(B,b)$ is a map $f : A \to B$ such that $f(a) = b$. We will treat the $k$-sphere $S^k$ as a pointed space with some choice of basepoint. Moreover we also treat $\Omega_d$ as a pointed space, using $\psi$ as the basepoint. The $k$-th homotopy group $\pi_k(\Omega_d)$ is the set of homotopy classes of based maps $f : S^k \to \Omega_d$. [Generally $\pi_0$ of a space need not be a group, but in this case since $\pi_0(\Omega_d)$ classifies SPT phases it will inherit an abelian group structure].

A \emph{path} in $\Omega_d$ from $\phi$ to $\phi'$ is a continuous map $\mu : [0,1] \to \Omega_d$ with $\mu(0) = \phi$ and $\mu(1) = \phi'$. We will also write
\begin{equation}
\phi \xrightarrow{\mu} \phi'.
\end{equation}
The $\Gspace$ action on $\Omega_d$ induces a $\Gspace$ action on paths: if a path $\phi \xrightarrow{\mu} \phi'$, for $\phi,\phi' \in \Omega_d$, is represented as a function $\mu : [0,1] \to \Omega_d$ with $\mu(0) = \phi, \mu(1) = \phi'$, then we have a path $g\phi \xrightarrow{g\mu} g\phi'$ with $(g\mu)(s) = g \mu(s)$.
A \emph{based loop} in $\Omega_d$ is a based map $\ell : S^1 \to \Omega_d$; observe that this is equivalent to a path from $\psi$ to itself.
% We can also define similar actions on higher homotopy groups.

%A \emph{2-path} in $\Omega_d$ between two paths $\xrightarrow{\mu} : \phi \to \phi'$ and $\mu' : \phi \to \phi'$ is a ``continuous path of paths''; that is, it is a function $\mathfrak{m} : [0,1] \times [0,1] \to \Omega_d$ with $\mathfrak{m}(s,0) = \mu(s)$ and $\mathfrak{m}(s,1) = \mu'(s)$. We will also write
%\begin{equation}
%\begin{tikzcd}
%\phi\arrow[r, bend left=50, "\mu"{above,name=U}, ""{name=U, below}]\arrow[r, bend right=50, ""{name=D}, "\mu'"{below}]& \phi' \arrow[Rightarrow,from=U,to=D, "\mathfrak{m}"{right}].
%\end{tikzcd}
%\end{equation}
%A \emph{based 2-loop} is a based map $S^2 \to \Omega_d$. Observe that this is equivalent to a 2-path from a path $\psi \xrightarrow{\mu} \phi$ (where $\psi$ is the basepoint, and $\phi$ is some other state) to itself. This also means that comparing two 2-paths between two paths $\psi \xrightarrow{\mu} \phi$ and $\psi \xrightarrow{\mu'} \phi$ gives rise to a based 2-loop.

Now, let $w_1(g) = [\psi] - [g\psi]$, where $[\cdot]$ denots the connected component in $\pi_0(\Omega_d)$ of a state. We call this the \textbf{first obstruction}. If the first obstruction vanishes, then there exists a continuous path
\begin{equation}
\label{eq:lambdapath}
\psi \xrightarrow{\lambda(g)} g\psi.
\end{equation}
We can define a $\Gspace$ action on $\pi_1(G)$ according to $[\ell] \mapsto [g * \ell]$, where $[\cdot]$ denotes the homotopy class of a based loop, and we have defined
\begin{equation}
\left(\psi \xrightarrow{g * \ell} \psi\right) = \left(\psi \xrightarrow{\lambda(g)} g \psi \xrightarrow{g\ell} g\psi \xrightarrow{\lambda^{\mathrm{op}}(g)} \psi \right),
\end{equation}
where $g\ell$ is the action on $\ell$ treating it as a special case of a path, and $\lambda^{\mathrm{op}}$ denotes the path traversed in the opposite direction.
Then we can define an element $w_1(g_1, g_2) \in\pi_1(\Omega_d)$ to be the homotopy class of the based loop
\begin{equation}
\label{second_obstruction}
u_2(g_1, g_2) = \,
\begin{tikzcd}
{} & {g_1 \psi} \arrow{dr}{g_1 \lambda(g_2)} & {} \\
{\psi} \arrow{ur}{\lambda(g_1)} & & {g_1 g_2 \psi} \arrow{ll}{\lambda^{\mathrm{op}}(g_1 g_2)} \\
\end{tikzcd}.
\end{equation}
 This defines a group cochain $w_2 \in \cC^2(\Gspace, \pi_1(\Omega_d))$. One can then show that $\delta w_2 = 0$ by observing that ${g_1 * w_2}(g_2, g_3) + w_2(g_1, g_2 g_3)$ and $w_2(g_1, g_2) + w_2(g_1 g_2, g_3)$ are both equal to the homotopy class of the based loop
\begin{equation}
\begin{tikzcd}
{g_1 \psi} \arrow{rr}{g_1 \lambda(g_2)} & & {g_1 g_2 \psi} \arrow{d}{g_1 g_2 \lambda(g_3)} \\
{\psi} \arrow{u}{\lambda(g_1)} & & {g_1 g_2 g_3 \psi} \arrow{ll}{\lambda^{\mathrm{op}}(g_1 g_2 g_3)}
\end{tikzcd}.
\end{equation}
One can also show that if we chose a different path $\lambda'(g)$ in \eqnref{eq:lambdapath} that this only causes a shift $w_1 \to w_1 + \delta k_1$, where $k_1(g)$ is the homotopy class of the based loop
\begin{equation}
\begin{tikzcd}
\psi\arrow[r, bend left=50, "\lambda'(g)"{above,name=U}, ""{name=U, below}]& g\psi \arrow[l, bend left=50, ""{name=D}, "\lambda^{\mathrm{op}}(g)"{below}].
\end{tikzcd}
\end{equation}
Hence, we obtain a class in $\cH^2(G_{\mathrm{space}}, \pi_1(\Omega_d))$ -- the \textbf{second
obstruction}. [If $\pi_1(\Omega_d)$ is non-Abelian, then this requires a definition of group cohomology with non-Abelian coefficients. In fact, one can give an argument that $\pi_1(\Omega_d)$ should be abelian\footnote{Since $\Omega_d$ should be equipped with a tensor product corresponding to stacking, we can use the Eckmann-Hilton argument\cite{BaezBeauty,Eckmann__62}}, but this is not actually essential for our proof.]

More generally, if the second obstruction vanishes then we can define a third obstruction, and so on. These higher obstructions would be useful when proving LSM results in higher spatial dimensions.

The important thing about these obstructions is that they \emph{all vanish} if the state $\psi$ is itself $\Gspace$ invariant. Indeed, clearly $w_1(g) = 0$ in that case, and we can take $\lambda(g)$ to be the trivial path, which ensures that $w_2(g_1, g_2) = 0$. Moreover, one can easily check that the first non-vanishing obstruction takes the same value for any two states $\psi,\psi'$ that are connected by a continuous path in $\Omega_d$. It follows that the obstructions must all vanish for any state $\psi$ that is connected by a continuous path in $\Omega_d$ to a $\Gspace$-invariant state.

  \subsection{The descent procedure}
  \label{subsec:descent}
%Now we fix an anomalous texture $p$ and consider any gapped ground state that is invariant under $G_{\mathrm{int}}$ and the $G_{\mathrm{int}}$ representation $U_{p}$ defined above. We can ask what SPT or SET the ground state is in with respect to $G_{\mathrm{int}}$ only. Suppose that that the ground state is a $G_{\mathrm{int}}$ SPT phase (or invertible phase). Furthermore, suppose that the inverse SPT phase can also be realized in a different system with $\Gspace$ symmetry, with spins \emph{not} carrying any projective representations of $G_{\mathrm{int}}$ (that is, no anomalous texture). For group-cohomology phases, the construction of [Chen,Gu,Liu,Wen] gives an explicit such realization; we conjecture that it is always possible for any bosonic SPT (or invertible) phase. (However, the analogous statement is not true for fermions, as the $p+ip$ superconductor example from Section  demonstrates.) Then we can adjoin this inverse SPT without changing the anomalous texture, and the resulting state is now in the trivial $G_{\mathrm{int}}$ SPT phase. Therefore, to rule out any invertible ground state (satisfying the condition mentioned), it is sufficient to rule out a ground state that is in the trivial $G_{\mathrm{int}}$ SPT phase. That is what we will do in this section, whenever the anomalous texture is not matchable by an invertible defect network.

In Theorem \ref{thm:mainthm}, the assumption is that the anomalous texture gives a non-trivial element of $H_{0}^{\Gspace}(X, P)$. As we noted in Section \ref{subsec:relationship}, there is an isomorphism
\begin{equation}
H_0^{\Gspace}(\mathbb{R}^d, P) \cong \cH^d(\Gspace, P).
\end{equation}
Thus, a convenient way to check if an anomalous texture gives a non-trivial element of the left-hand side is to compute its image in the right-hand side. To do this we can use the descent sequence described in Section \ref{subsec:relationship}. For concreteness, we will review it again here for the case $d=1$ or $d=2$.

We start from $\omega \in \cC^0(\Gspace, C_0(\mathbb{R}^2,P)) = C_0(\mathbb{R}^2,P)$ representing an anomalous texture. We define $\omega^{(0)} = \omega$.
We know that $\delta \omega^{(0)} = 0$ and $\partial \omega^{(0)} = 0$ (the latter because the boundary of a $0$-chain is always zero). Because the homology $H_0(\mathbb{R}^d,P)$ is trivial, this means that there exists $\alpha^{(0)} \in \cC^0(\Gspace, C_1(\mathbb{R}^2,P))$ such that $\partial \alpha^{(0)} = \omega^{(0)}$. Now define $\omega^{(1)} = \delta \alpha^{(0)} \in \cC^1(\Gspace, C_1(\mathbb{R}^d,P))$. Observe that by construction $\delta \omega^{(1)} = 0$, and moreover
\begin{equation}
\partial \omega ^{(1)} =\partial (\delta \alpha^{(0)}) = \delta(\partial \alpha^{(0)}) = \delta \omega^{(0)} = 0.
\end{equation}

For each $g \in G$, $\omega^{(1)}(g)$ defines an class in $H^1(\mathbb{R}^1,P)$. In the case $d=1$ we have $H_1(\mathbb{R}^1, P) \cong P^{\mathrm{or}}$, and so for each $g \in \Gspace$ we get an element $\mu(g) \in P^{\mathrm{or}}$ (recall the notation means that orientation-reversing elements act by complex conjugation on $P$). More concretely, in the cellular picture of $C_1(\mathbb{R}^d,P)$, one finds that $\omega^{(1)}(g) \in C_1(\mathbb{R}^2,P)$ assigns $\mu(g)$ to each 1-cell of $\mathbb{R}^2$. We can think of $\mu$ as a group cochain $\mu \in \cC^1(\Gspace, P^{\mathrm{or}})$, and hence we get an element of the group cohomology $\cH^1(\Gspace, P^{\mathrm{or}})$.

Meanwhile, for $d=2$ we have $H_1(\mathbb{R}^2,P) = 0$, and hence we can write $\omega^{(1)} = \partial \alpha^{(1)}$ for $\alpha^{(1)} \in \cC^1(\Gspace, C_2(\mathbb{R}^2,P)$.
Finally we define $\omega^{(2)} = \delta \alpha^{(1)} \in \cC^2(\Gspace, C_2(\mathbb{R}^2, P))$, and similarly to before we can show that $\delta \omega^{(2)} = 0$ and $\partial \omega^{(2)} = 0$. As before, since $H_2(\mathbb{R}^2,P) \cong P^{\mathrm{or}}$ we obtain an element $\mu(g_1, g_2) \in P^{\mathrm{or}}$, which defines a class in $\cH^2(\Gspace, P^{\mathrm{or}})$.

The goal now is to show that if the element of $\cH^d(\Gspace, P^{\mathrm{or}})$ obtained from this descent procedure is non-trivial, then the system does not admit a $G_{\mathrm{space}} \times G_{\mathrm{int}}$-invariant ground state which is in the trivial $G_{\mathrm{int}}$ SPT phase.

\subsection{Bringing it all together}

Now we can complete the argument. As described in Section \ref{subsec:hilbert}, we are free to assume that the Hilbert space and action of the symmetry are as defined in Section \ref{subsec:hilbert}. The idea now is to show that the descent sequence from Section \ref{subsec:descent} is closely related to the obstruction computation of Section \ref{subsec:obstructions} by using the results of Section \ref{subsec:pumps}.

Indeed, since $\omega = \partial \alpha^{(0)}$, we see that the state $\ket{\Psi_{\alpha^{(0)}}}$ constructed in Lemma \ref{lem:first} is invariant under the $\Gint$ action $U_\omega(g)$. We use $\ket{\Psi_{\alpha^{(0)}}}$ as the basepoint state $\psi$ from which the obstructions were computed in Section \ref{subsec:obstructions}. Then we also have $g\psi = \ket{\Psi_{g\alpha^{(0)}}}$. We will choose $\alpha^{(0)}$ such that $\ket{\Psi_{\alpha^{(0)}}}$ such that it is a product state upon grouping of sites (that we can do this is ensured by Lemma \ref{lem:grouping}). In particular, this implies that $\ket{\Psi_{\alpha^{(0)}}}$ is in the trivial $\Gint$ SPT phase.

If $d=1$, then one can check that if $g\alpha^{(0)} - \alpha^{(0)} \neq 0$, i.e. the descent sequence is producing a non-trivial element of $\cH^1(\Gspace, P^{\mathrm{or}})$, then this implies that $\ket{\Psi_{\alpha^{(0)}}}$ and $\ket{\Psi_{g\alpha^{(0)}}}$ are in different $\Gint$ SPT phases, and therefore the first obstruction is non-trivial.

Meanwhile, if $d=2$ then we have $g\alpha^{(0)} - \alpha^{(0)} = \delta \alpha^{(0)} = \partial \alpha^{(1)}$. Hence, Lemma \ref{lem:second} shows that the first obstruction vanishes, and we can define $\lambda(g)[t] = \ket{\Phi_{\alpha^{(0)}}^{\alpha^{(1)(g)}}(t)}$. Then we can apply Lemma \ref{lem:composition} with $\alpha = \alpha^{(0)}, \beta_1 = \alpha^{(1)}(g_1), \beta_2 = g_1 \alpha^{(1)}(g_2), \beta_3 = -\alpha^{(1)}(g_1 g_2)$, which gives $\beta_{\mathrm{tot}} = (\delta \alpha^{(1)})(g_1, g_2)$ and $p = \mu(g_1, g_2)$. So we see that if the descent sequence is producing a non-trivial element of $\cH^2(\Gspace, P^{\mathrm{or}})$, then the image of the second obstruction under the homomorphism $\cH^2(\Gspace, \pi_1(\Omega_2)) \to \cH^2(\Gspace, P^{\mathrm{or}})$ induced by the $\Gspace$-equivariant homomorphism $\pi_1(\Omega_2) \to P^{\mathrm{or}}$ postulated in Assumption 3 is non-trivial (here one needs to check that the statement of $\Gspace$-equivariance is indeed consistent with the definition of the action on $\pi_1(\Omega_d)$ defined in Section \ref{subsec:obstructions} and the constructions of Section \ref{subsec:pumps}). This implies that the second obstruction itself is non-trivial.

Now we can prove Theorem \ref{thm:mainthm}. Suppose that there exists a $\Gspace \times \Gint$-invariant ground state $\ket{\Psi_{\mathrm{inv}}}$ which is in the trivial $\Gint$ phase. Then since the state $\ket{\Psi_{\alpha^{(0)}}}$ is also in the trivial $\Gint$ phase, it follows that $\ket{\Psi_{\mathrm{inv}}}$ and $\ket{\Psi_{\alpha^{(0)}}}$ can be continuously connected in the space of $\Gint$-invariant ground states. But by the arguments of Section \ref{subsec:obstructions}, this is incompatible with non-triviality of the obstruction. This completes the proof of Theorem \ref{thm:mainthm}.

\section{Discussion}
\label{sec:discussion}
\subsection{Relation to prior works}
Recently there have been several advances in classification and construction of crystalline SPT phases and some discussion of their application to LSM theorems. The idea that the LSM theorem can be understood as an anomaly has presumably been around a long time, for instance \cite{Cheng_1511} explored the connection between so-called "weak" SPTs and LSM theorems by studying their anomalous textures on the boundary in the presence of dislocations.

In Ref.~\cite{Thorngren_1612}, the present authors gave a classification of crystalline SPTs based on topological response to crystal defects.
Real-space techniques for constructing and classifying crystalline SPTs \cite{Song_1604,Huang_1705}  were also developed around the same time and eventually Ref.~\cite{Else_1810} put this approach on the most general footing, and showed that it is equivalent to the framework of Ref.~\cite{Thorngren_1612}. Ref.~\cite{Song_1810} independently proved a subset of this result. Meanwhile, Ref.~\cite{Shiozaki_1810} supplied the key insight that the real-space picture can be most naturally formulated as an equivariant homology theory, but did not commit to a specific (mathematically-defined) homology theory, instead performing computations in particular cases based on physical arguments (we did similar computations in Section \ref{sec:anomalymatching}). Another approach based on invertible TQFTs, that also leads to an equivariant homology theory, appeared in Ref.~\cite{Freed_1901}.

The application of the real-space picture of crystalline SPTs to LSM theorems was mentioned in each of \cite{Shiozaki_1810,Else_1810,Song_1810}. In the case of Ref.~\cite{Else_1810,Song_1810}, it was not the main focus of these works, and no computations were performed there or systematic theory developed. In Ref.~\cite{Shiozaki_1810} some computations were performed (based on physical arguments), but but not an exhaustive calculation as we did in Section \ref{sec:exhaustive}, for example. Also, by contrast to Ref.~\cite{Shiozaki_1810}, which merely postulated the existence of some equivariant homology theory, we derived our results from the existing framework for crystalline SPTs \cite{Thorngren_1612,Else_1810}.

\subsection{Future directions}
In this paper, we have given a comprehensive framework to understand the constraints on ground states in quantum systems comprised microscopically of half-integer spins (or, more generally, projective symmetry action on sites). We focussed mainly on bosonic systems, but an important problem is to extend to fermion systems, and also to go beyond the in-cohomology approximation for bosonic systems (though we do not expect the latter to change the criterion for a traditional LSM theorem, as we discussed in Section \ref{sec:defnetclass}). In principle it is clear what one should do; the defect network/anomaly cancellation picture from Section \ref{sec:anomalymatching} should be valid generally, and in terms of abstract mathematical formalism the discussion in Appendix \ref{appendix:smoothstates} is also general, and one just needs to replace the Eilenberg-MacLane spectrum with something else (probably oriented cobordism for bosons, and spin cobordism for fermions \cite{Kapustin_1406}). However, the outstanding question (which is a question of mathematics, not physics) is how to turn this into
a practical computational method, analogous to what we did in Section \ref{sec:equivarianthomology} and Appendix \ref{appendix:computation}.

A traditional LSM result tells us when a system is allowed to have a symmetric gapped ground state that is not topologically ordered. Nevertheless, the anomaly matching picture makes clear that there are also non-trivial constraints on \emph{any} ground state that could arise, whether it be spontaneous symmetry breaking, or gapless, or topological ordered. In this paper, we have already partially laid the groundwork for such studies. For example, for a gapped ground state in either a bosonic or fermionic system, we have identified the relevant criterion: the topological phase must be described by a 0-degree anomalous defect network with the right anomalous texture. Meanwhile, for gapless ground states, generally spatial symmetries act like \emph{internal} symmetries on the fields in the low-energy field theory describing the system (see \cite{Metlitski_1707} for a recent discussion of this in a related context). Thus, provided that one knows how to compute the anomaly of an internal symmetry in some gapless field theory, we expect that this anomaly should match the one computed from the microscopic anomalous texture through the descent sequence constructed in Section \ref{subsec:relationship}.

Meanwhile, there is another class of results that constrain the ground state given microscopic data, which are sometimes also called LSM theorems \cite{Oshikawa_9911,Parameswaran_1212,Watanabe_1505,
Lu_1705_09298,Lee_1802}. These occur for a system for which there is, at least, a $\mathrm{U}(1)$ charge conservation symmetry and discrete translation symmetry. This allows one to define the \emph{filling}; that is, the average charge per unit cell. It is known, for example, that when the filling is not an integer, any gapped symmetric ground state must be topologically ordered. Unlike the LSM results we have discussed in this paper, this result is not obviously related to being on the boundary of a crystalline topological phase. For example, we cannot think of a one-dimensional system at fractional filling as the boundary of an SPT, because $\cH^3(\mathbb{Z} \times \UU(1), \UU(1)) = \cH^3(\UU(1), \UU(1)) = \mathbb{Z}$, i.e.\ there are no mixed $\mathbb{Z} \times \UU(1)$ SPTs in two dimensions. Hopefully, there will eventually be a general framework to understand these results, analogous to the one developed here.

\emph{Note added}.-- Simultaneously with this work, another preprint appeared \cite{Jiang_1907} discussing general approaches to LSM theorems, with a particular focus on SPT-LSM theorems. Our results agree with theirs where they overlap.

\begin{acknowledgments}
We thank Lukasz Fidkowski, Chao-Ming Jian, Hoi-Chun Po, and Michael Zaletel for helpful discussions. DVE was supported by the Microsoft Corporation and by the EPiQS Initiative of the Gordon and Betty Moore Foundation, Grant No. GBMF4303. RT was supported by the Zuckerman STEM Leadership Program and the NSF GRFP Grant Number DGE 1752814.
\end{acknowledgments}

\appendix

\section{Shapiro's Lemma and the classification of anomalous textures}\label{appendix:anomtexts}

In this appendix, we show a correspondence between two ways of thinking about an anomalous texture on a lattice $\Lambda$ with a symmetry group $G$. We do this by invoking general lemma, called Shapiro's Lemma \cite{Brown}, which will also be useful for us for other applications.

%By the definition in \ref{sec:symrep}, an anomalous texture is defined by a family of $G$ 2-cochains $\omega_s(g_1,g_2) \in C^2(G,\UU(1))$ parametrized by $s \in \Lambda$, satisying the associativity condition \eqref{an_associativity}:
%\begin{equation}\omega_{g_1^{-1} s}^{\sigma(g_1)} (g_2, g_3) \omega_s(g_1, g_2 g_3) = \omega_s(g_1, g_2) \omega_s(g_1 g_2, g_3),\end{equation}
%and subject to the equivalence relation \eqref{an_coboundary}:
%\begin{equation}\omega_s(g_1, g_2) \to \omega_s(g_1, g_2) \frac{\beta_{g_1^{-1} s}^{\sigma(g_1)} (g_2) \beta_s(g_1)}{\beta_s(g_1, g_2)}.\end{equation}
%We will consider the first as a cocycle condition and the second as a shift by a coboundary.

Let $S$ denote a (discrete) set with a $G$ action. Suppose $M$ is a $G$-module. We define the abelian group $C^k_G(S, M)$ of $S$-dependent $k$-cochains to comprise functions
\begin{equation}S \times (G)^k \to M.\end{equation}
For instance, $\omega_s$ above has $k = 2$ and $S = \Lambda$. We define the coboundary map
\begin{equation}
\delta_k : C^k_G(S, M) \to C^{k+1}_G( S, M).
\end{equation}

\begin{multline}
\label{group_coboundary}
\left(\delta\alpha\right)(s; g_1,\dots,g_{k+1})
\\ = g_1\cdot \alpha(g_1^{-1} s; g_2,\dots,g_{k+1})
+  (-1)^{i+1} \alpha(s; g_1,\dots,g_k).
\\ + \sum_{i=1}^p (-1)^{i} \alpha(s; g_1,\dots,g_{i-1},g_i g_{i+1},g_{i+2},\dots,g_{k+1}).
\end{multline}

We then define the cohomology
\begin{equation}
H^k_G(S, M) = \operatorname{ker} \delta_k / \operatorname{im} \delta_{k-1}.
\end{equation}
This is a special case of the equivariant cohomology as introduced in Appendix \ref{subsec:equivariantcohomology} (hence the notation), but that will not concern us here. Observe that, in particular, $H^2_G(\Lambda, \UU(1))$ agrees with the definition of the group of anomalous textures given in Section \ref{sec:symrep}.

Shapiro's Lemma \cite{Brown} is then the statement that
\begin{equation}
\label{shapiroslemma}
H^k_G(S, M) \cong \bigoplus_{[s] \in S/G} \cH^k(G_s, M),
\end{equation}
where the sum is over one representative for each $G$ orbit of $S$. This result implies the classification of anomalous textures discussed in Section \ref{sec:symrep}

It is easy to construct the isomorphism from the left-hand side of \eqnref{shapiroslemma} to the right-hand side, we just restrict the cocycles in $C^k_G(S, M)$ to $\{s\} \times G_s$  for some $s$ in each orbit. Constructing the inverse isomorphism is much trickier, and we will now spend some time to do this in the case $k=2$.

\subsubsection*{The inverse isomorphism}\label{ashapiroinverse}

First of all, let us introduce a simplicial complex $S//G$ defined as follows. The vertices of $S//G$ are given by the elements $s \in S$. The edges $s \to s'$ are given by the elements $g\in G$ for which $g(s) = s'$. Then we add a 2-simplex for every composable triple $g_1, g_2, g_3$ with $g_1 g_2 = g_3$. Then we add higher simplices for all higher relations in the group amongst composable edges.

The 2-simplices in $S$ are given by triples $(s,g_1,g_2)$ with $s \in \Lambda$, $g_1,g_2 \in G$. One sees that an anomalous texture is just a class $[\omega] \in H^2(S//G,\UU(1))$.

$S//G$ is known as the homotopy quotient, and one can check that its simplicial cohomology $\cH^{\bullet}(S//G,M)$ is the same as $\cH^{\bullet}_G(S, M)$. Shapiro's Lemma \eqref{shapiroslemma} then follows from the homotopy equivalence
\begin{equation}\label{e:anomtext-homotopy-equiv}
S//G = \bigsqcup_{s} BG_s,
\end{equation}
where $\bigsqcup_s$ denotes a disjoint union, ranging over a representative $s$ for each $G$ orbit of $S$. Here $BG_s$ is the classifying space of $G_s$, $BG_s \simeq \star//G_s$.

The homotopy equivalence \eqref{e:anomtext-homotopy-equiv} has a particularly simple form. For a fixed $s \in S$, we have a simplicial embedding $BG_s \hookrightarrow S//G$ where the unique vertex of $BG_s$ is the $s$ vertex of $S//G$. The $g \in G_s$ edges of $BG_s$ are the edges in $S//G$ which go from $s \to s$ (since all such $g$ by definition fix $s$). All higher simplices of $BG_s$ are there to impose relations in $G_s$ and so are present as well in $S//G$.

%We see therefore that given an anomalous texture $\omega_s(g_1,g_2) \in Z^2(\Lambda//G,U(1))$, we obtain the corresponding $[\alpha_s] \in H^2(BG_s,\UU(1)) = \cH^2(G_s,\UU(1))$ by restricting our cocycle $\alpha_s(g_1,g_2) = \omega_s(g_1,g_2)$ for $g_1,g_2 \in G_s$.

%This shows also how to obtain an $\alpha_s$ for every $s \in \Lambda$, not only for our reference points. Of course, all $\alpha_s$ for $s$ in the same orbit will be related by conjugation in $G$, and each one determines the rest.

Let us now discuss how to construct the inverse isomorphism in \eqnref{shapiroslemma}. Let us fix an $s \in S$ and consider its orbit $O(s)$. (We can do the procedure we are about to describe separately for each orbit.) We start from a cocycle $\alpha_s \in \cZ^2(G_s, M)$, and we want to construct a cocycle $\omega : O(s) \times G \times G \to M$] satisfying $\delta \omega = 0$ and which restricts to $\alpha_s$ on $\{s\} \times G_s \times G_s$. Geometrically, what we we need is to find a retraction $O(s)//G \to BG_s$, where $O(s)$ is the orbit containing $s \in S$. One way to do it is to choose, for each $s' \in O(s)$, an element $g(s,s') \in G$ with $g(s,s')(s) = s'$ and $g(s,s) = 1$. Geometrically, we collapse $O(s)//G$ along the tree defined by these elements. Algebraically, we define $\omega_s$ inductively using the cocycle condition:
\begin{itemize}
\item $\omega_s(g_1,g_2) = \alpha_s(g_1,g_2)$ when $g_1, g_2 \in G_s$.
\item $\omega_s(g_1,g_2) = 1$ whenever $g_2 g_1 = g(s,s')$ for some $s'$. This requires $\alpha_s$ satisfy a normalization condition $\alpha_s(g,g^{-1}) = 1$, which is always possible by rephasing the operators in the projective representation to which it corresponds.
\item When $g_1 \in G_s$ and $g_2:s \to s'$, we study a tetrahedron with backbone
\begin{equation}s \xrightarrow{g_1^{-1}g_2^{-1}g(s,s')} s \xrightarrow{g_1} s \xrightarrow{g_2} s'\end{equation}
and find by the cocycle condition and the first two conditions
\begin{equation}\omega_s(g_1,g_2) = \alpha_s(g_1^{-1}g_2^{-1}g(s,s'),g_1).\end{equation}
\item When $g_1(s) = s'$, $g_2(s') = s''$, we study a tetrahedron with backbone
\begin{equation}s \xrightarrow{g_1^{-1}g_2^{-1}g(s,s'')} s \xrightarrow{g_1} s' \xrightarrow{g_2} s''\end{equation}
which reduces us to the previous case:
\begin{equation}\omega_s(g_1,g_2) = \omega_s(g_1^{-1}g_2^{-1}g(s,s''),g_1)\end{equation}
from which we obtain
\begin{equation}\omega_s(g_1,g_2) = \alpha_s(g(s,s'')g_2 g(s,s'),g_1^{-1}g_2^{-1}g(s,s'')).\end{equation}
\item Finally in the general case of $g_1(s') = s''$, $g_2(s'') = s'''$ but all vertices still in the orbit of $s$, we use a tetrahedron with backbone
\begin{equation}s \xrightarrow{g_1^{-1}g_2^{-1}g(s,s''')} s' \xrightarrow{g_1} s'' \xrightarrow{g_2} s''',\end{equation}
which reduces us to the previous case:
\begin{equation}\omega_{s'}(g_1,g_2) = \omega_s(g_1^{-1}g_2^{-1}g(s,s'''),g_1).\end{equation}
Moving two steps more we find
\begin{equation}\omega_{s'}(g_1,g_2) = \alpha_s(x,y)\end{equation}
where
\begin{equation}x = g(s,s''')^{-1}g_2^{-1}g(s,s'')^{-1}g(s,s''')^{-1}g_2 g_1 g(s,s')\end{equation}
\begin{equation}y = g(s,s'')g_2 g(s,s''').\end{equation}
\end{itemize}

\section{Equivariant chains and lattice wavefunctions}
\label{appendix:lattice_wfns}
Let $G$ be a finite group (see below for some generalizations to infinite discrete groups). We want to show that certain equivariant chains introduced in Section \ref{sec:equivarianthomology} give rise to concrete lattice wavefunctions.

We work in terms of a lattice system defined on the vertices of a $G$-invariant triangulation $X$ with $G$-invariant branching structure, where each vertex carries a Hilbert space with a basis labelled by $G$ (see below for what one can do if $G$ is infinite). The representation $U(g)$ of $G$ on the lattice Hilbert space corresponds to permuting the vertices according to the action of $G$ on $X$ combined with the on-site action $\ket{h} \to \ket{gh}$ and acting with $\mathfrak{C}^{\sigma(g)}$, where $\mathfrak{C}$ is complex conjugation in the $G$-labelled basis.

Recall that an equivariant $-1$-chain $\alpha \in C_{-1}^G(X,\mathrm{U}(1))$ consists of data $(\alpha_0, \cdots, \alpha_d)$ where $\alpha_k \in \cC^{k+1}(G, C_k(X,\mathrm{U}(1))))$. For the purposes of this section, we work with the homogeneous group cochains from  Section \ref{subsec:groupcohomology}, and we will also write the group law in $\mathrm{U}(1)$ multiplicatively rather than additively to reflect the identification of $\mathrm{U}(1)$ with the complex numbers of unit modulus.

Then we can construct a corresponding wavefunction
\begin{equation}
\ket{\Psi_\alpha} = \sum_{\gvec} \left(\prod_{k=0}^d w_i^\alpha(\gvec)\right) \ket{\gvec},
\end{equation}
where $\gvec$ is a shorthand for $g_1, \cdots, g_N$ (where $N$ is the number of vertices), and
\begin{equation}
\label{eq:wk}
w_k^\alpha(\gvec) = \prod_{ s_1\cdots s_{k+1} = \sigma_k \in X_k} \alpha_k(1, g_{s_1}, \cdots, g_{s_k})_{\sigma_k}
\end{equation}
where the sum is over all $k$-simplices $\sigma_k$, and $\{ s_1, \cdots, s_{k+1} \}$ are the vertices of the simplex, ordered according to the branching structure. Moreover, the orientation of $\sigma_k$, which is needed to define the sign of $\alpha_{\sigma_k}$ is also determined by this ordering (see the definition of $k$-chain in Section \ref{subsec:cellular_homology}).

Next we show that if $\alpha$ is an equivariant \emph{cycle}, then $\ket{\Psi_\alpha}$ is invariant under $U(g)$. Indeed, we compute
\begin{equation}
\frac{\bra{\gvec} U(g) \ket{\Psi_\alpha}}{\braket{\gvec}{\Psi_\alpha}}
= \prod_{k=0}^d v_i^{\alpha}(g; \gvec),
\end{equation}
where
\begin{align}
& v_i^{\alpha}(g;\gvec) \\
& = \prod_{ s_1\cdots s_{k+1} = \sigma_k \in X_k} \frac{\alpha_k^{\sigma(g)}(1, g^{-1} g_{s_1}, \cdots, g^{-1}g_{s_{k+1}})_{g^{-1}\sigma_k}}{\alpha_k(1, g_{s_1}, \cdots, g_{s_{k+1}})_{\sigma_k}} \\
&= \prod_{ s_1\cdots s_{k+1} = \sigma_k \in X_k} \frac{\alpha_k(g, g_{s_1}, \cdots, g_{s_k})_{\sigma_k}}{\alpha_k(1, g_{s_1}, \cdots, g_{s_{k+1}})_{\sigma_k}}  \\
&= \prod_{ s_1\cdots s_{k+1} = \sigma_k \in X_k} (\delta \alpha_k)(g, 1, g_{s_1}, \cdots, g_{s_{k+1}})_{\sigma_k} \nonumber \\
& \quad\quad\quad \times \prod_{ s_1\cdots s_{k+1} = \sigma_k \in X_k}(\partial \alpha_k)^{-1}( g, 1, g_{s_1}, \cdots, g_{s_k})_{\sigma_{k-1}}, \label{finalGaction}
\end{align}
where we have used the homogeneous condition on the cochains and the definitions of $\delta$ and $\partial$. We see that if $\alpha$ is an equivariant cycle, then by definition $\delta \alpha_k (\partial \alpha_{k-1})^{-1} = 1$, hence $\ket{\Psi_\alpha}$ is invariant under $U(g)$ (to compare with $D$ defined according in \eqnref{equivariant_boundary} we have rewrite $\UU(1)$ multiplicatively).

Next we consider the case where $\alpha = D\beta$ for some equivariant $0$-chain $\beta = (\beta_0, \cdots, \beta_d)$. Then one can check that
\begin{widetext}
\begin{equation}
\ket{\Psi^{D\beta}} = \sum_{\gvec} \left(\prod_{k=0}^{d} \prod_{ s_1\cdots s_{k+1} = \sigma_k \in X_k}\beta_k(\sigma_k; g_{s_1}, \cdots, g_{s_{k+1}}) \right) \ket{\gvec},
\end{equation}
which is related to $\ket{\Psi^0}$ by the finite-depth quantum circuit
\begin{equation}
\mathcal{U}^{\beta} = \sum_{\gvec} \left(\prod_{k=0}^{d} \prod_{ s_1\cdots s_{k+1} = \sigma_k \in X_k}\beta_k(\sigma_k; g_{s_1}, \cdots, g_{s_{k+1}}) \right) \ket{\gvec} \bra{\gvec},
\end{equation}
which from the homogeneous condition on $\beta_k$ is manifestly $G$-symmetric at each layer. Similarly, one can show that $\ket{\Psi^\alpha}$ and $\ket{\Psi^{\alpha + D\beta}}$ are related by a symmetric finite-depth quantum circuit.

Finally, we can also show that if we have an anomalous texture represented by an equivariant $-2$-cycle $\gamma \in Z_{-2}^G(X,\UU(1))$ such that $D\alpha = \gamma$ for some equivariant $-1$-chain $\alpha$, then it can be cancelled by an in-cohomology wavefunction as above. First we define the symmetry action corresponding to the anomalous texture according to $U_\gamma(g) = T_\gamma(g) U(g)$, where $U(g)$ is as defined above, and
\begin{equation}
T_\gamma(g) = \sum_{\gvec} \left(\prod_{k=0}^d\prod_{ s_1\cdots s_{k+1} = \sigma_k \in X_k} \gamma(g, 1, g_{s_1}, g_{s_2}, \cdots, g_{s_k})_{\sigma_k} \right) \ket{\gvec}\bra{\gvec},
\end{equation}
Comparing with \eqnref{finalGaction}, we see that indeed if $D\alpha = \gamma$ then $U_\gamma(g) \ket{\Psi_\alpha} = \ket{\Psi_\alpha}$.

\end{widetext}

We end this appendix with some comments about infinite symmetry groups. In particular, if $G$ is an infinite group acting on $\mathbb{R}^d$ that contains translations $\mathbb{Z}^d$ as a normal subgroup, such that $G_{\mathrm{pt}} := G/\bZ^d$ is finite, then we can also perform an analogous construction. We exploit the ``rolling" isomorphism $H_k^G(\mathbb{R}^d, \mathrm{U}(1)) \cong H_k^{G_{\mathrm{pt}}}(\mathbb{T}^d, \mathrm{U}(1))$, where $\mathbb{T}^d = \mathbb{R}^d / \mathbb{Z}^d$ is a $d$-dimensional torus.

Therefore, given a $G$-equivariant -1-cycle on $\bR^d$, we produce a $G_{\rm pt}$-equivariant -1-cycle on the torus, hence by our construction above a wavefunction on the torus with a $|G_{\rm pt}|$-dimensional Hilbert space at each vertex. We can also use this idea to define a $G$-symmetric wavefunction on $\bR^d$. We take the on-site Hilbert spaces to still be labelled by $G_{\rm pt}$, with $U(g)  \ket{h} = \ket{\varphi(g) h}$ where $\varphi : G \to G_{\rm pt}$ is the quotient homomorphism. Then we replace \eqnref{eq:wk} with
\begin{equation}
w_k^\alpha(\gvec) = \prod_{ s_1\cdots s_{k+1} = \sigma_k \in X_k} \alpha(1, g_{s_1}, \cdots, g_{s_k})_{\phi(\sigma_k)}
\end{equation}
where $\phi : \mathbb{R}^d \to \mathbb{T}^d$ is the quotient map. Then all the above manipulations proceed as before. In a similar way, we can also discuss deformations, anomalous textures, and ground state wavefunctions in the presence of anomalous textures on $\mathbb{R}^d$ while retaining a finite on-site Hilbert space labelled by $G_{\rm pt}$.

\section{Aspects of equivariant homology and cohomology}\label{sec:eq-hom-append}

%The definitions presented here are standard and may be found in any number of references on the subject. Mostly this section is to fix notation. We cannot aspire to give a full introduction to this vast subject.

\subsection{Equivariant cohomology and duality}
\label{subsec:equivariantcohomology}
In this section, we relate equivariant homology to the Borel equivariant \emph{co}homology used to classify crystalline SPT phases in \cite{Thorngren_1612,Else_1810}. We will use a discrete coefficient group, which is typically $A = \bZ$, and in Section \ref{subsec:coefficients} we will discuss the relationship to $\UU(1)$ coefficients.

First we recall the definition of cellular cohomology, which is dual to cellular homology defined in Section \ref{subsec:cellular_homology}. We denote by $C_k(X,A)$ the abelian group of chains with coefficients in $A$ on a space $X$ with cell decomposition, and we define a \emph{(cellular) $k$-cochain} to be a map
\begin{equation}\alpha: C_k(X,\bZ) \to A\end{equation}
and these form a group denoted $C^k(X,A)$. We denote the pairing of $\alpha$ with a $k$-cycle $\Gamma \in C_k(X,\bZ)$ by
\begin{equation}\int_\Gamma \alpha.\end{equation}
This map is linear in both $\Gamma$ and $\alpha$, like familiar integration, and the set of these values over $\Gamma$ defines $\alpha$, by definition. We define a map
\begin{equation}d:C^k(X,A) \to C^{k+1}(X,A)\end{equation}
called the (cellular) coboundary map by
 \begin{equation}\int_\Gamma d\alpha = \int_{\partial \Gamma} \alpha,\end{equation}
 compare Stokes' theorem. $d^2 = 0$ as before. We denote the kernel of $d$ as $Z^k(X,A)$, whose elements are \emph{(cellular) $k$-cocycles}, and the image of $d$ in $Z^k(X,A)$ as $B^k(X,A)$, the group of \emph{exact (cellular) $k$-cocycles}. We obtain a group
 \begin{equation}H^k(X,A) = Z^k(X,A)/B^k(X,A)\end{equation}
 called the $k$th cohomology of $X$ with coefficients in $A$.

 We can also relate cellular cohomology to group cohomology. There is a simplicial complex called the \emph{classifying space} $BG$ for which the cellular $k$-cochains on $BG$ valued in $A$ are exactly the group $k$-cochains valued in $A$, and the cellular coboundary map equals the group coboundary with trivial action on the coefficients. Thus,
\begin{equation}H^k(BG,A) = \cH^k(G,A).\end{equation}
The existence of $BG$ allows topological proofs of many theorems in group cohomology. It is also possible to generalize this to account for non-trivial $G$-modules $M$, but on the left-hand-side we have to use local coefficients \cite{Hatcher}.

Given a $G$-equivariant cell complex of $X$, we can also define the double complex
\begin{equation}\cC^p(G,C^q(X,A)),\end{equation}
using the dual action of $G$ on $C^q(X,A)$ induced by its action on $X$. Taking $k = p+q$, we can define a total differential
\begin{equation}\Delta = \delta + (-1)^k d\end{equation}
on this double complex as before. The cohomology of $\Delta$ defines the equivariant cohomology $H^{p+q}_G(X,A)$. Like group cohomology, this cohomology is equivalent to cellular cohomology of a classifying space, known as the \emph{homotopy quotient} $X//G$:
\begin{equation}H^k_G(X,A) = H^k(X//G,A).\end{equation}
A construction for $X//G$ in the case that $X$ is discrete is discussed in Appendix \ref{ashapiroinverse}. A general description can be found in \cite{Thorngren_1612}. Equivariant theories constructed using a classifying space are called Borel equivariant cohomology.

% \de{Do we use any of these pairings? I thought we mainly just want to show that we have Poincare duality.}
% This cohomology has a graded product structure when $A = R$ is a ring. There is also a pairing
% \begin{equation}\cC^{p_1}(G,C^q(X,A)) \otimes \cC^{p_2}(G,C_q(X,\bZ)) \to \cC^{p_1 + p_2}(G,A)\end{equation}
% induced by cup product in the $G$ variables and the pairing between cellular homology and cohomology. Let us denote this pairing
% \begin{equation}\int_\Gamma \alpha, \qquad \alpha \in \cC^{p_1}(G,C^q(X,A)) \quad \Gamma \in \cC^{p_2}(G,C_q(X,\bZ)).\end{equation}
% It's easy to verify
% \begin{equation}\delta \int_\Gamma \alpha = \int_{\Gamma} \delta \alpha + (-1)^{p_1} \int_{\delta \Gamma} \alpha\end{equation}
% \begin{equation}\int_\Gamma d\alpha  + (-1)^{p_1} \int_{\partial \Gamma} = 0,\end{equation}
% so we obtain a pairing (after careful tracking of degrees)
% \begin{equation}H^{p}_G(X,A) \otimes H_{q}^G( X, \bZ) \to \cH^{p-q}(G,A).\end{equation}

Now suppose $X$ is a closed $n$-manifold. It has a dual cell complex $X^\vee$ whose $k$-cells meet the $n-k$-cells of $X$ transversely at their so-called barycenters. Two cells meeting in such a way are dual and are denoted $\sigma \in X_{n-k}$, $\sigma^\vee \in X_k^\vee$. If $X$ has an orientation, we can use it to associate orientations of $\sigma$ with orientations of $\sigma^\vee$. This gives us an isomorphism
\begin{equation}I:C_k(X,A) \to C^{n-k}(X^\vee,A)\end{equation}
by sending $\sigma$ with its orientation to the indicator cocycle on $\sigma^\vee$ with its dual orientation and extending to an $A$-linear function. This map satisfies
\begin{equation}I(\partial \sigma) = d I(\sigma).\end{equation}
It therefore gives us an isomorphism of bicomplexes by mapping the coefficients:
\begin{equation}\cC^j(G,C^k(X^\vee,A^{\rm or})) \to \cC^j(G,C_{n-k}(X,A))\end{equation}
\begin{equation}Id = \partial I, \qquad I\delta = \delta I,\end{equation}
where the superscript $or$ on the left-hand-side indicates that orientation-reversing elements of $G$ swap the sign of the coefficients, because we used an orientation to define the map, which might not be $G$-invariant. This twist is necessary for $I\delta = \delta I$, since $\delta$ involves the $G$-action. This descends to an isomorphism on the homology of the total complexes:
\begin{equation}\label{eqn:poincareduality} H^k_G(X,A^{\rm or}) = H_{n-k}^G(X,A).\end{equation}
Note that $H^k_G(X^\vee,A^{\rm or}) = H^k_G(X,A^{\rm or})$ since $X$ and $X^\vee$ are isomorphic $G$-spaces (cohomology does not depend on the cell complex).

In the case that $X$ is not compact, but still lacks boundary, then we can still form the dual cell complex as above. It is clear that it has the same properties for Borel-Moore homology and cohomology, so again \eqref{eqn:poincareduality} holds.

%  This defines a fundamental cycle
% \begin{equation}[X] \in \cZ^0(G,Z_n(X,\bZ)) < H_n^G(X,\bZ).\end{equation}
% This lets us define another pairing (this is the Poincar\'e duality pairing of cohomology with itself)
% \begin{equation}H^p_G(X, R) \otimes H^{n-p}_G(X,R) \to \cH^0(G,R) = R^G,\end{equation}
% given by
% \begin{equation}\langle \alpha, \beta\rangle = \int_{[M]} \alpha \cup \beta.\end{equation}
% More generally, taking $\beta \in H^q_G(X,R)$, we get a map
% \begin{equation}H^p_G(X,R) \to {\rm Hom}(H^*_G(X,R),\cH^{p-n+*}(G,R)),\end{equation}
% where the right hand side is a (shifted) hom space of graded rings. This map is an isomorphism whenever $H^*_G(X,R)$ is a free $\cH^*(G,R)$-module\cite{mathoverflow}.

\subsection{Relative equivariant homology and cohomology}
\label{subsec:relativecohomology}

Suppose $X$ is a cell complex and $Y$ is a subcomplex. We define the group of relative $A$-valued $k$-cochains $C^k(X,Y,A)$ to be the subgroup of $C^k(X,A)$ which are zero on $C_k(Y,\bZ)$. Clearly the coboundary map $d$ preserves this property, ie. it defines a map $C^k(X,Y,A) \to C^{k+1}(X,Y,A)$. This allows us to define the relative cohomology $H^k(X,Y,A)$ as the cohomology of $d$ restricted to these relative $k$-cochains.

If there is a cellular $G$-action on $X$ such that $Y$ is $G$-invariant, then we can likewise define the bicomplex of relative equivariant cochains
\[\cC^p(G,C^q(X,Y,A))\]
with the restricted coboundary map $d$ and the usual group coboundary map $\delta$. The cohomology of the total differential $\Delta = \delta + (-1)^{p+q} d$ defines the relative equivariant cohomology $H^k_G(X,Y,A)$.

There is also relative equivariant homology. We take $Z_r^G(X,Y,A)$ to be the equivariant $r$-chains satisfying the relaxed cycle condition $D\beta \in C_r^G(Y,A)$, where we have used the inclusion map $C_r^G(Y,A) \to C_r^G(X,A)$. We have $C_r^G(Y,A) \subset Z_r^G(X,Y,A)$ and we define
\begin{equation}\label{eqn:relativehomology}
  H_r^G(X,Y,A) = Z_r^G(X,Y,A) / B_r^G(X,A) \oplus C_r^G(Y,A).
\end{equation}

This allows us to form a version of Poincar\'e duality for spaces with boundary, known as Poincar\'e-Lefschetz duality. Indeed, if $X$ is an $n$-manifold with boundary and a cell complex, one can still construct a dual cell complex by forming the coned space $C\partial X \cup X$, which is a boundaryless cell complex, forming the dual as usual, although the dual region of the cone point won't be a cell unless $\partial X$ is a sphere. We simply remove it to find $X^\vee$. This has the nice property that $\partial(X^\vee) = (\partial X)^\vee$. It is easy to check that if $\sigma \in X_k$, then we have
\begin{equation}I(\partial \sigma \cap \partial X) = i^* I(\sigma),\end{equation}
where $i:\partial X \to X$ is the inclusion. The properties are discussed in detail in Chapter 1 of \cite{ryanthesis}. This means that in the case with boundary, we have the more general Poincar\'e-Lefschetz duality
\begin{equation}H^k_G(X,A^{\rm or}) = H_{n-k}^G(X,\partial X,A)\end{equation}
between equivariant cohomology and relative equivariant homology. Likewise we also have
\begin{equation}H^k_G(X,\partial X,A^{\rm or}) = H_{n-k}^G(X,A).\end{equation}

\subsection{Continuous groups and $U(1)$ vs. $\bZ$ coefficients}
\label{subsec:coefficients}

In this section, we describe how to work with continuous coefficient groups $A$, especially $A = \UU(1)$, and how the calculations relate to $A = \bZ$.

All the groups $G$ we study are locally compact Lie groups. For these we can define the (left) Haar measure on the identity component and translate it to all other components using the left action of $G$ on itself; we can do the same thing for the coefficient group $A$. This defines a product measure on $G \times \cdots \times G$ and we take our cochains used to define group cohomology in Section \ref{subsec:groupcohomology} to be \emph{measurable} cochains $G \times \cdots \times G \to A$ for any topological abelian group $A$, thus defining the so-called \emph{measurable group cohomology}. That this is the appropriate version of group cohomology for discussing SPT phases has long been known \cite{Chen_1106}.
Note that if either of $A$ or $G$ is discrete, then all cochains are measurable.

This version of group cohomology has the property that for any compact Lie group $G$, we have
\begin{equation}
\label{eq:Rvanishing}
\cH^n(G,\mathbb{R}) = 0, \qquad \forall n > 0.
\end{equation}
From this, one can use the long exact sequence in cohomology induced by the short exact sequence of coefficient groups
\begin{equation}
\label{eq:shortexactcoeff}
0 \to \mathbb{Z} \to \mathbb{R} \to \mathrm{U}(1) \to 0
\end{equation}
 to show that
\begin{equation}\label{eqn:coefficients}
\cH^n(G,\UU(1)) = \cH^{n+1}(G,\bZ) \qquad \forall n > 0.
\end{equation}
This map is given by lifting a $\UU(1) = \bR/\bZ$-valued $n$-cocycle to a $\bR$ cochain; applying the group differential $\delta$ then gives $\mathbb{Z}$-valued $(n+1)$-cocycle. This map is known as the Bockstein.
In turn, by definition of $BG$, we have $\cH^{n+1}(G,\bZ) = H^{n+1}(BG,\bZ)$, so we obtain
\begin{equation}
  \cH^n(G,\UU(1)) = H^{n+1}(BG,\bZ) \qquad \forall n > 0.
\end{equation}

%More generally, if $A$ is a \emph{connected} topological abelian group with a continuous $G$ action, we have a natural isomorphism
%\begin{equation}\label{eqn:coefficients2}
%\cH^n(G,A) = \cH^{n+1}(G,\pi_1 A) \qquad \forall n>0.
%\end{equation}

Let us now extend these statements to equivariant homology and cohomology.
We define measurable equivariant homology and cohomology by taking our equivariant chains and cochains to be measurable functions $G \times \cdots \times G \to C_k(X,\UU(1))$ or $C^k(X,\UU(1))$, respectively, where the chains and cochains on $X$ are the ``discrete" ones which take constant values over cells. We will show that for a group $G$ with a an action on $X$ such that the isotropy group $G_\Sigma$ of any cell $\Sigma$ is compact, then we have a generalization of \eqnref{eqn:coefficients}, namely:
\begin{equation}
  H^n_G(X,\UU(1)) = H^{n+1}_G(X,\bZ) \qquad \forall n > \dim X
\end{equation}
\begin{equation}
  H_n^G(X,\UU(1)) = H_{n-1}^G(X,\bZ) \qquad \forall n < 0.
\end{equation}
For brevity we describe the first isomorphism, but the second is derived by the same technique.

The point is to again use the short exact sequence
\begin{equation}
 0 \to \bZ \to \bR \to \UU(1) \to 0,
\end{equation}
which gives a long exact sequence of cohomology groups which includes for each $n$,
\begin{multline}\label{eqn:longexactcoeff}
  H^n_G(X,\bR) \to H^n_G(X,\UU(1)) \\\to H^{n+1}_G(X,\bZ) \to H^{n+1}_G(X,\bR).
\end{multline}
The middle map is the Bockstein we're interested in. Now, $H^n_G(X,\bR)$ may be computed by the isotropy spectral sequence of Appendix \ref{appendix:spectralsequences}, with $E_1$ page given by the sum of $\cH^k(G_\sigma,\bR)$ for $n-k$-cells $\sigma$, one for each $G$-orbit, where $G_\sigma$ is the isotropy group of $\sigma$, which is compact by assumption. If $n > \dim X$, then $k > 0$, so we can use \eqnref{eq:Rvanishing} and find that the $E_1$ page vanishes in these degrees. Thus, $H^n_G(X,\bR) = H^{n+1}_G(X,\bR) = 0$ and so the Bockstein is an isomorphism, as claimed.

Finally, let us also note that similar results hold for relative equivariant homology and cohomology, and can be proven by similar techniques.

\section{The K\"unneth splitting for $\Gspace \times G_{int}$}
\label{sec:kunneth}
We consider a symmetry of the form $G = \Gspace \times G_{\mathrm{int}}$, where $G_{\mathrm{int}}$ is a (possibly anti-unitary) internal symmetry.

If $X$ is a space with a $G_{\rm spatial}$ action and a trivial $G_{\rm int}$ action, then the homotopy quotient splits as
\begin{equation}X//G = X//G_{\rm spatial} \times BG_{\rm int}.\end{equation}
This gives us K\"unneth splittings for various cohomology groups we're interested in. For instance, if we take $\UU(1)$ (with possibly $G$ action), we have
\begin{equation}H^k_G(X, \UU(1)) \simeq \bigoplus_{p+q = k} H^p_{G_{\rm spatial}}(X, \cH^q(G_{\rm int},\UU(1))),\end{equation}
with the induced $G_{\rm spatial}$ action on $\cH^q(G_{\rm int},\UU(1))$.

We can apply this to anomalous textures as follows. At a given site $s$, let $G_{\mathrm{spatial,s}} \leq \Gspace$ be the subgroup of $\Gspace$ that leaves $s$ invariant. Then the full isotropy group at $s$ is $G_s = G_{\mathrm{spatial},s} \times G_{\mathrm{int}}$. The data of the anomalous texture at site $s$ is an element of $\cH^2(G_{\mathrm{int}} \times G_{\mathrm{s}},\UU(1))$, which we can expand using the K\"unneth formula as
\begin{align}
\cH^2(G_{\mathrm{int}} &\times G_{\mathrm{s}},\UU(1)) = \nonumber \\
& \cH^2(\Gspace,\UU(1)) \nonumber \\ &\oplus \cH^1(\Gspace, \cH^1(G_{\mathrm{int}},\UU(1))) \nonumber \\& \oplus \cH^2(G_{\mathrm{int}},\UU(1)). \label{sitekunneth}
\end{align}

It is also possible to prove a K\"unneth splitting for equivariant homology from the equivariant chains definition. This is done by noting that an equivalent complex to $C^p(G, C_q(X,M))$ is
\begin{equation} = \bigoplus_{p_1 + p_2 = p}\cC^{p_1}(G_{\rm spatial},C_q(X,C^{p_2}(G_{\rm int},M))),\end{equation}
where now
\begin{equation}\delta = \delta_{\rm spatial} + (-1)^{p_1}\delta_{\rm int},\end{equation}
where the right-hand-side are the differentials of the $G_{\rm int}$ and $G_{\rm spatial}$ parts, separately. We thus obtain a triple complex. It's easy to show that the total cohomology of this triple complex reduces to the shifted sum of cohomologies of double complexes obtained by taking cohomology with respect to $\delta_{\rm int}$ (ie. the spectral sequence degenerates at the 2nd page). The $n$th double complex is
\begin{equation}\cC^{p}(G_{\rm spatial},C_q(X,\cH^{p+n}(G_{\rm int},M))),\end{equation}
which contributes to equivariant homology in degree $p+n-q$. Thus we find
\begin{equation}H_n^G(X, M) = \bigoplus_{k} H_{n+k}^{G_{\rm spatial}}(X, \cH^k(G_{\rm int},M)).\end{equation}

\section{Smooth states and anomaly in-flow}
\label{appendix:smoothstates}
In Refs.~\cite{Thorngren_1612,Else_1809} we introduced the picture of ``smooth states''. The idea is that the space of ground states in $d$ dimensions is assumed to be described by some space $\Theta_d$, which (in the case where we are discussing invertible phases) satisfies the property that the based loop space of $\Theta_d$ is homotopy equivalent to $\Theta_{d-1}$ (that is, $\Theta_\bullet$ forms an $\Omega$-spectrum) \cite{KitaevIPAM, Freed_1604,Xiong_1701,Gaiotto_1712}. A smooth state on a d-dimensional manifold $X$ with a $G$ action is then\footnote{In the case where $G$ contains orientation-reversing or anti-unitary symmetries, this definition needs to be extended to take into account a twist \cite{Else_1809}.
For simplicity, we assume in this section that $G$ does not contain such symmetries, but the arguments can easily be extended to the general case.} a continuous map $X//G \to \Theta_d$, where $X//G = (X \times EG)/G$ is the ``homotopy quotient'' of $X$ by $G$, where $EG$ is a contractible space on which $G$ acts freely. The homotopy classes of such maps defines the ``generalized cohomology'' $h^d(X//G)$.

As we showed in Ref.~\cite{Else_1809}, smooth states are very closely connected to anomaly-free defect networks, in the sense that they have the same classification under an appropriate equivalence relation. If the defect networks are defined in terms of a cell decomposition of $X$, then the argument makes use of the \emph{dual} cell decomposition $X^\vee$, such that a dual $k$-cell corresponds to a $d-k$-cell in the original cell decomposition. Following similar arguments to Ref.~\cite{Else_1809}, we can also see that degree-0 anomalous defect networks corresponds to maps $X_{d-1}^\vee//G \to \Theta_d$, where $X_{d-1}^\vee$ is the $(d-1)$-skeleton of the dual cell decomposition---that is, the union of all the dual $k$-cells for $k \leq d-1$.

Moreover, the arguments of Ref.~\cite{Else_1809} also show that degree-0 anomalous smooth states correspond to maps $
X_{d-1}^{\vee}//G \to \Theta_d$, which are classified by $h^d(X_{d-1}^{\vee}//G)$, whereas an anomalous texture corresponds to a map $X//G \to \Theta_{d+1}$\footnote{$X$ and $X^\vee$ are identical as $G$-spaces; it doesn't matter which we use in computing the cohomology.} that reduces to the trivial map on $X_{d-1}^\vee//G$; homotopy classes of such maps correspond to the \emph{relative} generalized cohomology $h^{d+1}(X//G, X_{d-1}^\vee//G)$. We can then invoke the long exact sequence of a pair (which follows from the axioms of generalized cohomology, or equivalently from the $\Omega$-spectrum property of $\Theta_{\bullet}$), of which a portion looks like
\begin{multline}
\cdots \to h^{d}(X_{d-1}^\vee//G) \xrightarrow{s_1} h^{d+1}(X//G, X_{d-1}^\vee//G) \\ \xrightarrow{s_2} h^{d+1}(X//G) \to \cdots
\label{longexact}
\end{multline}
The arrow $s_1$ is telling us the anomalous texture associated to a degree-0 anomalous smooth state, and the arrow $s_2$ (inclusion), combined with the isomorphism $h^{d+1}(X//G) \cong h^{d+1}([X \times \mathbb{R}]//G)$, is telling us the invertible crystalline topological phase in $d+1$ spatial dimensions which hosts our anomalous texture appears at the boundary. The fact that this sequence is exact is precisely the claim we made in Section \ref{s:def-bulk-boundary1}, namely that an anomalous texture can be cancelled by an invertible-substrate defect network if and only if corresponds to a trivial crystalline topological phase in $d+1$ dimensions.

In general, the correct choice of $\Theta_\bullet$ is not known, although for invertible states in bosonic systems it is conjectured to be the spectrum of oriented cobordism \cite{Kapustin_1403}. However, the results of Section \ref{sec:equivarianthomology} are derived by making the ``ordinary cohomology'' approximation; that is, we approximate $\Theta_d$ by the Eilenberg-MacLane space $K(\mathbb{Z},d+2)$.
 The model of defect networks that results from making this approximation corresponds to what we call the ``in-cohomology'' defect networks. Note that
for phases with internal symmetry $G$, which are classified by homotopy classes of maps $BG \to \Theta_d$, the ordinary cohomology classification gives a classification of $H^{d+2}(BG, \mathbb{Z})$ (where $BG = EG/G$ is the classifying space), which in most cases of interest is isomorphic to the group cohomology with $\mathrm{U}(1)$ coefficients, $\cH^{d+1}(G, \mathrm{U}(1))$ (see Appendix \ref{subsec:coefficients}). This classification is known not to be complete even for bosonic systems \cite{Vishwanath_1209,Wang_1302,Burnell_1302,Kapustin_1403}; for example, it fails to capture the Kitaev $E_8$ state \cite{Kitaev_0506,Lu_1205} in two spatial dimensions, which does not require any symmetry to protect it ($G=1$). Therefore, by making this approximation we are restricting ourself to defect networks that are not built out of such ``beyond (ordinary) cohomology'' components.

Let us now show, however, that if we make this ordinary cohomology approximation, then the results of Section \ref{sec:equivarianthomology} follow. First of all, in the ordinary cohomology approximation, one finds that homotopy classes of maps $X//G \to \Theta_d$ are classified by the ordinary cohomology $H^{d+2}(X//G, \mathbb{Z})$. From a cellular description of $X//G$ one can show that this is equivalent to the equivariant cohomology $H^{d+2}_G(X, \mathbb{Z})$ defined in Appendix \ref{subsec:equivariantcohomology}, which moreover (see Appendix \ref{subsec:coefficients}) is equivalent in the cases of interest to us to $H^{d+1}_G(X, \mathrm{U}(1))$. By Poincar\'e duality (see Appendix \ref{subsec:equivariantcohomology}), this is also equivalent to the equivariant homology $H_{-1}^G(X, \mathrm{U}(1))$. This gives Theorem \ref{thm:firsthomologythm}.

We also find in the ordinary cohomology approximation that  $h^{d+1}(X//G, X^{d-1}//G)$ reduces to the equivariant relative cohomology $H^{d+3}_G(X,X_{d-1}, \mathbb{Z})$ introduced in Appendix \ref{subsec:relativecohomology}. By duality, this is equivalent to equivariant homology $H_{-2}(X_0, \mathbb{Z})$, where $X_0$ is the 0-skeleton of the original cellulation of $X$ (as opposed to the dual one), and the map $s_2$ in \eqnref{longexact} corresponds in homology to the map
\begin{equation}
H_{-2}^G(X_0, \mathbb{Z}) \to H_{-2}^G(X, \mathbb{Z}).
\end{equation}
induced by the inclusion $X_0 \to X$.
Interpreting this map in terms of equivariant chains, and again invoking the results of Appendix \ref{subsec:coefficients},
 we obtain Theorem \ref{thm:secondhomologythm}.

Finally, let us note that we can also repeat the arguments above for $k$-skeletal defect networks, which correspond to smooth states that restrict to the constant map on $X^{d-k-1}//G$. One finds that Theorems \ref{thm:firsthomologythm} and \ref{thm:secondhomologythm}  still hold when $X$ is replaced with the $k$-skeleton $X_k$.

By similar arguments to the above, we can also derive the general statement that in equivariant homology, the boundary anomaly of a defect network on a $d+1$-dimensional space $X$ with a $G$-symmetric boundary is given by boundary map
\begin{equation}\partial:H_{-1}^G(X,\partial X,\UU(1)) \to H_{-2}^G(\partial X,\UU(1)),\end{equation}
which is Poincar\'e-Lefschetz dual to the ordinary restriction map in equivariant cohomology
\begin{equation}i^*:H^{d+2}_G(X,\UU(1)) \to H^{d+2}_G(\partial X,\UU(1)),\end{equation}
where $i:\partial X \to X$ is the inclusion map.

In the case of a half-cylindrical space $X \times \bR_{\ge 0}$, these restriction maps are isomorphisms because $X \times \bR_{\ge 0}$ equivariantly contracts to $X$. That is,
\begin{align}
H_{-2}^G(X, \UU(1)) &\cong H_{-1}^G(X \times \mathbb{R}_{\ge 0}, X, \UU(1)),
\end{align}
which in this case is also isomorphic to
$H_{-1}^G(X \times \bR,\UU(1))$ since we are using Borel-Moore chains.

\section{Spectral sequences}\label{appendix:spectralsequences}

In this appendix, we develop some computational techniques for equivariant homology and cohomology, based on the double complex introduced in Section \ref{subsec:equivarianthomology}:
\begin{equation}Q^{p}_q = C^p(G,\cC_q(X,A)),\end{equation}
\begin{equation}D = \partial + (-1)^{p+q} \delta.\end{equation}
A reference for this is Chapter 7 of \cite{Brown}. We focus on equivariant homology but it is easy to dualize $C_q$ to $C^q$ and work with the double complex with $\delta$ and $d$, obtaining spectral sequences for equivariant cohomology as well.

A spectral sequence is a method for computing the cohomology of $D$ in terms of cohomologies of $\partial$ and $\delta$. Any double complex like $Q$ admits two different spectral sequences, depending on whether we first take cohomology with respect to $\delta$ or $\partial$ first. In the first case, we get
\begin{equation}E_1^{p,q} = \ker \delta/{\rm im\ } \delta = \cH^p(G,C_q(X,A)).\end{equation}
In the second
\begin{equation}\tilde E_1^{p,q} = \ker \partial/{\rm im\ } \partial = \cC^p(G,H_q(X,A)).\end{equation}
The former can be simplified by splitting the coefficients along the orbits $[\sigma]$ of $q$-cells of $X$ under the $G$-action, with isotropy group $G_\sigma$, and using Shapiro's lemma of Appendix \ref{appendix:anomtexts}. We have
\begin{equation}E_1^{p,q} = \bigoplus_{[\sigma] \in X_q/G} \cH^p(G_\sigma,A).\end{equation}
For this reason we call the spectral sequence beginning with $E_1$ the \emph{isotropy spectral sequence}. The one beginning with $\tilde E_1$ we will refer to as the \emph{Serre spectral sequence} since it is Poincar\'e dual to the spectral sequence of the fibration $X \to X//G \to BG$. The isotropy spectral sequence is more convenient for simplifying calculations in contractible space $X = \bR^d$, since in the other case already at the 1st ``page" we have $\tilde E_1^{p,d} = \cC^p(G,A)$, all others zero. Thus we will mostly focus on defining the differentials for the isotropy spectral sequence.

Let $[\omega_0] \in E_1^{p,q}$, ie. $\omega_0 \in C^p(G,C_q(X,A))$ and $\delta \omega_0 = 0$. As in Section \ref{subsec:equivarianthomology} we define $r = q-p$. $\omega_0$ does not necessarily define an equivariant cycle, since
\begin{equation}D\omega_0 = \partial \omega_0 + (-1)^r\delta \omega_0 = \partial \omega_0.\end{equation}
As we have described in Section \ref{subsec:equivarianthomology}, all equivariant cycles can be written as a sum
\begin{equation}\omega_0 - \omega_1 + \cdots\end{equation}
with
\begin{equation}\omega_j \in \cC^{p-j}(G,C_{q-j}(X,A)).\end{equation}
The idea of the spectral sequence is to iteratively compute the $\omega_j$'s to construct an equivariant cycle from $\omega_0$.

That is, we want to find an $\omega_1 \in \cC^{p-1}(G,C_{q-1}(X,A))$ such that
\begin{equation}\label{1stdiffcons}
  D(\omega_0 - \omega_1) \in \cC^{p-2}(G,C_{q-1}(X,A))
\end{equation}
To have \eqref{1stdiffcons} we need
\begin{equation}\delta \omega_1 = \partial \omega_0.\end{equation}
Since $\partial \delta = \delta \partial$, $\delta D\omega_0 = 0$, we get a class
\begin{equation}[D\omega_0] \in E_1^{p+1,q},\end{equation}
and since $E_1^{p+1,q}$ is the cohomology with respect to $\delta$, we see that the existence of $\omega_1$ satisfying \eqref{1stdiffcons} is equivalent to
\begin{equation}[D\omega_0] = 0.\end{equation}
This motivates the definition of a map known as the \emph{1st differential} of the spectral sequence:
\begin{equation}d_1:E_1^{p,q} \to E_1^{p+1,q},\end{equation}
given by $[D\omega_0] = [\partial\omega_0]$, noting that $D\delta = \delta D$ so we get a well-defined map in cohomology, ie. $[D\omega_0]$ only depends on $[\omega_0]$. Likewise we define
\begin{equation}\tilde d_1:\tilde E_1^{p,q} \to \tilde E^{p,q-1}\end{equation}
\begin{equation}\tilde d_1 [\tilde \omega_0] = [\delta \tilde\omega_0].\end{equation}
Observe the $\tilde E_1$ differentials go the other direction.

Summarizing so far, an element of $E_1$ has a chance to grow to become an equivariant cycle iff it has vanishing $d_1$. On the other hand, it could be that
\begin{equation}\omega_0 = \partial \omega_{-1}\end{equation}
for some $\omega_{-1} \in E_1^{p,q+1}$, ie. $d_1 \omega_{-1} = \omega_0$. In this case $\omega_0$ gives rise to an equivariant boundary. So if we are interested only in the equivariant homology class that $\omega_0$ potentially gives rise to, we are interested in its cohomology class with respect to $d_1$. The same is true for the isotropy spectral sequence.

This motivates the \emph{2nd page} of the spectral sequence by
\begin{equation}E_2^{p,q} = \ker d_1/{\rm im\ } d_1\end{equation}
and likewise for $\tilde E_2^{p,q}$. These are equivalence classes of $\omega_0$ for which there exists an $\omega_1$ satisfying \eqref{1stdiffcons}. We find
\begin{equation}\tilde E_2^{p,q} = \cH^p(G,H_q(X,A)),\end{equation}
which matches the Serre spectral sequence of $X \to X//G \to BG$, as claimed. On the other hand the 2nd page of the isotropy group cannot be phrased in this simple form, although it can be phrased using sheaf cohomology.

To continue the spectral sequence, we observe
\begin{equation}D(\omega_0 - \omega_1) = \partial \omega_1\end{equation}
and we want to find an $\omega_2$ with
\begin{equation}\delta \omega_2 = \partial \omega_1\end{equation}
so that
\begin{equation}D(\omega_0 - \omega_1 + \omega_2) \in \cC^{p-3}(G,C_{q-2}(X,A)).\end{equation}
We see we can relax this slightly to finding an $-\omega_1' + \omega_2$ with
\begin{equation}D(-\omega_1' + \omega_2) = \partial \omega_1.\end{equation}
Then we can form
\begin{equation}\omega_0 - \omega_1 - \omega_1' + \omega_2.\end{equation}
We can find such a pair $(\omega_1',\omega_2)$ iff
\begin{equation}[\partial \omega_1] = 0 \in E_2^{p-2,q-1}.\end{equation}
Note that $\partial \omega_1$ satisfies \eqref{1stdiffcons} by
\begin{equation}D (\partial \omega_1 + 0) = \partial \delta \omega_1 = \delta^2 \omega_0 = 0,\end{equation}
so it indeed defines an element of $E_2^{p-2,q-1}$. This motivates the definition of the \emph{2nd differential}
\begin{equation}d_2:E_2^{p,q} \to E_2^{p-2,q-1},\end{equation}
which one easily checks is well-defined, in particular it does not depend on the choice of $\omega_1$. Furthermore, $d_2^2 = 0$. Likewise, if there is a pair whose 2nd differential is $\omega_0$, then $\omega_0$ defines a nullhomologous equivariant cycle.

Therefore, pairs $[\omega_0,\omega_1]$ for which an $\omega_2$ extension exists which has the possibility to define a nonzero equivariant homology class are given by the cohomology of $d_2$, which defines the 3rd page
\begin{equation}E_3^{p,q} = \ker d_2/{\rm im\ } d_2.\end{equation}
One continues this way to define all the (infinitely many) pages of the spectral sequence. At the $n$th page the elements are equivalence classes of $n$-tuples $[\omega_0,\omega_1,\ldots,\omega_{n-1}]$ such that there exists an $\omega_n \in \cC^{p-n}(G,C_{q-n}(X,A))$ with
\begin{equation}\label{eqndiffcons}
  D(\omega_0 - \omega_1 + \cdots - (-1)^n \omega_{n-1}) = \delta \omega_n.
\end{equation}
The $n$th differential is then given by
\begin{equation}d_n[\omega_0,\ldots,\omega_{n-1}] = [\partial \omega_n],\end{equation}
where $\partial \omega_n$ may be extended to an $n$-tuple satisfying \eqref{eqndiffcons} by padding with zeros, since by construction $D\partial \omega_n = 0$. This also makes it clear $d_n^2 = 0$.

For any given $\omega_0 \in \cC^{p}(G,C_q(X,A))$, the spectral sequence stops having ``outgoing" differentials, that is, $d_k$'s coming from $\omega_0$ after $q$ steps, since $C_{<0}(X,A) = 0$ eventually there are no more $\omega_j$'s to construct. Likewise for a $d$-dimensional space $X$, $C_{>d}(X,A)$ so there are also no ``incoming" differentials after $d$ steps. Thus we say the spectral sequence converges absolutely.

\subsection*{The descent sequence}

Let us discuss an application from this point of view, which is the computation of the LSM anomaly associated to an anomalous texture. An anomalous texture is described by $\omega_0 \in \cZ^2(G,C_0(\bR^d,\UU(1)))$ which we identify as an element on the $E_1$ page of the isotropy spectral sequence. We observe $D\omega_0 = 0$, so $\omega_0$ defines an element of $H_{-2}^G(\bR^d,\UU(1))$ which we would like to identify as zero or nonzero.

Equivalently, all of the differentials $d_{>1}$ take $\omega_0$ out of the domain of the spectral sequence, since $C_{<0}(\bR^d,\UU(1)) = 0$, and we would like to see if there is a differential with $\omega_0$ in its image. The general method we describe has to be performed at each page of the spectral sequence in the computation of the cohomology of $d_n$. That is, we have described above how to compute $\ker d_n$ but now we must describe how to compute ${\rm im\ }d_n$. One can in principle do this page-by-page, but there is a more convenient iterative method we refer to as the \emph{descent sequence} which in the case of a nonzero LSM anomaly will moreover give us the isomorphism $H_{-2}^G(\bR^d,\UU(1)) = \cH^{d+2}(G,\UU(1))$.

First, $\omega_0$ is in the image of $d_1$ if we can find an $\eta_0 \in \cC^2(G,C_1(\bR^d,\UU(1)))$ with $\delta \eta_0 = 0$, $\partial \eta_0 = \omega_0$. In this case, the anomalous texture is trivial in lattice homotopy, described in Section \ref{subsec:latticehomotopy}.

Next, $\omega_0$ is in the image of $d_2$ if there is a pair $\eta_0 \in \cC^2(G,C_1(\bR^d,\UU(1)))$, $\eta_1 \in \cC^3(G,C_2(\bR^d,\UU(1)))$ with
\begin{equation}\partial \eta_0 = \omega_0\end{equation}
\begin{equation}\partial \eta_1 = \delta \eta_0\end{equation}
\begin{equation}\delta \eta_1 = 0.\end{equation}
We either find a solution or this process continues until we cannot find a $\partial \eta_{n+1} = \delta \eta_{n}$, meaning that $\delta \eta_{n}$ is nontrivial in homology. For $X = \bR^d$ this only happens for $n = d$ in which case we find
\[\delta \eta_d \in \cZ^{d+2}(G,\cZ_d(\bR^d,\UU(1))).\]
By Poincar\'e duality, or simply by looking at the value on a single $d$-cell of $\bR^d$, we obtain a group cocycle
\[\alpha \in \cZ^{d+2}(G,\UU(1)),\]
which by construction defines an equivalent class in equivariant homology as $\omega_0$, being related by
\[D(\eta_0 - \eta_1 + \cdots + (-1)^d \eta_d).\]
Also note that by shifting $\eta_d$ by a cycle (which is precisely its ambiguity) we shift $\alpha$ by a group coboundary, so we have thus obtained a map
\[\cH^2(G,Z_0(\bR^d,\UU(1))) \to \cH^{d+2}(G,\UU(1))\]
which is the equivariant pushforward described in Section \ref{subsec:relationship}. In Appendices \ref{a:classic-lsm} and \ref{a:point-group} we compute this map explicitly using this technique.

 \section{Computing the LSM criterion}
 \label{appendix:computation}
In this Appendix, we will give some more details on how the LSM criterion for quantum magnets can be checked on a computer. The idea is to work in terms of the equivariant chains introduced in Section \ref{sec:equivarianthomology}. An equivariant chain gives a ``traditional'' LSM theorem (i.e.\ it guarantees non-invertible ground state), when it defines trivial element in the equivariant homology $H_{-2}^G(X, U(1))$. For computational purposes it is more convenient  to work in terms of $H_{-3}^G(X, \mathbb{Z})$, which is equivalent as shown in Appendix \ref{subsec:coefficients}.

Observe that when $X = \mathbb{R}^d$ and the symmetry includes a space group, then both the cell decomposition of $X$ and the group are infinite. So we need a way to convert the problem into one that only involves a finite amount of data. To do this, we can use the fact that there if $G$ contains translations as a normal subgroup (as does any space group), $\mathbb{Z}^d \leq G$, then there is an isomorphism

\begin{equation}
H_{\bullet}^G(\mathbb{R}^d, \mathbb{Z}) \cong
H_{\bullet}^{G_{\mathrm{pt}}}(\mathbb{T}^d, \mathbb{Z}),
\end{equation}
where $G_{\mathrm{pt}} = G/\mathbb{Z}^d$. and $\mathbb{T}^d = \mathbb{R}^d/\mathbb{Z}^d$ is the $d$-torus.
This is the equivariant homology version of the ``rolling and unrolling'' principle discussed in \cite{Thorngren_1612}. Therefore, in order to obtain a finite problem we can work in terms of anomalous textures on the torus with respect to $G_{\mathrm{pt}}$.

Once expressed in this way, determining whether a given equivariant chain with $\mathbb{Z}$ coefficients is exact reduces to finite-dimensional sparse linear algebra over the ring of integers, which can be solved using the routines contained in the software packages Sage \cite{Sage} or Magma \cite{Magma}. We will now discuss various techniques for making this computation more efficient.

Firstly, we exploit the K\"unneth decomposition discussed in Section \ref{subsec:eqhomkunneth} whenever possible. In particular, the results of the exhaustive computational search discussed in Section \ref{sec:exhaustive} (specifically, cases \ref{nosoc} and \ref{soc_tr}) come from considering the $k=2$ factor in \eqnref{tophomology_kunneth} from Section \ref{subsec:eqhomkunneth}, which always gives $H_{-2}^{\Gspace}(X, \mathbb{Z}_2)$ in these cases. In Case 1, there is no need to consider the other factors at all, since the assumed form of the anomalous texture ensures that it will never map into any non-trivial element of these factors. For Case 2, however, even if there is no traditional LSM coming from the $k=2$ factor, one might think that there could be a traditional LSM coming from the other two factors. However, it turns out that this can never happen in Case 2 either. The reason is that the symmetry representation of a spin-orbit coupled quantum paramagnet can always be lifted to a representation of $O(3)  \times \Gspace$. The anomalous texture of this enlarged symmetry group is non-trivial only in the third factor of \eqnref{sitekunneth} (which is identical in form to the third factor for the original symmetry group, hence gives a traditional LSM if and only it did for the original symmetry group). But if there is an invertible ground state which is symmetric under the enlarged symmetry group, then we can simply perturb it by adding a small spin-orbit coupling to the Hamiltonian, giving an admissible ground state for the original symmetry group.

Next, we observe that if all the $d$-cells in a $d$-dimensional space $X$ acted upon by a group $G$ have $G_\Sigma = 0$ (as is generally the case when $G$ is a pure space group), there is an important simplification, as follows. Let $\beta \in C_r^G(X, A)$ be an equivariant $r$-chain whose only nonzero component is $\beta_0 \in C^{-r}(G, C_0(X,A))$ (as is always the case when we want $\beta$ to represent an anomalous texture). Let $\lambda$ be an equivariant $r+1$-chain $\lambda \in C_{r+1}^G(X, A)$ such that $D \lambda = \beta$, and let $\lambda_d$ be its component in $C^{d-r-1}(G, C_d(X,A))$. The fact that $\beta$ has only a $\beta_0$ component implies that $\delta \lambda_d = 0$.

By Shapiro's Lemma (see Appendix \ref{appendix:anomtexts}; we choose $S$ there to be a set of labels for the $d$-cells of $X$), this implies that we can write $\lambda_d = \delta \mu_d$ for some $\mu_d \in C^{d-r-2}(G, C_d(X, A))$. Defining an equivariant $r+2$-chain $\mu$ with $d$-th component $\mu_d$ (and the rest zero), we find that $\lambda - D\mu$ has component $(\lambda - D\mu)_d = 0$. Moreover, $D(\lambda - D\mu) = D \lambda = \beta$ (since $D^2 = 0$). It follows that, without loss of generality, in looking for equivariant 1-chains which trivialize $\beta$ such that $D \lambda = \beta$, we are free to restrict ourself to $\lambda$ having component $\lambda_d = 0$. From the perspective of the isotropy spectral sequence of Appendix \ref{appendix:spectralsequences}, this works because $E_1^{d+2,d} = 0$.

 A final speed-up can be obtained if we recall that the definition of group cochains and group coboundaries in Section \ref{subsec:groupcohomology} is only one of many possible equivalent definitions \cite{Brown}. In general, for any group $G$, let
\begin{equation}
\label{resolution}
\cdots \to F_n \to F_{n-1} \to \cdots \to F_0 \to \mathbb{Z} \to 0
\end{equation}
be a free resolution of $\mathbb{Z}$ treated as a module over the group ring $\mathbb{Z}[G]$ (that is, an exact sequence of free $\mathbb{Z}[G]$ modules). Then we can define a group $n$-cochain of $G$ with coefficients in $M$ (where $M$ is an abelian group equipped with compatible $G$-action, or equivalently a $\mathbb{Z}[G]$ module) to be an element of $\mathrm{Hom}(F_n, M)$, that is, the group of $\mathbb{Z}[G]$-module homomorphisms from $F_n$ to $M$. The coboundary operator on group cochain is induced from the boundary maps in \eqnref{resolution}. The ``standard'' definition of group cochains and coboundary operator given in Section \ref{subsec:groupcohomology} comes from a particular choice  resolution known as the ``bar resolution''. Any two free resolutions of $\mathbb{Z}$ are chain homotopic, and as such the group cohomology $\cH^{\bullet}(G,M)$ and the equivariant homology $H_{\bullet}^G( X, M)$ are isomorphic regardless of the choice of free resolution $F_\bullet$. The bar resolution is not particularly suited for computations, because the dimension of $F_n$ grows like $|G|^n$.
 Therefore, in our computations, we used a routine (which is based on the algorithm described in Ref.~\cite{Ellis__04}) from the HAP package \cite{Hap} of the GAP computer algebra system \cite{Gap} to produce more efficient resolutions.

\section{Descent sequence for the classic LSM theorem}\label{a:classic-lsm}

In this appendix we use the descent sequence of Section \ref{subsec:relationship} to illustrate the computation of the LSM anomaly associated with the classical setting of a lattice of projective internal-symmetry representations.

Let us consider $X = \bR$ with a cell decomposition with vertices at the integer coordinates and edges on the open intervals between them. Let $G = G_{int} \times \bZ$ have an internal component $G_{int}$ (which could be a Lie group) and a translation component $\bZ$. Furthermore, let each site be endowed with the same projective representation $V$ of $G_{int}$, with cocycle $\alpha \in \cZ^2(G_{int},\bR/\bZ^T)$, where $\bR/\bZ^T$ denotes twisting by anti-unitary elements of $G_{int}$. Let us suppose also that translations act by simply permuting the site spaces.

This situation is described by an anomalous texture
\begin{equation}\sum_{s\in \Lambda} \omega_s(g_1,g_2)[s] = \sum_{s \in \Lambda}e^{2\pi i\alpha(\bar g_1, \bar g_2)}[s] \in \cZ^2(G,Z_0(X,\UU(1)^T)),\end{equation}
which we will write additively:
\begin{equation}\log \omega := \sum_{j \in \bZ} \alpha(\bar g_1,\bar g_2) [j] \in \cZ^2(G,Z_0(X,\bR/\bZ^T)),\end{equation}
where $[j]$ denotes the 0-chain defined by the vertex at coordinate $j$ and $\bar g$ denote the quotient map $G \to G_{int}$. We have $\partial \log \omega = 0$ and $\delta \log \omega = 0$.

We wish to construct from $\omega_s$ a cocycle in $\cZ^3(G,\UU(1)^T)$ which captures the class of the 2+1D crystalline SPT phase for which this anomalous texture forms a symmetric boundary condition. This is the setup of the classic LSM theorem in one space dimension (originally with $G_{int} = SO(3)$ and $\alpha$ corresponding to a half-integer spin representation $V$), and the non-triviality of the associated 2+1D phase captures the ground state constraint of the LSM theorem.

We define
\begin{equation}\lambda_1(g_1,g_2) = \sum_{j \in \bZ} j\alpha(\bar g_1,\bar g_2)[j,j+1] \in \cC^2(G,C_1(X,\bR/\bZ^T)),\end{equation}
where $[j,j+1]$ denotes the 1-chain of the oriented edge $j \to j+1$. $\lambda_1$ is constructed so that
\begin{equation}\partial \lambda_1 = \log \omega.\end{equation}
We can think of $\lambda_1$ as an anomalous defect network with a $G_{int}$ SPT $j [\alpha] \in \cH^2(G_{int},\bR/\bZ^T)$ along each edge $[j,j+1]$. From this it follows
\begin{equation}\log \omega - (\partial + \delta)\lambda_1 = - \delta \lambda_1 \in \cZ^3(G,Z_1(X,\bR/\bZ^T)).\end{equation}
It remains to compute this cocycle.

We do this directly from the defining equation \eqref{} for $\delta$:
\begin{equation}(\delta \lambda_1)(g_1,g_2,g_3) = g_1 \cdot \lambda_1(g_2,g_3) - \lambda_1(g_1 g_2,g_3)\end{equation}\begin{equation} + \lambda_1(g_1,g_2 g_3) - \lambda_1(g_1,g_2).\end{equation}
First, observe that if $g_1$ is internal, that is, it does not include any translation component and so acts trivially on $C_1(X,\bZ)$, then
\begin{equation}(\delta \lambda_1)(g_1,g_2,g_3) = \sum_{j \in \bZ}j (\delta \alpha)(\bar g_1,\bar g_2,\bar g_3) [j,j+1] = 0,\end{equation}
by virtue of $\delta \alpha = 0$. On the other hand, if $g_1$ involves a translation by $l$, then we have $g_1 [j,j+1] = [j+l,j+l+1]$, and so
\begin{equation}(\delta \lambda_1)(g_1,g_2,g_3) = \sum_{j \in \bZ} (j-l) (-1)^{p(g_1)}\alpha(\bar g_2, \bar g_3)[j,j+1]\end{equation}
\begin{equation}+ j(- \alpha(\bar g_1 \bar g_2, \bar g_3) + \alpha(\bar g_1,\bar g_2 \bar g_3) - \alpha(\bar g_1,\bar g_2))[j,j+1]\end{equation}
\begin{equation}= \sum_{j \in \bZ} - l \alpha(\bar g_2,\bar g_3)[j,j+1],\end{equation}
where we have used $\delta \alpha = 0$ again, and $p(g)$ is 0 mod 2 if $g$ is unitary and 1 mod 2 if $g$ is anti-unitary. Observe $\partial \delta \lambda_1 = 0$, as expected. If we write $\tau \in Z^1(G,\bZ)$ for the 1-cocycle where $\tau(g) = l$, the number of unit translations in the symmetry element $g$, then we have
\begin{equation}\label{e:clsmstep1}
  -\delta\lambda_1 = (\tau \cup \alpha) \sum_{j \in \bZ}[j,j+1],
\end{equation}
where we have used the cup product
\begin{equation}(\tau \cup \alpha)(g_1,g_2,g_3) = \tau(g_1)\alpha(\bar g_2,\bar g_3) \in \cZ^3(G,\bR/\bZ^T).\end{equation}
The class $[\tau \cup \alpha] \in \cH^3(G,\bR/\bZ^T)$ captures the LSM anomaly of this anomalous texture. As expected, this is precisely the form of the topological response for a stack of 1+1D $G_{int}$ SPTs classified by $\alpha$ (with projective representation $V$ at one boundary), as identified by \cite{Thorngren_1612}.

Let us show the equivalence between the equivariant homology class of the anomalous texture $[\omega_s]$ and the group cohomology class $[\tau \cup \alpha]$. Suppose there is a 2-cochain $\eta \in \cC^2(G,\bR/\bZ^T)$ with $\delta \eta = \tau \cup \alpha$. Then we may consider
\begin{equation}\tilde \eta = \sum_{j \in \bZ} \eta(g_1,g_2) [j,j+1] \in \cC^2(G,\bR/\bZ^T),\end{equation}
which satisfies $\partial \tilde \eta = 0$, $\delta \tilde \eta = - \delta \lambda_1$. It follows
\begin{equation}\log \omega = (\partial + \delta)(\lambda_1 - \tilde \eta),\end{equation}
so $\log \omega$ is trivial in equivariant homology. Conversely, if $(\partial + \delta)\rho = \log \omega$, then
\begin{equation}(\partial + \delta)(\rho - \lambda_1) = -\delta \lambda_1 = (\tau \cup \alpha) \sum_{j \in \bZ}[j,j+1].\end{equation}
However, $C_2(X,\bZ) = 0$, since $X$ is one-dimensional, so $\partial(\rho - \lambda_1) = 0$, from which it follows that
\begin{equation}\rho - \lambda_1 = \eta \sum_{j \in \bZ} [j,j+1]\end{equation}
with $ \eta \in \cC^2(G,\bR/\bZ^T)$ with
\begin{equation}(\tau \cup \alpha) \sum_{j \in \bZ}[j,j+1] = (\partial + \delta)(\lambda_1 - \rho) = (\delta \eta) \sum_{j\in \bZ}[j,j+1],\end{equation}
hence $\delta  \eta = \tau \cup \alpha$.

We end this section noting that this discussion extends straightforwardly to higher dimensions. Indeed, let's consider $X = \bR^2$ with a cell decomposition adapted to $\Lambda$ where the 2-cells occupy a whole unit cell. We first construct $\lambda_1 \in \cC^2(G,C_1(X,\bR/\bZ^T))$ as above, except only using the first coordinate of $X$:
\begin{equation}\lambda_1(g_1,g_2) = \sum_{j,k \in \bZ} j \alpha(\bar g_1,\bar g_2)[(j,k),(j+1,k)],\end{equation}
where the lattice coordinates are written in an integer basis as $(j,k)$, and $[(j,k),(j+1,k)]$ is the 1-chain corresponding to the oriented edge from $(j,k)$ to $(j+1,k)$. Then, defining $\tau_1(g) \in \cZ^1(G,\bZ)$ to be the number of unit translations $g$ does along the first coordinate, we find
\begin{equation}\log \omega - (\partial + \delta)\lambda_1 = \sum_{j,k \in \bZ} (\tau_1 \cup \alpha) [(j,k),(j+1,k)].\end{equation}
Then we define
\begin{equation}\lambda_2(g_1,g_2,g_3) = \sum_{j,k \in \bZ} k(\tau_1 \cup \alpha)(g_1,g_2,g_3)\square(j,k),\end{equation}
where $\square(j,k)$ is the 2-chain of the unit cell with $(j,k)$ in the lower left corner. $\lambda_2$ is constructed so
\begin{equation}\log \omega - (\partial + \delta)\lambda_1 = \partial \lambda_2.\end{equation}
We find by a computation analogous to the above that
\begin{equation}\log \omega - (\partial + \delta)(\lambda_1 + \lambda_2) = (\tau_2 \cup \tau_1 \cup \alpha)\sum_{j,k \in \bZ}\square(j,k),\end{equation}
where $\tau_2(g) \in \bZ^1(G,\bZ)$ is defined like $\tau_1$ as the number of unit translations $g$ does along the second coordinate. Thus the LSM anomaly is $\tau_2 \cup \tau_1 \cup \alpha \in \cH^4(G,\bR/\bZ^T)$. This captures the topological response of a 3+1D array of 1+1D SPTs labelled by $\alpha$, as identified by \cite{Thorngren_1612}. In general dimensions we will find the LSM anomaly $\tau_d \cup \ldots \cup \tau_1 \cup \alpha$. This class is nontrivial iff $\alpha$ is, so we reproduce the classic LSM constraint.

\section{Descent sequence for point-group LSM theorems}\label{a:point-group}

In this appendix we use the descent sequence of Section \ref{subsec:relationship} to compute the LSM anomaly associated to point groups pinning a projective internal symmetry representation.

We consider $X = \bR^d$ with a point-like anomalous texture at the origin. We take our symmetry group $G$ to act by linear orthogonal transformations $G \to O(d)$, some of which may be internal. The anomalous texture is thus captured by $\omega_0 \in \cH^2(G,\UU(1))$. As in Appendix \ref{a:classic-lsm}, the resulting LSM anomaly will be of the form
\begin{equation}\label{e:pointgrouplsm}
LSM(\omega_0) = e(X) \cup \omega_0 \in \cH^{d+2}(G,\UU(1)^{\rm or}),
\end{equation}
where
\begin{equation}e(X) \in \cH^d(G,\bZ^{\rm or})\end{equation}
is a special class associated to any linear representation $G \to O(d)$ called the Euler class. The twisting indicates that orientation-reversing elements of $G$ negate $\bZ$.

To construct the Euler class, we let $G \to O(d)$ define an $\bR^d$-vector bundle $E$ over the classifying space $BG$. We then iteratively construct a generic section of this vector bundle $s_k \to E$ over each $k$-skeleton of $BG$ up to $k = d$.

In the first step, this section over the 0-skeleton assigns a point $s_0(\star) \neq 0 \in \bR^d$ to the basepoint of $BG$. Then we proceed to the 1-skeleton, on which it assigns a path $s_1(g)$ from $s_0(\star)$ to $g \cdot s_0(\star)$ to each $g \in G$. If $d = 1$, then this path will generically cross $0 \in \bR$ on some edges. We count this crossing with a sign according to some local orientation of $E$ and it defines a cocycle $e(E,s) \in \cZ^1(BG,\bZ^{\rm or})$.

If $d > 1$, this path does not generically cross $0$ and we continue. Always in the stage of extending the section over the $d$-skeleton we encounter some unavoidable zeros, and count them with a local orientation of $E$ to obtain the Euler class $e(E,s) \in \cZ^d(BG,\bZ^{\rm or})$. It can be shown that this class is independent of the chosen section.

We will show how this construction is implemented in real space by the descent sequence. This will prove the formula \eqref{e:pointgrouplsm}.

We assume for simplicity that the cell complex of $X$ is composed of open cones with the origin at the tip. It is always possible to find such an equivariant cell complex. Then, by intersecting the unit sphere with this cell complex we obtain a cellulation of $S^{d-1} \in \bR^d$. The sections $s_k$ we construct will be cellular maps into this unit sphere.

The first step in the spectral sequence is to choose a ray $r_1$ from the origin and place $\omega_0$ on it to form a $-1$-chain
\begin{equation}c_1(g_1,g_2) = \omega_0(g_1,g_2) [r_1] \in C^2(G,C_1(X,\UU(1))),\end{equation}
where $[r_1] \in C_1(X,\bZ)$ is the 1-chain associated to $r_1$ with its orientation pointing out of the origin. By construction
\begin{equation}\partial c_1 = \omega_0 [0].\end{equation}
We associate to this step in the construction a section $s_0:BG_0 \to S^{d-1}_0$ (subscript denotes the $0$-skeleton) by sending the basepoint $\star \in BG$ to the intersection of $r_1$ with the unit sphere.

The next step is to study $\delta c_1$. We find, using the cocycle condition of $\omega_0$,
\begin{equation}\delta c_1(g_1,g_2,g_3) = \omega_0(g_2,g_3)(g_1 \cdot [r_1] - [r_1]).\end{equation}
If we are in 1d, we are finished, and we see that
\begin{equation}g_1 \cdot [r_1] - [r_1] = R(g_1) [l_1],\end{equation}
where $R(g_1) = 1$ if $g_1$ acts as a reflection and zero otherwise. On the other hand, we can interpret $(g_1 \cdot [r_1] - [r_1]) \cap B^1$ (restricting to the unit interval) as a path from $s_0(\star)$ to $g_1 \cdot s_0(\star)$. We extend our a section $s_1:BG_1 \to B^1_1$ using this path. We find that this section vanishes along an edge $[g] \in BG_1$ iff $R(g) = 1$. Thus, $R(g)$ is the Euler class of this representation and we deduce \eqref{e:pointgrouplsm} for $d = 1$.

Let's suppose $d > 1$. In the case $g_1 \cdot [r_1] - [r_1] \neq 0$, they are linearly independent and we can choose a sector of the plane spanned by $g_1 \cdot r_1$ and $r_1$. Decorating this sector with $\omega(g_2,g_3)$ defines a -1-chain $c_2(g_1,g_2,g_3) \in  C^3(G,C_2(X,\UU(1)))$ with
\begin{equation}\partial c_2 = \delta c_1,\end{equation}
in accordance with the descent sequence. We use this define a section $s_1:BG_1 \to S^{d-1}_1$ by intersecting the chosen sector for $c_2(g,-,-)$ with the unit sphere. Clearly this section is nonvanishing and extends $s_0$.

We continue likewise in this way, now studying
\begin{equation}\delta c_2(g_1,g_2,g_3,g_4) = [f(g_1,g_2)] \omega(g_3,g_4),\end{equation}
where $[f(g_1,g_2)] \in C^2(G,C_2(X,\UU(1)))$. Either these fill space, in which case we are done (and we find the Euler class of $s_1$) or they are boundaries of certain polyhedral cones which define $c_3(g_1,g_2,g_3,g_4)$. The intersections of these cones with the unit sphere define the next section, $s_2$ in the series. Always, once we reach the dimension of space, we find \eqref{e:pointgrouplsm}.

\bibliography{ref-autobib,ref-manual}

\end{document}